\documentclass[fleqn,useAMS,usenatbib]{mn2e}

\usepackage{amssymb}
\usepackage{amsmath}
\usepackage{times}
\usepackage{graphicx}
\usepackage{wasysym}
\usepackage{natbib}
\usepackage{graphicx}
\usepackage{psfrag}

\usepackage[utf8]{inputenc}

\DeclareFontFamily{U}{gbsn5d}{}
\DeclareFontFamily{U}{gbsn66}{}
\DeclareFontShape{U}{gbsn5d}{m}{n}{<-> gbsnu5d}{}
\DeclareFontShape{U}{gbsn66}{m}{n}{<-> gbsnu66}{}
\def\aap{A\&A}
\def\apj{ApJ}

\def\apjl{ApJ}
\def\pasa{PASA}
\def\mnras{MNRAS}
\def\araa{ARA\&A}

\def\nat{Nat}

\def\apjs{ApJS}
\def\pasp{PASP}

\newcommand{\mincir}{\raise
-2.truept\hbox{\rlap{\hbox{$\sim$}}\raise5.truept 
\hbox{$<$}\ }}
\newcommand{\magcir}{\raise
-2.truept\hbox{\rlap{\hbox{$\sim$}}\raise5.truept
\hbox{$>$}\ }}
\newcommand{\minmag}{\raise-2.truept\hbox{\rlap{\hbox{$<$}}\raise
6.truept\hbox
{$>$}\ }}

\newcommand{\be}{\begin{equation}}
\newcommand{\ee}{\end{equation}}
\newcommand{\ba}{\begin{eqnarray}}
\newcommand{\ea}{\end{eqnarray}}
\newcommand{\brr}{\begin{array}}
 
\newcommand{\err}{\end{array}}
\newcommand{\bc}{\begin{center}}
\newcommand{\ec}{\end{center}}

\DeclareMathAlphabet{\mathsc}{OT1}{cmr}{m}{sc}
\def\testbx{bx}%
\DeclareRobustCommand{\ion}[2]{%
\relax\ifmmode
\ifx\testbx\f@series
{\mathbf{#1\,\mathsc{#2}}}\else
{\mathrm{#1\,\mathsc{#2}}}\fi
\else\textup{#1\,{\mdseries\textsc{#2}}}%
\fi}

\title[{\rm GAME I}]{Gamma Analytical Modeling Evolution (GAME) I: The physical implications of deriving the stellar mass functions from z=0 to z=8. 
}

\author[A. Katsianis et al.]
{Antonios Katsianis$^{1}$\thanks{E-mail:kata@sysu.edu.cn,}
Qingshan Wang$^{1}$,
Xiaohu Yang$^{2, 3, 4}$,  
Xian Zhong Zheng$^{4, 5}$,
\and
Pedro Cataldi$^{6}$,
Nicola Napolitano$^{7, 8}$,
Weishan Zhu$^{1}$,
Nicolas Tejos$^{9}$,
\and
Weiguang Cui$^{10, 11, 12}$,
Cheng Li$^{13}$,
Weipeng Lin$^{1}$,
Long-long Feng$^{1}$,
\and 
Junde Li$^{2, 3}$,
Ying Tang$^{1}$,
Yuchang Li$^{1}$,
Hangxin Pu$^{1}$ \\
\\
$^{1}$School of Physics and Astronomy, Sun Yat-sen University, Zhuhai Campus, 2 Daxue Road, Xiangzhou District, Zhuhai, China \\
$^{2}$Department of Astronomy, Shanghai Key Laboratory for Particle Physics and Cosmology, Shanghai Jiao Tong University, \\ 800 Dongchuan RD. Minhang
District, Shanghai 200240, China \\
$^{3}$Key Laboratory for Particle Physics, Astrophysics and Cosmology, Ministry of Education, Shanghai Jiao Tong University, \\ 800 Dongchuan RD. Minhang
District, Shanghai 200240, China \\
$^{4}$Tsung-Dao Lee Institute and Key Laboratory for Particle Physics, Astrophysics and Cosmology, Ministry of Education, \\ 
Shanghai Jiao Tong University, Shanghai 200240, China \\
$^{5}$ Purple Mountain Observatory, Chinese Academy of Sciences, Nanjing 210023, China  \\
$^{6}$ Instituto de Astronomía y F\'{i}sica del Espacio, CONICET-UBA, Casilla de Correos 67, Suc. 28, 1428, Buenos Aires, Argentina \\
$^{7}$Department of Physics E. Pancini, University Federico II, Via Cinthia 21, I-80126, Naples, Italy\\
$^{8}$INAF – Osservatorio Astronomico di Capodimonte, Salita Moiariello 16, I-80131 Napoli, Italy\\
$^{9}$Instituto de F\'{i}sica, Pontificia Universidad Cat\'{o}lica de Valpara\'{i}so, Casilla 4059, Valpara\'{ı}so, Chile \\
$^{10}$Departamento de F\'{i}sica Te\'{o}rica, M-8, Universidad Aut\'{o}noma de Madrid, Cantoblanco E-28049, Madrid, Spain \\
$^{11}$Centro de Investigaci\'{o}n Avanzada en F\'{i}sica Fundamental (CIAFF), Universidad Aut\'{o}noma de Madrid, Cantoblanco, E-28049 Madrid, Spain \\
$^{12}$Institute for Astronomy, University of Edinburgh, Royal Observatory, Edinburgh EH9 3HJ, United Kingdom \\
$^{13}$Department of Astronomy, Tsinghua University, Beijing 100084, People’s Republic of China}

\begin{document}

\maketitle

\begin{abstract}
The $\Gamma$ growth model is an effective parameterization employed across various scientific disciplines and scales to depict growth. It has been demonstrated that the cosmic star formation rate density (CSFRD) can also be described broadly by this pattern, i.e. $\frac{dM(T)}{dT} =  M_{z,0}\, \times  \frac{\beta^{\alpha}}{\Gamma(\alpha)} \,  T^{\alpha-1}  e^{-\beta \, T }$ M$_{\odot}$ Gyr$^{-1}$, where $M_{z,0}$ is the stellar mass at $z$ = 0, $\alpha = 3.0$, $\beta = 0.5 $ Gyr$^{-1}$ and $T$ describes time. We use the identical $\Gamma$ growth pattern given by the CSFRD to extend the present day (z = 0) stellar mass bins $M_{\ast}(T)$ of the Galaxy Stellar Mass Function (GSMF) and investigate if we are able to reproduce observations for the high redshift GSMFs. Surprisingly, our scheme describes  successfully the evolution of the GSMF over 13.5 Gyrs, especially for objects with intermediate and low masses. We observe some  deviations that manifest {\it solely} at very high redshifts ($z > 1.5$, i.e. more than 9.5 Gyr ago) and {\it specifically} for very small and exceedingly massive objects. We discuss the possible solutions (e.g. impacts of mergers) for these offsets. Our formalism suggests that the evolution of the GSMF is set by simple (few parameters) and physically motivated arguments. The parameters $\beta$ and $\alpha$ are theoretically consistent within a multi-scale context and are determined from the dynamical time scale ($\beta$) and the radial distribution of the accreting matter ($\alpha$). We demonstrate that both our formalism and state-of-the-art simulations are consistent with recent GSMFs derived from JWST data at high redshifts.
\end{abstract}

\begin{keywords}
cosmology: theory -- galaxies: formation -- galaxies: evolution
\end{keywords}

\section{Introduction}
\label{sec_intro}
%https://fastercapital.com/topics/application-of-gamma-distribution-in-real-life.html/1

Analytical growth models and probability density functions are commonly used in numerous fields of science, from biology \citep[Population growth of Bacteria]{Alonso2014}, medicine \citep[carcinogenic events]{Belikov2017}, economy \citep[Food demand]{Mori2015}, epidemiology \citep[Spread of infectious Diseases]{Vazquez2021,ZiffandZiff}, Pharmakokinetics \citep[concentration of drugs in blood]{Weiss1987,Wesolowski2022} and engineering \citep[failures of a system]{Maghsoodloo2014} to astronomy \citep{1976ApJ...203..297S,Diemer2017,Katsianis2021b}. The aim of these models is to succinctly and accurately characterize the average growth of population numbers, sizes, and masses with minimal parameters when many uncertain factors are involved and it is particularly difficult to track individually the members of the population \citep{Sameh2011,Brilhante2023}. Among these models the $\Gamma$ growth pattern is one of the most interesting \citep{Singer2020}. This  growth motif, used on multiple scales and scenarios, can be expressed mathematically as follows: a combination of a power-law growth ($T^{a-1}$) at early times of evolution accompanied by an exponential decrease at later times ($e^{-\beta \times T}$). Power-law growth is commonly seen when accretion / merger / accumulation of close-by resources occurs, with $\alpha$ being determined by the density of close-by resources \citep{MUKHOPADHYAY2022481}. The exponential decrease reflects a growth-dependent saturation \citep{Ameli2015,Berthelot2011,MANDAL2021104621} or/and a growth-dependent consumption of the available resources \citep{SEMKOW2009415,Wachtmeister2017}. Finally, the normalization is simply mathematically set ($\Gamma$ function) by the final value after the growth has stopped (V$_{final}$ $\times  \frac{\beta^{\alpha}}{\Gamma(\alpha)}$). Putting all the processes described above together, we obtain the rate of growth at time $T$ as: 
\begin{eqnarray}
\label{eq:PGgastconsumptionrateFactorial}
\frac{{\rm d \, V}(T)}{dT} =  {\rm V}_{final}\, \times  \frac{\beta^{\alpha}}{\Gamma(\alpha)} \,  T^{\alpha-1}  e^{-\beta \, T } \,. 
\end{eqnarray} 

 %Besides its simplicity this pattern has been employed to describe broadly  ... as it embodies the main ..... giving a better understanding for the timescales and Physics involved  like $\beta$ (usually the timescale related to the consumption of the available resources or generally speaking the growth dependent deterioration) and $\alpha$ (set usually by accretion, merging or networks) of a particular system.

In our Universe, the growth of a galaxy is driven by the accretion of the surrounding material or/and merging with other nearby galaxies. Usually due to the non-linearity nature of the collapse, the complex hierarchical assembly, and the numerous uncertain physical processes that shape galaxies and their contents \citep{Lopez2020,Yang2021,Xu2023,Schiebelbein-Zwack2024,Lu2024},  the growth of galaxies/dark matter halos has been followed by using sophisticated cosmological simulations and semi-analytic models \citep{Jiang2019wang,Wang2020abc,Fong2022,Crain2023,Mutch2024}. This computational approach led to a large variety of successes including realistic galaxy stellar mass functions, star formation rate functions, gas phase metallicities and galaxy scaling relations \citep{Katsianis2016,Blanc2019,Pfeffer2023,Cannarozzo2023,Lohmann2023,Tang2024,Qiao2024}. However, further comparative studies and analyses of cutting-edge cosmological simulations, such as EAGLE, IllustrisTNG, and Simba \citep{Schaye2015,Pillepich2018,dave2019} have also highlighted the limitations of this methodology \citep{Katsianis2021b,CorchoCab2021,Habouzit2022,Weaver2023,Picouet2023,Kanehisa2024,Eisert2024,Lewis2024}. Some big drawbacks in hydrodynamical simulations are: 
\begin{itemize}
\item  The subgrid recipes used are empirical and do not usually arise from a well defined physical justification or ab initio theory \citep{Katsianis2021b,Yongseok2023}. In addition, there is no connection between how the small scales (which are surely not resolved in cosmological simulations) are connected to the macroscopic scales \citep{Rosen2014,Crain2023}.
 \item The subgrid schemes and the physics implied by hydrodynamic simulations  are quite different among different research groups. This alone demonstrates the uncertainty of our current understanding for galaxy formation and evolution.  The continuous re-calibration of the numerous parameters needed to describe star formation, SNe feedback/winds and AGN feedback after the emergence of updated observations (ongoing for 2 decades) is very concerning, as it underlines that we are not having yet a predictive formalism but rather different paradigms with many uncertainties that need to be constantly updated via further re-tuning and imposing of additional mechanisms, e.g. galactic winds and AGN feedback (that introduce even more parameters). 
 \item  Even if there is a consensus on some results (such as stellar mass functions produced by various collaborations) from different models developed by different groups, since these results are derived using significantly different physics, they encompass numerous parameters and substantial complexity, there is no assurance as to which model is ``correct" or has reached the observable results for the ``right physical reasons" (i.e., there is a lack of Philosophical or Physical basis to determine the correctness of these solutions).
 \item  Severe resolution effects are demonstrated. It is {\it guaranteed} that the parameters of the subgrid models are not known a priori or driven by first principles, and their numerical values are resolution dependent. Some ``successes" occur only at the particular resolution at which the parameters were calibrated \citep{Zhao2020,Katsianis2021,Pakmor2023,Zeng2024}.
 \item There have been detailed comparisons for a large list of properties of galaxies that have not been reproduced by state-of-the-art cosmological simulations. We note that not all models fail equally on the following quantities/relations/distributions but a brief list of notable failures involve  morphologies \citep{Haslbauer2022,Kartaltepe2023},  HI contents in galaxies \citep{Li2022}, quanched fractions at low \citep{Meng2023} and high redshifts \citep{Weller2024}, specific star formation rate functions at z = 0 \citep{Katsianis2021}, the AGN luminosity function \citep{Habouzit2022} and the black hole – halo mass relation \citep{Voit2024}. 
\end{itemize} 
We stress that besides the above shortcomings, cosmological simulations are valuable tools for exploring different prescriptions and their influence simultaneously on various properties of galaxies \citep{Katsianis2014,Pillepich2018,Esteban2023,Lovell2023,Kulier2023,Huang2023,Schare2024}. It is also important to keep in mind that after developing a strong theoretical driven backbone, further development via cosmological simulations or semi-analytical models is very useful for several research areas, including the creation of mock observations \citep{Baes2020,Lagos2020,Trcka2022,Fortuni2023,Yizhou2024,Bottrell2024}.

Leaving simulations, and returning to more analytical approaches, there have been different attempts to model the growth of galaxies \citep{Lapo2017} or to describe their average star formation histories via various simple empirical parameterizations like the double power law \citep{Behroozi2013} and the log-normal \citep{Gladders2013,Abramson2016}. However, according to \citet{Ciesla2017}, these patterns do not properly model the early cosmic SFH while the slope of the declining part is too steep at lower redshifts. Interestingly, upon considering limitations of star formation indicators\footnote{It has been demonstrated that UV, IR, H$\alpha$, SED fitting, radio and X-ray SFR indicators produce different results. Compiling the results obtained with different methodologies without considering this limitation can heavily impact the inferred shape of both the SFR-$M_{\ast}$ relation and the CSFRD \citep{TescariKaW2013,Katsianis2015,Davies2019,Leja2020,Trcka2020,Fishbach2023,Kouroumpatzakis2023,Qiao2024,Das2024}}, the observed cosmic star formation rate density (CSFRD) can be broadly described by a simple $\Gamma$/factorial growth pattern as follows \citep{Katsianis2021b,Katsianis2023} :
\begin{eqnarray}
\label{eq:PGgastconsumptionrate}
\frac{dM(T)}{dT} =  M_{z0}\, \times  \frac{\beta^{\alpha}}{\Gamma(\alpha)} \,  T^{\alpha-1}  e^{-\beta \, T } M_{\odot} \,  {\rm Gyr^{-1}},  
\end{eqnarray}
and its integral,
\begin{eqnarray}
\label{eq:2}
{M(T)} =  M_{z0} \, \times \frac{\gamma(\alpha, \beta T)}{\Gamma(\alpha)} \, M_{\odot},
\end{eqnarray}
where $M(T)$ is the mass at time $T$ (in Gyrs), $M_{z0}$ is the mass at $z = 0$,  $\alpha = 3.0$,  $\beta = 0.5 $ Gyr$^{-1}$ and $T$ is the average age of galaxies (equal to the age of the Universe minus the age of the Universe when the first stars emerged at $\sim$ 0.17 Gyr). The reason for starting the growth of galaxies at $0.17$ Gyr after the birth of the Universe is because the first stars are not expected to emerge at T = 0 (i.e. immediately after the big bang) as the conditions for star formation and galaxy formation are not yet met (Dark matter halos are not massive enough to host galaxies at the early Universe and the temperature of the available gas has to decrease in order for star formation to begin).  We adopt a lag phase\footnote{We borrow this term from Pharmacokinetics. The absorption of the injected drug to the bloodstream of a patient does not start immediately after the injection and the concentration of the medicine in the blood can be written as \citep{Weiss1987}: $C(t) = A \, (t - T_{Lag})^k \, e^{-b \, (t-T_{Lag})}$, which resembles a $\Gamma$ growth pattern being delayed by $T_{Lag}$.}  (delay) for star formation to commence with respect the big bang of $T_{Lag} = 0.17$ Gyr consistent with the $\Lambda$CDM framework \citep{zheng2012,Oesch2018,Villanueva2018,Laporte2021,Haslbauer2023,Katsianis2023}. The above description is just a typical $\Gamma$ motif (Eq.  \ref{eq:PGgastconsumptionrateFactorial}), it commences an effort to connect different scales/fields of science and describes the growth of the cosmic budget of stellar mass and cosmic mass accretion history in a clear way / few parameters. Due to its simplicity, multi-disciplinary applicability and straightforward physics, it is of interest to validate if this pattern can describe additional observables in extragalactic astrophysics and at which extent galaxies with different masses deviate from this motif. In addition, it is important to gain further insights of the timescales and basic principles that govern galaxy growth via studies that focus on the $\alpha$ and $\beta$ parameters. Finally, considering the uncertainties of the subgrid Physics employed by cosmological simulations (related to Star formation/feedback) and their overwhelming resolution limitations, analytical modeling can provide a complementary approach in order  to understand galaxy formation and evolution.

In this work, we set a very simple formalism to describe galaxy growth employing a minimum number of parameters. This paper is organized as follows. In section \ref{GAME0Ms0} we discuss the importance of the z = 0 GSMF and discuss the results of different authors. In Section \ref{GAME0} we discuss the backbone of our model, i.e., the $\Gamma$ growth pattern. Our objective is to describe the evolution of the GSMF from $z = 0$ to $z = 8$. Like most theoretical/numerical motivated efforts to describe galaxy evolution, we start our effort from the stellar mass function/mass bins at redshift zero (subsection \ref{GAME0Ms0}). We then employ Equation \ref{eq:2} to track the growth history of the stellar mass bins at previous redshifts/eras.  We demonstrate the remarkable {\it} success of this simple scheme by comparing these results with both observations and state-of-the-art simulations (Section \ref{Comparison0}). In Section \ref{Why?}, we illustrate that the principles of accretion and gravitational collapse naturally lead to a $\Gamma$ growth pattern.  The parameters of our model are derived within a multi-scale context making our scheme physically motivated. In Section \ref{GAMEba} we present an improved version of our model that is able to reproduce even better the observations by slightly tuning only the parameter $\beta$. In Section \ref{sec_concl} we present our conclusions.  In appendix \ref{comparisonEmpirical} we compare further the predictions of our formalism with the empirical models of UniverseMachine and \citet{Leja2020}. In appendix \ref{Quantify} we quantify the success of each model considered in this work with respect our compliation of observations using as metrics the Root Mean Square Distance (RMSD) and the Akaike information criterion (AIC) .  RMSD evaluates the precision of each model to reproduce the observed GSMF while the AIC evaluates the quality of the model taking into account the number of free parameters employed by the model. Our results suggest that galaxy stellar growth is mostly governed by straightforward Physics and can be described with a minimal number of parameters. The growth of the stellar mass of galaxies can be seen as a conventional $\Gamma$  growth model, meaning it comprehensively involves a growth rate that relies on resource consumption, fatigue buildup, and the early aggregation of nearby materials. In our work, we adopt a \citet{chabrier03} Initial Mass Function (IMF) and \citet{Planck2018} cosmology.

\begin{figure*}
\centering
\includegraphics[scale=0.55]{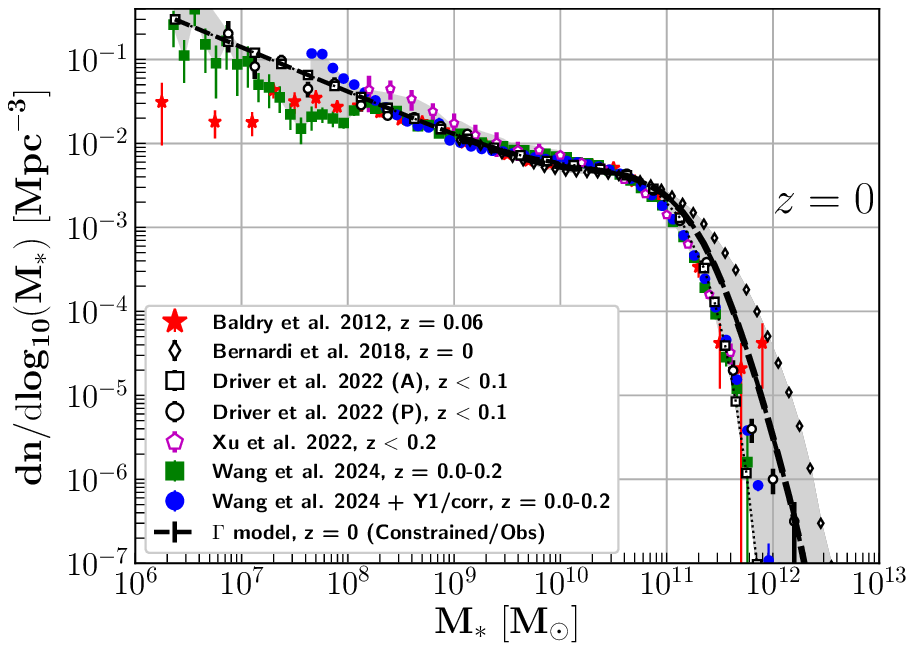}
\includegraphics[scale=0.55]{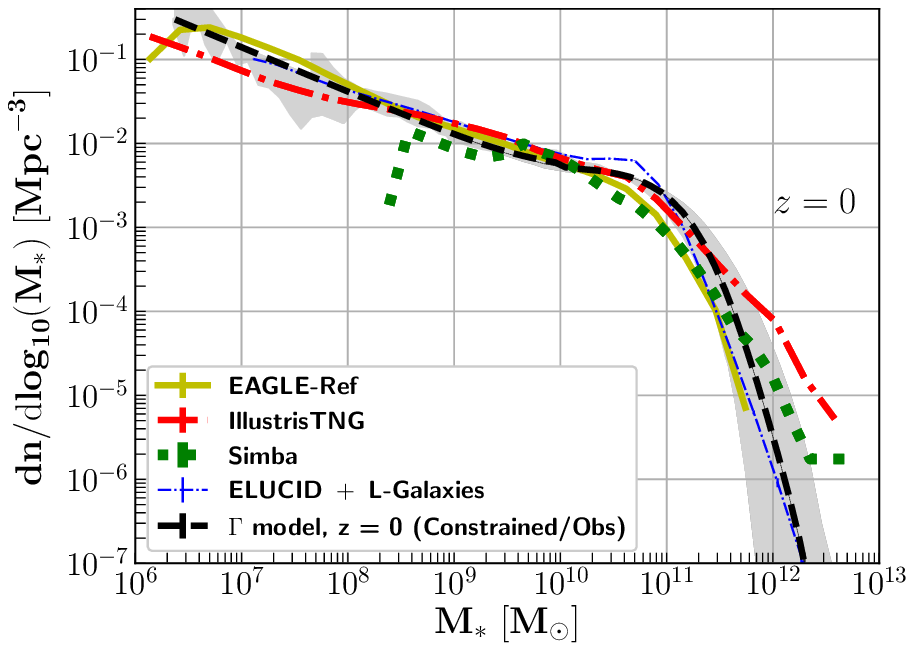}
\vspace{-1.0cm}
\caption{Left: The stellar mass functions from \citet{Baldry2012}, \citet{Driver2022}, \citet{Bernardi2018} and \citet{Wang2024}. The gray area represents the uncertainty of the observations at z = 0 and thus the uncertainty of our formalism. According to the updated observations the results of \citet{Baldry2012} were unable to capture successfully the low mass end. Right: The stellar mass function at z = 0 in the state-of-the-art models of TNG (red dot-dashed line), EAGLE  (orange solid line) Elucid+L-galaxies (blue dashed line) and Simba (green dotted line). Our reference GSMF (black dashed line), used to constrain our model, is in broad agreement with these results.}
\label{GSMF0obs0}
\end{figure*}

\section{The galaxy stellar mass function at $z = 0$}
\label{GAME0Ms0}

In order to describe the uncertain physical mechanisms and to constrain the large number of free parameters related to them, it is a common practice to tune theoretical models of galaxy evolution with respect to key observations. For example, the stellar mass function at $z = 0$ \citep{Bell2003,Yang2009,Weinberg2017,Pillepich2018} has been commonly used to constrain the free parameters in cosmological simulations (e.g.  33 free parameters only for the SF, AGN and SNe feedback for IllustrisTNG). Without this tuning, simulations tend to not align with the observed stellar mass function and cosmic star formation rate density. In our work,  similarly with comsological simulations we also choose {\it the GSMF at $z = 0$} as the key observable to be reproduced.

Thus, we start with the GSMF of \citet{Baldry2012} represented by the red stars in Fig.  \ref{GSMF0obs0}. The authors used an area of 143 deg$^2$ from the Galaxy And Mass Assembly (GAMA) survey. The GSMF is determined from a sample of 5210 galaxies, while the results are described by a double Schechter ($\Gamma$) functions. We also employ the GSMF of \citet{Bernardi2018} described as black diamonds in Fig. \ref{GSMF0obs0}. The authors started with a Chabrier IMF but used spatially resolved kinematics of nearby galaxies to argue that the IMF may be more bottom heavy for galaxies with large velocity dispersion. The authors adopted a correction for the GSMF that affected the high-mass end of the distribution and estimated a dynamical corrected stellar mass function for a substantial sample of $6 \times 10^5$ SDSS galaxies.  Furthermore, we include the \citet{Driver2022} GSMF represented by the dotted black line (analytical form, Driver et al. (A)) and black circles (stepwise determination, Driver et al. (P)) in Fig. \ref{GSMF0obs0}. The authors used GAMA DR4, combined with previous surveys, and this resulted in 330542 galaxies across five sky regions covering $\sim 250$ deg$^2$.  \citet{Driver2022} used a double Schechter function to describe the data following \citet{Baldry2012}.  In addition, we include the observations of \citet{Xu2022}, represented by the magenta pentagons. The study of the authors allows access to lower stellar masses and argues that a notable increase in smaller galaxies at $z = 0.1$ is present with respect previous studies. Finally, we include in our analysis the results of \citet{Wang2024}. The authors present the GSMF using spectroscopic observations from the Data Release 9 (DESI LS DR9), Survey Validation 3 (SV3), and Year 1 (Y1) data sets. This involves a very large number of 138,315,649 galaxies from redshift 0 to $z_{max} = 0.6$. The above observations are represented by the green squares (SV3 dataset) and the blue circles (Y1) in Fig. \ref{GSMF0obs0}. \citet{Wang2024} noted as well that the slope of the GSMF at the low mass end is steeper than in previous studies\footnote{\citet{Chen2019} proposed a method based on the conditional stellar mass functions in dark matter halos, which provided an unbiased estimate of the global GSMF using SDSS. In agreement with \citet{Wang2024} the authors also show that the GSMF has a significant upturn at the low mass end, something that has been missed in many earlier measurements of the $z = 0$ GSMF.} once corrections due to photometric redshift errors and local void effects are taken into account (blue points of Fig. \ref{GSMF0obs0}).

We note that all these efforts to determine the GSMF at $z = 0$ have some discrepancies, especially at the high and low mass ends. This is due to the fact that the chosen SED fitting code, assumed SFH, number of galaxies available, and adopted metallicity impact the results \citep{Trayford2020,Wang2022,Liq2022}. The above shortcomings are even more prominent at high redshifts and for other quantities such as SFRs \citep{Leja2020,Katsianis2020}. Additional effects from cosmic variance and completeness at the high- and low-mass ends also have an imprint. Generally, uncertainties of a factor of 2 - 3 (0.3-0.5 dex) are quite common at z = 0 but become even larger at high redshifts. Thus, we decide to use the average of these observed present day GSMFs described by the black dashed line described in Fig. \ref{GSMF0obs0} and display the spread of all these observations using the gray-shaded region to grasp the uncertainty. 

The above choice of the GSMF that we pick as our constraint is in good agreement with the simulated results obtained from the EAGLE simulations (represented by the solid orange line) and the ELUCID+L-galaxies semi-analytic model (represented by blue dotted-dashed line), both presented in the right panel of Fig. \ref{GSMF0obs0}. Thus, our starting point can be considered to be very similar to the constraints that other state-of-the-art models have adopted. It is also very similar to the commonly adopted constraint from \citet{Baldry2012} GSMF, with the advantage that it addresses limitations at high masses $M_{\ast,0} > 10^{12} M_{\odot}$ while it does not suffer significantly from incompleteness for objects with $M_{\star} <  10^{8} M_{\odot} $. We also note that EAGLE has lower values for the GSMF at the knee of the distribution ($M_{\ast} \sim  10^{11} M_{\odot}$) a problem already noted in \citet{McAlpine2016} and confirmed against the updated observations. This limitations is found for the Simba simulations as well. The knee of these observations is better captured by IllustrisTNG. However, the latter tends to have more massive galaxies, a problem that could be alleviated by considering stars only within a radius of 30 kpc \citep{Pillepich2018} or defining galaxies by using surface brightness limit segmentation procedure to mimic observations \citep{Tang2021,Tang2023}.

\section{The GAME$_0$ model}
\label{GAME0}

\subsection{Going to higher redshfts/earlier eras with the GAME$_0$ model}
\label{GAME0Msparam}

%In this subsection we describe our simple formalism. In section \ref{GAME0vs} we demonstrate its success with respect observations, while in section \ref{Why?} we explain in detail why the form {\it and parameters} reflect the Physics of gravitational collapse. 
Beginning with a mass bin of the stellar mass function at $z=0$, $M_{\ast,z0}$, we aim to employ the basic $\Gamma$ growth model, identical to the CSFRD (in form and parameters), and estimate the mass value of this mass bin at higher redshifts/earlier eras. Thus, in our formalism we adopt that all the factors that contribute to galaxy growth aggregate to follow a $\Gamma$  growth pattern of the form \citep{Katsianis2021b,Katsianis2023}: 
%\xy{I would suggest to use the form of Eq. 2 to describe the evolution of galaxies, so that the passive evolution effect can be incorporated into the equation. See Eq. 14 in Yang et al. 2013}:
%
\begin{eqnarray}
\label{eq:MAHmodel}
\begin{split}
{M_{\ast}(T)} &=&  M_{\ast,final} \, \times \frac{\gamma(\alpha, \beta \times T)}{\Gamma(\alpha)}\\
&=&  \frac{M_{\ast, z0}}{f_{0}} \, \times \frac{\gamma(\alpha, \beta \times T)}{\Gamma(\alpha)},
\end{split}
\end{eqnarray}
where $M_{\ast,z0}$ is the stellar mass at $z=0$,  $\alpha = 3.0$,  $\beta = 0.5 $ Gyr$^{-1}$,  $T$ describes time in Gyr after the birth of galaxies, and $f_{0}$ is the fraction of the stellar mass $M_{\ast, final}$ achieved at $T => \infty$. We adopt $T$ equal to the age of the Universe minus 0.17 Gyr, i.e. ($T$ = Age $-$ 0.17) following \citet{Katsianis2023}, while we can adopt that on average most of the stellar mass in galaxies (95 $\%$) has already been built at redshift $z=0$ (i.e. $f_{0} \sim 1$) as suggested by the studies of \citet{Sobral2013}, \citet{Man2018} and \citet{Katsianis2023}.  For example, the mass of a galaxy with present  (z = 0) stellar mass  $M_{\star} =  10^{10} \, M_{\odot}$, at z =4, when the age of the Universe is T = 1.53 Gyr is calculated as: 
\begin{eqnarray}
\label{eq:MAHmodelexa}
\begin{split}
{M_{\ast}(1.53 \, {\rm Gyr})} &=&  10^{10} \times \frac{\gamma(3.0, 0.5 \times (1.53-0.17) )}{\Gamma(3.0)} \, M_{\odot} \\ 
&=& 319883185 ~~~ = ~~~ 10^{8.50} \, M_{\odot}\,.
\end{split} 
\end{eqnarray}

Since we do not change the values of the $\alpha$ and the $\beta$ parameters (which we remind to the reader have a very strong physical motivation described in Section \ref{Why?}) with respect the cosmic mass accretion history and cosmic star formation rate density (0 parameters re-tuned) or add any additional mechanisms to particularly replicate an observable related to galaxies in the above scheme we label this model as Gamma Analytical Modeling Evolution 0 i.e. GAME$_{0}$. We remind we do not tune our results to reproduce any other high redshift GSMF.  Is this simple model going to describe the high redshift bins of the GSMF at $z = 0-8$ ?

It is important to make clear that this methodology is anticipated to just provide an approximation of the mean mass growth trajectory of galaxies within each mass bin.
\begin{itemize}
\item The approach followed in GAME$_{0}$ maintains the same $\alpha$, $\beta$ and $T$ for all mass bins regardless of mass. This is a simplification, yet it is crucial to assess when and for which mass bins this simplification becomes inadequate. The parameters $\alpha$ and $\beta$ could change among different mass bins since galaxies of different masses can follow different evolutionary paths (e.g. Mergers, tidal interactions, or environmental quenching). In addition, Important mechanisms (outlined at section \ref{GAME1vsSimulations}) for galaxy formation and evolution should change the parameters $\alpha$ and $\beta$ at different redshifts. Using distinct values for the $\beta$ parameter at different mass bins/redshifts is detailed in Section \ref{GAMEba}.
\item galaxies at z = 0, especially the most massive ones, are anticipated to have other smaller progenitors. The numbers of present day (z = 0) galaxies at a mass bin can separate into galaxies with lower masses at high redshifts \citep{Kaviraj2009,Williams2014}. This factor could result in underestimating/overestimating the low mass/high mass end of the predicted GSMF from GAME$_{0}$ at high redshifts.
\item We remind the reader that the mass growth of a central galaxy consists of many different additive components that aggregate to create the final growth evolution \citep{Yang2012,Yang2013}: (1) in situ star formation from accreted gas, (2) the accretion of satellite galaxies (cannibalism), (3) the mass loss due to the evolution of stars and (4) major mergers. While the evolution of satellite galaxies is driven by additional mechanisms, such as dynamical friction, tidal stripping, ram pressure stripping, and harassment \citep{Vadenbosch2008,Vera-Ciro2012,Kanehisa2024}. 
\end{itemize}
$\Gamma$ growths and distributions are a natural outcome that occurs when many random factors aggregate/contribute to create the final growth trajectory and this is one of the main reasons $\Gamma$ growth patterns emerge so often in nature beyond Astronomy \citep{Katsianis2021b,Vazquez2021}. In GAME$_{0}$ we do not isolate/separate the different contributions to stellar growth we just employ the aggregated $\Gamma$ pattern suggested by the CSFRD. If our approach provides results consistent with high redshift observations of the GSMF without considering or particularly isolating all the above factors of growth it means that either these contributions aggregate with each other to create the $\Gamma$ pattern or they are not individually heavily impacting the average evolution of the bins of the GSMF.

\subsection{Comparing GAME$_0$ predictions with observations and simulations}
\label{Comparison0}

\begin{figure*}
\centering
\includegraphics[scale=0.55]{SFRF000SAMstar.ps}
\vspace{-1.0cm}
\includegraphics[scale=0.55]{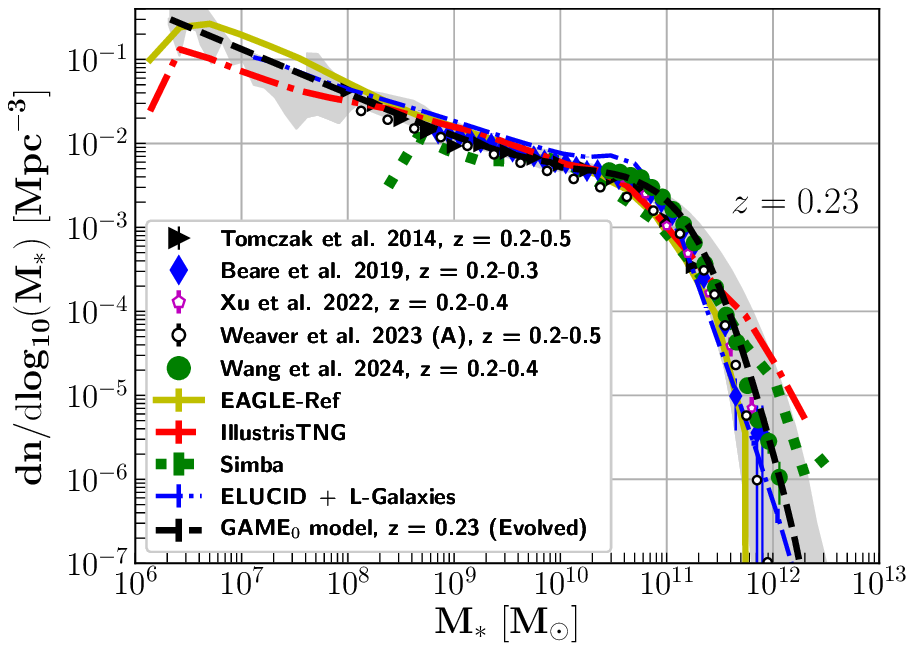}
\includegraphics[scale=0.55]{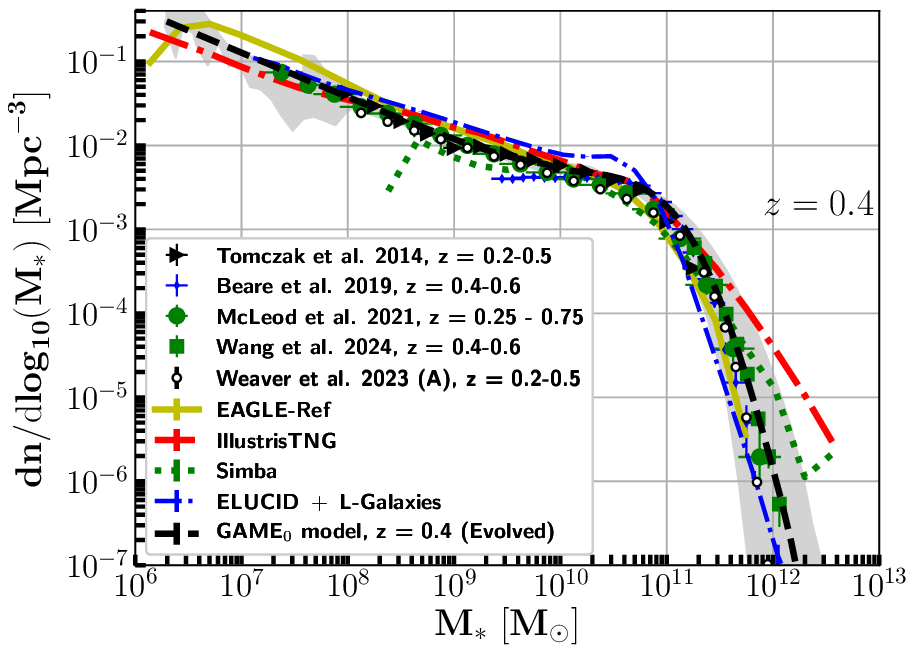}
\includegraphics[scale=0.55]{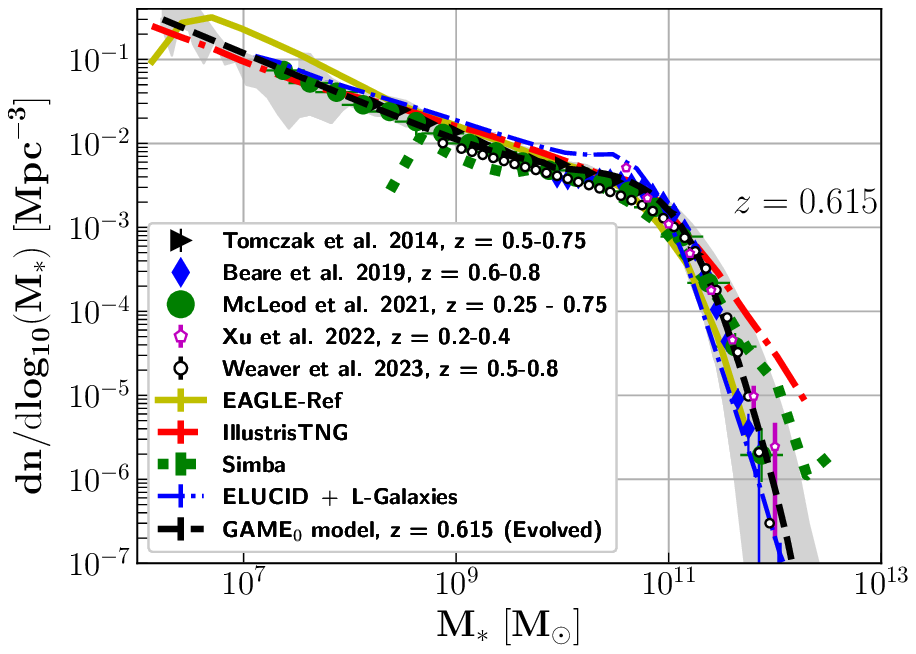}
\vspace{-1.0cm}
\caption{Comparison of the GAME$_{0}$ model (black dashed line) and observations up to T = 7.8 Gyr, z = 0.615 (6 Gyr ago). The comparison with respect observations complements the simplicity of our formalism. We consider that the number of free parameters for the model is 0 since the $\alpha$ and $\beta$ values are well motivated both by theory (subsection \ref{Why?}) and observations (CSFRD). GAME$_{0}$ is the start / backbone for our family of models. We provide comparison with IllustrisTNG (red dot-dashed line), EAGLE  (orange solid line) Elucid+L-galaxies (blue dashed line) and Simba (green dotted line).}
\label{GSMF0obs1}
\end{figure*}

In Fig. \ref{GSMF0obs1} we present the evolution of the GSMF predicted by the GAME$_{0}$ model (represented by the black dashed line) from redshift $z = 0$ (present day) to $z = 0.615$ (6 Gyr ago, represented by the black dashed line).  Alongside we present the results from the EAGLE simulations (represented by the solid orange line), the ELUCID+L-galaxies semi-analytic model (represented by the blue dotted-dashed line), the IllustrisTNG simulations (represented by the red dot-dashed line), and the Simba simulations (represented by the green dotted line). 

We have to note that at redshift $z = 0$ (top left panel of Fig. \ref{GSMF0obs1}), different models have produced somewhat different results, but these are within the uncertainties of the observations. This discrepancy among different state-of-the-art models has also been pointed out in the work of \citet{Weaver2023} and is more severe at higher redshifts and other distributions of properties of galaxies such as the star formation rate function \citep{Picouet2023}.

In the top right panel of Fig. \ref{GSMF0obs1} we compare GAME$_{0}$ with the observations of \citet{Tomczak2013}, \citet{Beare2019}, \citet{Xu2022}, \citet{Weaver2023} and \citet{Wang2024} at redshift 0.23 (2.85 Gyr ago). There is excellent agreement between GAME$_{0}$ and the observations. At redshifts 0.4 (bottom left panel, 4.4 Gyr ago) and 0.615 (bottom right panel, 6 Gyr ago), we compare our results with the observations of \citet{Tomczak2013}, \citet{Beare2019}, and \citet{McLeod2021}. Once again a good agreement persists between GAME$_{0}$ and the observations. It seems that within 6 Gyr (45$\%$ of the age of the Universe), the evolution of the GSMF can be captured very well by our simple formalism. We note that our simple assumption for the different mass bins to evolve similarly and have the same $\alpha$ and $\beta$ does not actually break at $z = 0 - 0.615$. Notably, across all redshift bins, the GSMFs exhibit an approximately uniform evolution for nearly half the Universe's age, adhering to a straightforward $\Gamma$  growth model (equation \ref{eq:MAHmodel}). Indicative of this success is the excellent RMSD with respect observations at this redshift range detailed in appendix \ref{Quantify}.

The comparison with simulations is also encouraging, but also intriguing.  In each panel we present the evolution of the gray region originating from the uncertainty of  the GSMF $z = 0$  (discussed in section \ref{GAME0Ms0}). This region reflects the uncertainty (originating from the uncertainties of the observed z = 0 GSMF) inherent in our GAME$_{0}$ extrapolation.  Simulations from different groups reproduce the results within this area of uncertainty. Our results are in good agreement with EAGLE (orange solid line), but perform better at the knee of the distribution. The high mass ends are well reproduced with respect to observations, while we note that there are high values of the GSMF for the IllustrisTNG (red dotted-dashed line) and Simba (green dotted line) simulations (similar results are obtained by \citet{Weaver2023}). Finally, we see that the GSMF from the ELUCID+L-Galaxies  configuration (blue dashed line) does not perform well at the knee of the distribution. Overall, the different simulations presented in this work via their different and complicated / uncertain / resolution dependent subgrid prescriptions (see section \ref{GAME1vsSimulations} for more details on the severe limitations and achievements of cosmological simulations) and exotic feedback mechanisms (which add even more complexity with the high number of parameters that are required to describe them) {\it acting differently at different mass bins} are also able to broadly capture the evolution of the GSMF at $z = 0-0.615$. However, we just demonstrated that the observations for the GSMF are captured even better by a much simpler scheme and with much fewer complexity. In appendix \ref{Quantify} we demonstrate that according to our RMSD analysis GAME$_{0}$ is having an excellent comparison with respect observations at $z = 0-0.615$ for low, intermediate (table \ref{B1}) and massive galaxies (table \ref{B3}). The average RMSD of GAME$_{0}$ at this redshift range is 0.13 dex, while EAGLE (0.19 dex), IllustrisTNG (0.24 dex) and Simba (0.25 dex) peform slightly worse.  In addition, in tables \ref{B5} and \ref{B7} we demonstrate that this success has excellent AIC values displaying a good balance between precision and numbers of parameters considered. Can GAME$_0$ (i.e., the $\Gamma$ growth model) broadly capture the key physics of star formation on galaxy scales to approximately represent galaxy growth at $z = 0-0.615$? Fig. \ref{GSMF0obs1} is a great start and suggests that it can.

An additional aspect of GAME$_{0}$ reproducing the GSMF at this large redshift range ($z=0$ - 0.615, 45$\%$ of the age of the Universe) is that  we do {\it not} require an ``exotic"  mechanism to act extensively solely at the low and high mass ends to replicate the observed evolution of the GSMF. It actually seems that the $\alpha$/$\beta$ parameters do not deviate among different mass bins (including the two extreme mass regimes). Nevertheless, we acknowledge that processes like SN feedback and AGN feedback could be significant in shaping galaxy evolution, including the distribution of metals in and around galaxies and the regulation of supermassive black hole growth \citep{Blanc2019,Tejos2021,Fernandez-Figueroa2022,Masafumi2023,Lin2023}\footnote{We stress that employing a strong AGN feedback prescription to replicate the GSMF at the high-mass end has never been a panacea and has its drawbacks. Simulations like IllustrisTNG, EAGLE or Simba that adopt AGN feedback to quench star formation at the high-mass end are unable to reproduce other key observations at z = 0 - 4 like the AGN luminosity function \citep{Habouzit2022}, the specific star formation rate function \citep{Katsianis2021}, the quenched part of the GSMF \citep{Weaver2023}. the quenched fraction \citep{Dickey2021,Meng2023}, the black hole (BH)-halo mass relation \citep{Voit2024} and observations of the Coma cluster in constrained zoom in simulations \citep{Luo2024}. }. However, we argue that mechanisms acting exclusively only at the low and high mass ends or different adjustments to the $\alpha$/$\beta$ parameters among the different mass bins just for the sake to replicate {\it specifically the} GSMF evolution are not essential at these low redshifts (Fig. \ref{GSMF0obs1}). This finding is in agreement with observations of the star formation rate function at z = 0-1 that demonstrate a nearly parallel evolution of all the SFR bins of the SFRF that is not particularly different at the low- and high-star-forming ends with respect to the knee of the distribution (\citet{Katsianis2021b}, their Fig. 2, bottom left panel). 
%Thus, we believe that a straightforward and much simpler formalism that is able to reproduce the observed GSMF at $z = 0-1$ is quite compelling.
\begin{figure*}
\centering
\includegraphics[scale=0.55]{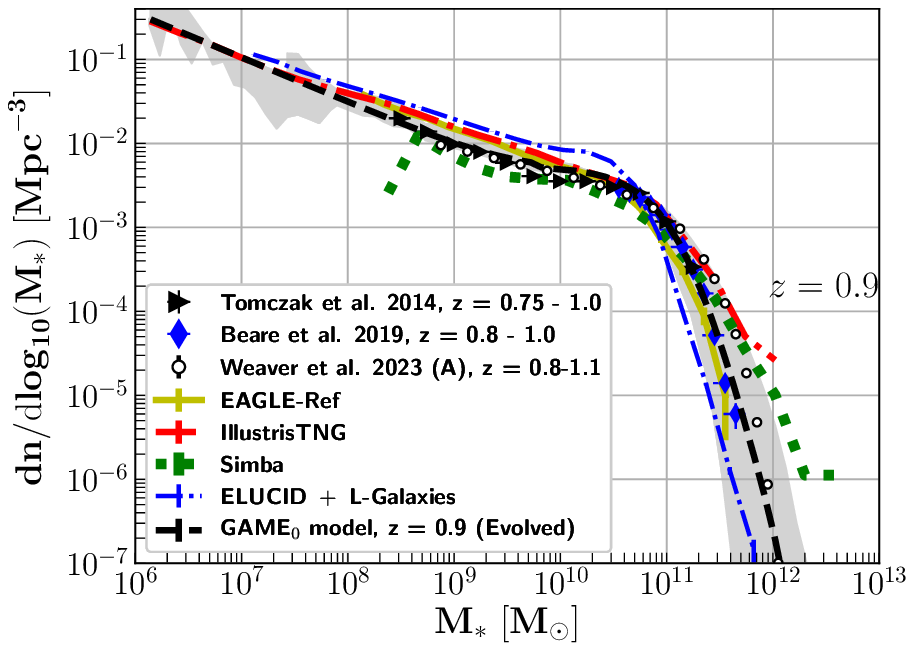}
\vspace{-1.0cm}
\includegraphics[scale=0.55]{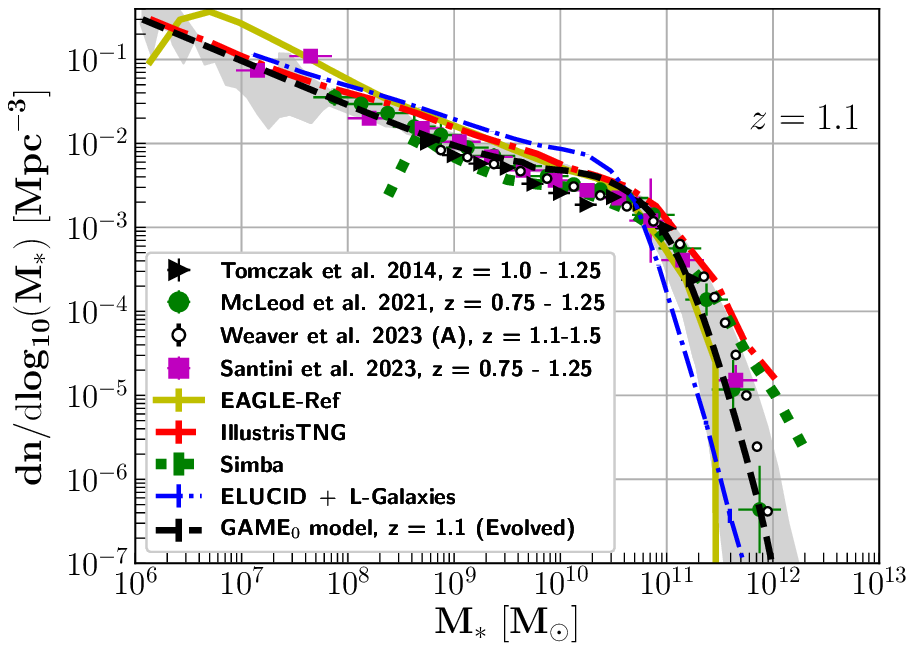}
\includegraphics[scale=0.55]{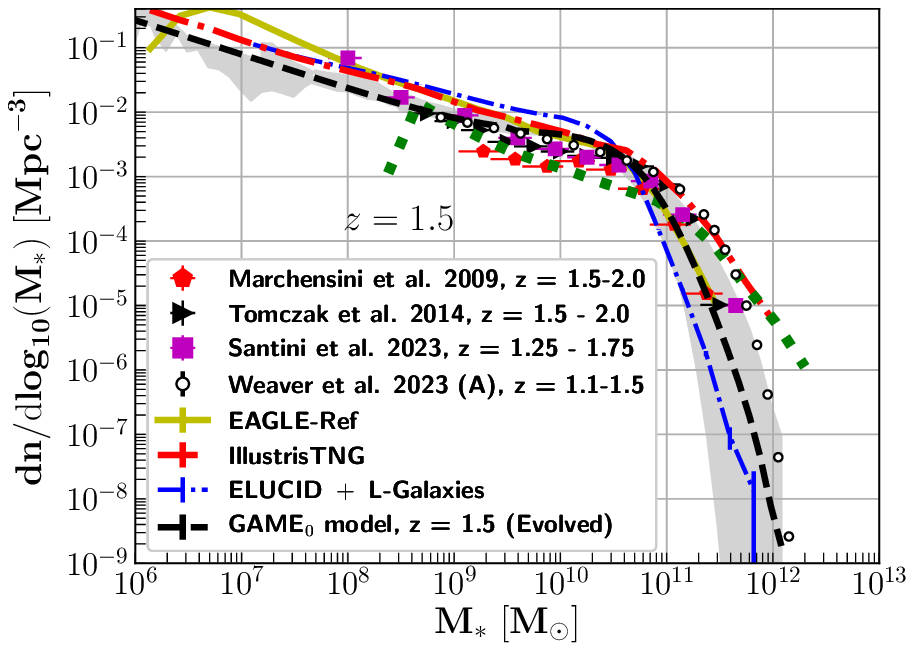}
\vspace{-1.0cm}
\includegraphics[scale=0.55]{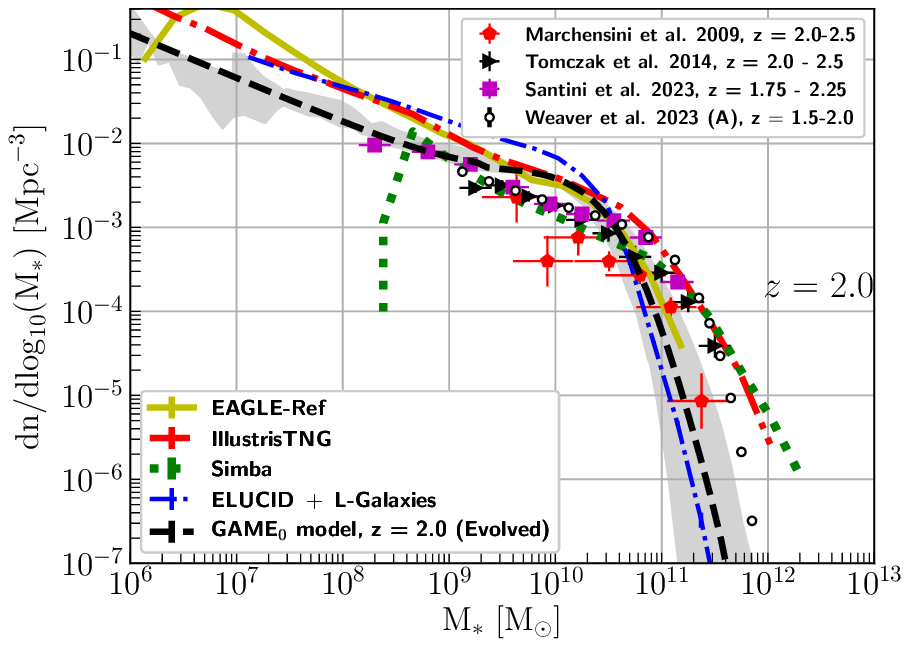}
\includegraphics[scale=0.55]{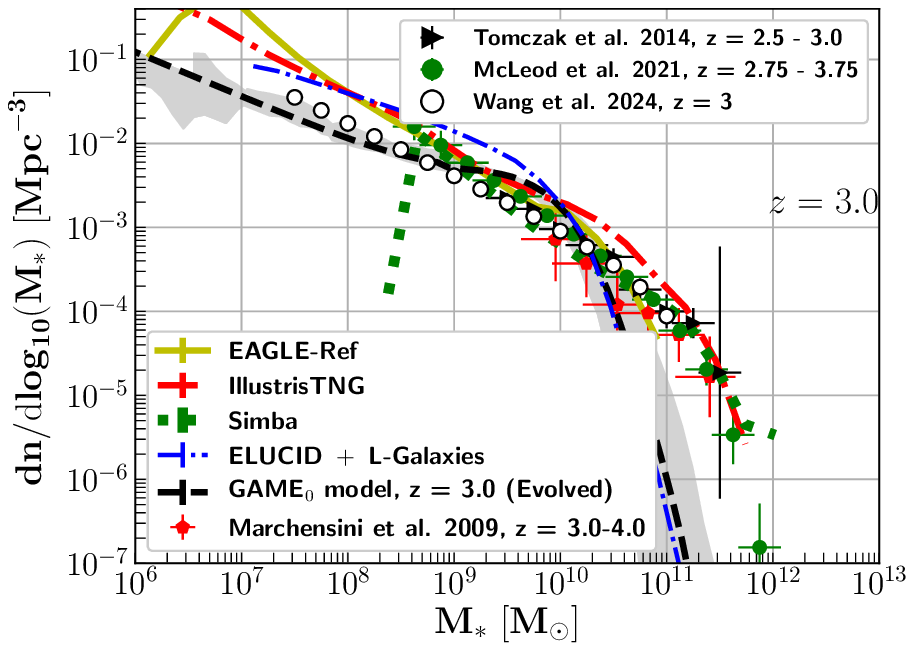}
\includegraphics[scale=0.55]{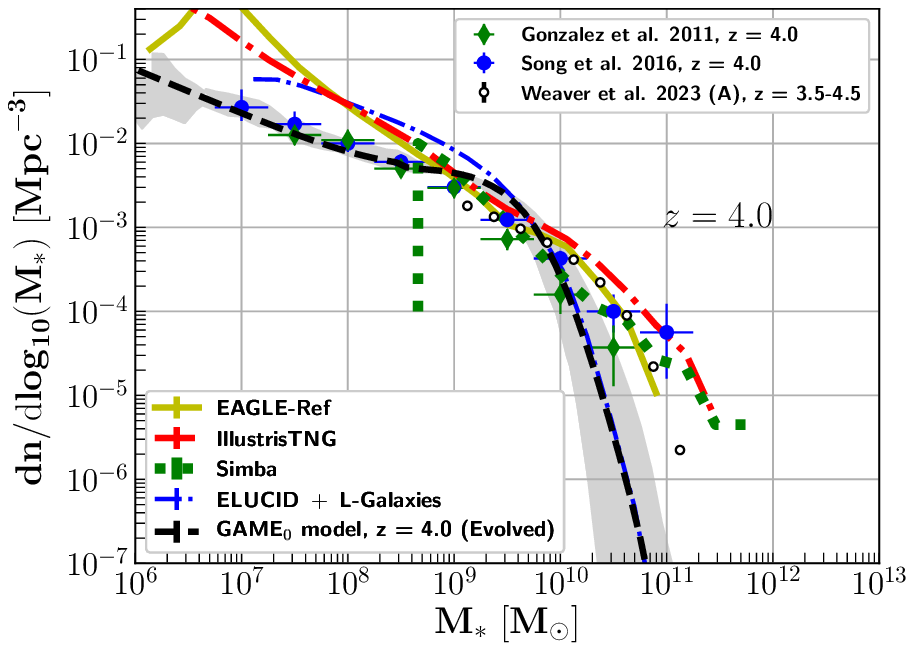}
\vspace{-1.0cm}
\caption{Comparison of GAME$_{0}$ (black dashed line) and observations up to T = 1.54 Gyr (z = 4). The comparison with respect observations up to T = 4.3 Gyr (9.5 Gyr ago) particularly highlights the success of the simple scheme we adopt. We note that GAME$_{0}$ reproduce GSMFs in agreement with observations for 70$\%$ of the age of the Universe. However at redshift 2-4 it is evident that the high mass end is not well reproduced. Cosmological simulations represented by IllustrisTNG (red dot-dashed line), EAGLE  (orange solid line) and Simba (green dotted line) are performing well at this mass regime.}
\label{GSMF0obs2}
\end{figure*}

Before venturing to higher redshifts and explaining in depth the physics, limitations and success of GAME$_{0}$, we note the following contribution of GAME$_{0}$. It is evident that observations are usually unable to probe galaxies with stellar masses of $M_{\ast} = 10^{6} -  10^{8}  \, M_{\odot}$ at $z>0$. However, GAME can compensate for observations at this redshift regime and give us an estimate of how the observed GSMF evolves at higher redshifts and at the low mass end. We note that cosmological simulations like IllustrisTNG and EAGLE suffer from severe resolution effects for objects with masses less than $M_{\star} =  10^{8}  \, M_{\odot}$, thus comparisons at these regimes with observations are usually avoided in the literature.

\begin{figure*}
\centering
\includegraphics[scale=0.55]{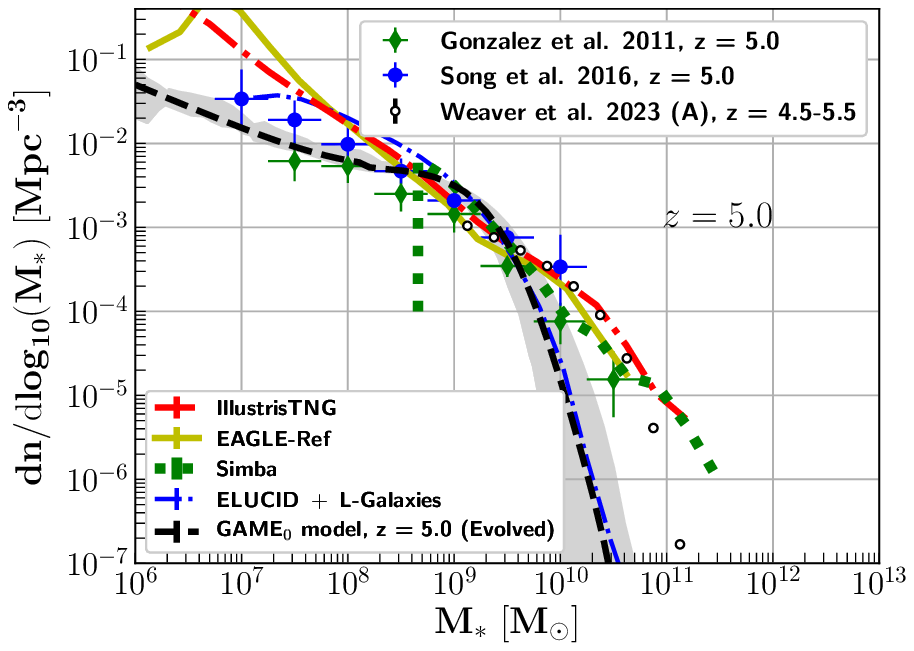}
\vspace{-1.0cm}
\includegraphics[scale=0.55]{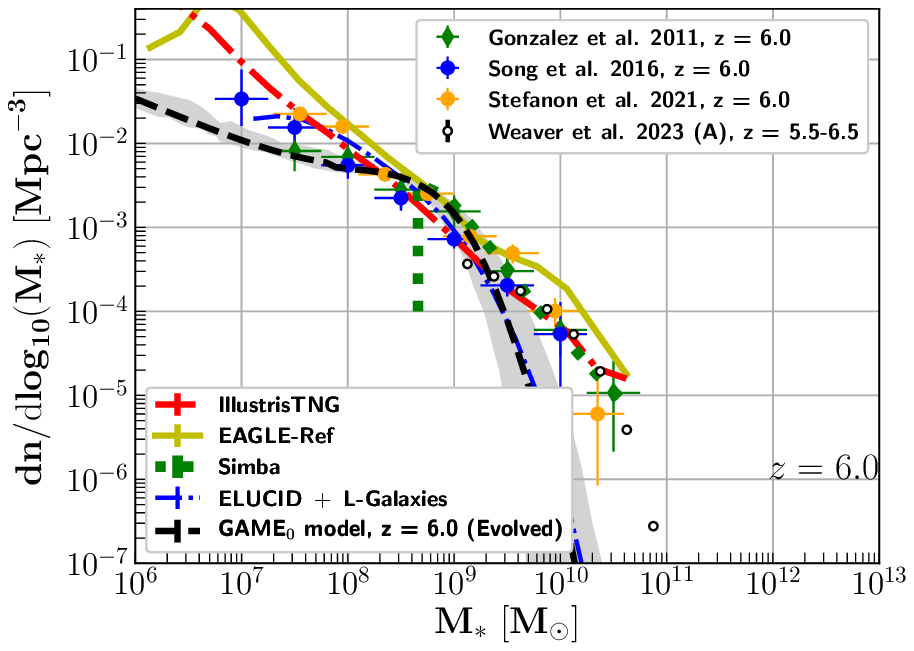}
\includegraphics[scale=0.55]{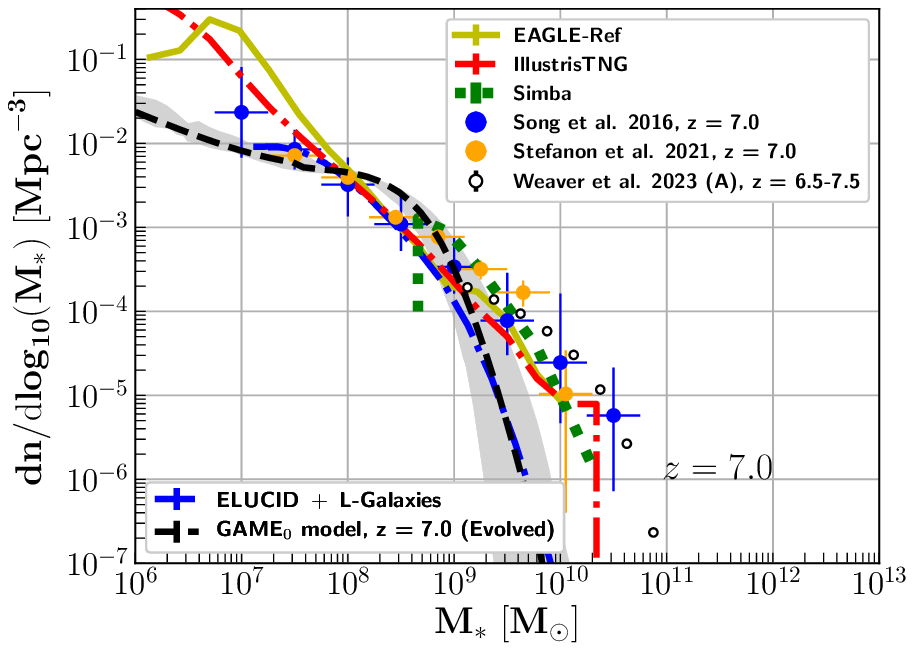}
\includegraphics[scale=0.55]{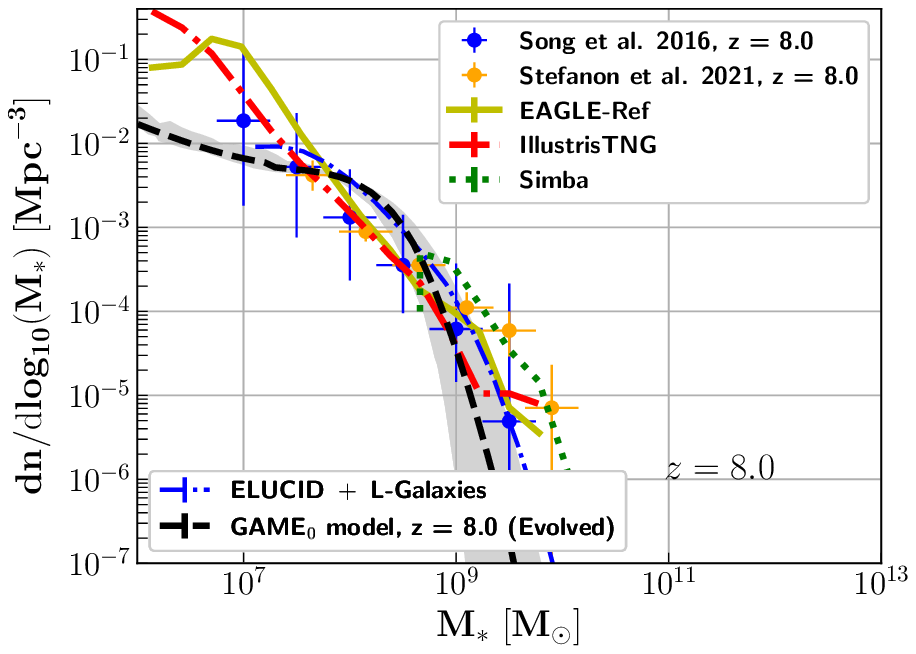}
\vspace{-1.0cm}
\caption{Comparison of GAME$_{0}$ (dashed black line)) and observations up to T = Gyr (z =8). We note that GAME$_{0}$ produces GSMFs that are in good agreement with respect observations, especially for low mass and intermediate mass galaxies. IllustisTNG (red dot-dashed line), EAGLE  (orange solid line) and Simba (green dotted line) are performing well at the high mass end.}
\label{figEpicG0}
\end{figure*}

Simulated galaxies in cosmological simulations with $M_{\ast} = 10^{6} -  10^{8}  \, M_{\odot}$  are described by very few gas/star particles ($< 100$). Thus, there is a concensus in the community that the physics of star formation and the feedback/outflows/interactions cannot be well described by such a low number of particles and the predictions of cosmological simulations at this extreme mass regime are disputed. However, in recent years, it is evident that resolution effects affect galaxies regardless of stellar mass or SFR \citep{Schaye2015,Zhao2020}\footnote{Comparisons between simulations with different resolution but with the same subgrid physics have actually shown that {\it all} simulated galaxies regardless of stellar mass (even at the high-mass end and the knee of the distribution) and SFR (from 0.01 to 100 $M_{\odot}/yr$) will be impacted by limitations of resolution effect, i.e. there is no strong convergence among different resolutions regardless of the stellar masses of the simulated objects, even the most massive ones \citep{Schaye2015,Zhao2020}. It is evident that the feedback prescriptions are not well understood theoretically while observations cannot easily provide the outflow properties. Changing the resolution of a simulation changes the frequency of the individual feedback events, the energy, mass, and momentum that is injected; thus a re-tuning of the parameters is required upon changing resolution in order to capture "successfully" the observations, even for the high mass end.}. 

Thus, since {\it most simulated objects will be impacted by resolution effects} we decide to test the performance of the simulations in the regime $M_{\star} = 10^{6} -  10^{8}  \, M_{\odot}$ too. Interestingly, the low-mass end of the GSMF is well reproduced by TNG and EAGLE at $z = 0-0.615$. The limitations related to the small number of gas/star particles for the low-mass simulated galaxies ($M_{\star} = 10^{6} -  10^{8}  \, M_{\odot}$) seem to be of secondary importance. But if the interactions among particles are not a decisive factor, what primarily decides if a cosmological simulation reproduces results consistent with observations (see section \ref{GAME1vsSimulations})?

In Fig. \ref{GSMF0obs2}, we present the evolution of the GSMF predicted from the GAME$_{0}$ model (represented by the black dashed line) from redshift z = 0.9 (7.48 Gyr ago) to z = 4 (12.2 Gyr Ago). Starting at z = 0.9 (top left panel of Fig. \ref{GSMF0obs2}), we see an excellent agreement between our scheme and the observations of \citet{Tomczak2013}, \citet{Beare2019}, and \citet{Weaver2023}. Cosmological simulations and ELUCID+Lgalaxies tend to over-predict slightly the GSMF. The good consistency between the GAME$_{0}$ model and observations, including \citet{McLeod2021}, \citet{Santini2023} and \citet{Marchesini09} at z = 1.1 and z = 1.5 (9.5 Gyr ago) is once again remarkable. We have just demonstrated that GAME$_{0}$ (i.e. a $\Gamma$  growth pattern where the parameters stand the same for all galaxies regardless of stellar mass bin) is able to reproduce the observed GSMF for 70$\%$ of the age of the Universe with very few parameters !  We remind though, in our quest of describing the GSMF via this formalism we are also seeking for any deviations from this motif that can highlight additional key mechanisms appearing at a particular mass or redshift beyond those implied by a pure $\Gamma$ Growth. 

At redshift 2 (10.5 Gyr ago) we see for the first time that GAME$_{0}$ performs poorly at the high-mass end. However, our simple model still provides a good description of the GSMF at the low and intermediate masses (confirmed in appendix \ref{Quantify}) with a good performing RMSD with respect observations of 0.2 dex This limitation of GAME$_{0}$ has its roots on the shortcomings outlined in subsection \ref{GAME0Msparam}. We confirm that cosmological hydrodyamic simulations perform well with respect observations at the high mass end. Besides hydrodynamic simulations we note that the evolution of the GSMF at z =0-3 can also be reproduced by empirical models like UniverseMachine \citet{Behroozi2019} and \citet{Leja2020}.  

UniverseMachine is a sophisticated approach with the objective to  determine precisely the star formation histories of individual galaxies within the $\Lambda$CDM paradigm. \citet{Behroozi2019} have employed a total of 44 parameters (table 2 of their work) which have been constrained against a large set of observations, like the stellar mass function at $z \sim 0-4$,  the cosmic star formation rate density at $z =  0-10$ and the UV luminosity functions at z =4-10 (table 1 of their work).  In addition, \citet{Leja2020} studied the evolution of the GSMF from z = 0 to 3.0 using the Prospector SED-fitting code. The study resolved tensions between the observed star formation rates and stellar masses of galaxies, and  presented a continuity model which successfully describes their observed GSMF.  The authors also presented a continuity model which successfully describes their observed GSMF. The model requires as an input the parameters of double schechter functions at 3 different redshifts (11 parameters) and was constrained at z = 0.2, 1.5 and 3.  Both empirical approached have employed significantly more parameters/tuning and a far higher level of complexity than GAME$_{0}$ as their motivation is different than ours (more details can be found in appendix \ref{comparisonEmpirical}). Both  models are able to reproduce successfully the observed GSMF at $z < 3$, but this can be partially attributed to the fact that the observed GSMF at $z = 0-3$ has been used as a constraint for their development.

Going to redshift 4 once again we see GAME$_{0}$ performing excellent for galaxies up to the characteristic mass, M*, (i.e. excluding the high mass end).  In addition, GAME$_{0}$ is in excellent agreement with the semianalytic model of ELUCID + L galaxies, specifically for high masses.  Simulations tend to predict larger numbers of low-mass objects than observations. However, we note that the simulations perform quite well at the high mass end, something that GAME$_{0}$ cannot do. (Indicative of this limitation for the rare high mass galaxies is that the RMSD with respect observations can reach discrepancies of even 1 dex).  In Fig. \ref{figEpicG0} we extend the GAME$_0$ scheme to extremely high redshifts (z = 5-8). We compare our predictions with the observations of \citet{Song2016} and \citet{Stefanon2021} demonstrating the good performance of the model for low and intermediate masses.  We note that cosmological simulations also have relatively good agreement with observations. However, there is a tendency for these models to over-predict the GSMF at the low-mass end, while they perform well at the high-mass end. We also again note good agreement of GAME$_{0}$ with the semi-analytic model of ELUCID + L galaxies. In Section \ref{GAMEba} we extend the comparison with respect to recent James Webb Space Telescope (JWST) data \citep{Navarro-Carrera2024,Wang2024b}. Besides that GAME$_{0}$ cannot capture the evolution of the high mass end of the GSMF it seems that any calibration needed to obtain more precisely the total distribution, having GAME$_{0}$ as a backbone, does not need to be extensive as already GAME$_{0}$ performs very well at low/intermediate masses. These facts will guide us to further improve our formalism (Section \ref{GAMEba}) and modify the parameter $\beta$ accordingly, giving it a physical motivated mass/redshift dependence.  

In conclusion, GAME$_{0}$, {\it besides not being tuned to do so}, is capable of reasonably reproducing the observed GSMF for almost the whole history of the universe. In fact, our model predictions are consistent with 88$\%$ of the data points provided by the observational studies. The impressive average RMSD of the model with respect observations up to the characteristic mass (i.e. excluding the rare high mass galaxies), at z =0-8 is 0.17 dex (Fig. \ref{GSMF0obsRMSDuptoMstar}), slightly overperforming with respect EALGE (0.34 dex) and IllustrisTNG (0.24 dex). We also demonstrate that on average galaxy growth ($M_{\star}$) and the evolution of the GSMF are comprehensive, simple, and do not require many parameters or complicated/exotic mechanism to follow for the last 9.5 billion years. However, further analysis is required to confirm if individual galaxies grow also similarly to this framework. We note that besides their slightly worse performance for the GSMF cosmological hydrodynamic simulations employ complicated mechanisms  with the additional objectives to reproduce entire distributions and a variety of structural, dynamical, morphological and chemical galaxy properties (sometimes successfully other times not), while GAME is currently focusing solely on reproducing the number counts in bins of stellar mass. However, it is important to also note that most of the free parameters in simulations are dedicated to reproduce the CSFRD or the GSMF at z = 0. In addition, before improving the comparison between GAME and the observed GSMF at high redshifts/high masses by slightly re-tuning $\beta$ we have to note that observational studies have their shortcomings too, and they are not necessarily describing the ``true'' GSMF. It is always possible that GAME$_{0}$ would not even need further tuning  if an other IMF or SFH were used for the observations \citep{Jevrabova2018}. In the next section, we focus on why GAME$_{0}$ was so successful (Section \ref{Why?}) and why {\it the choice of parameters and the $\Gamma$ form have a strong multi-scale physical motivation.}

\section{The Physics of GAME$_0$, its parameters and its $\Gamma$ form.}
\label{Why?}

Star formation is regulated by multiple factors (gas accretion, feedback, metallicity, gas cooling and heating, magnetic fields, turbulence, radiation which suppress the collapse of gas, galaxy environment, density, dust content). Besides differences among different models of star formation and the impact of each factor it is widely accepted that star formation is governed by the cold gas reservoirs available. Is it possible to follow the average growth pattern by understanding how these reservoirs grow via accretion and gravitational collapse ? In order to better understand the physics of GAME$_{0}$, connect the different scales involving star formation, and justify the choices of the parameters $\alpha$ and $\beta$ (the theoretical and observational motivation) our approach is to break the $\Gamma$ form ($\frac{dM(T)}{dT}=  M_{z,0}\times\frac{\beta^{\alpha}}{\Gamma(\alpha)}T^{\alpha-1}e^{-\beta \, T }$) into its main components.

\begin{figure*}
\centering
\includegraphics[scale=0.55]{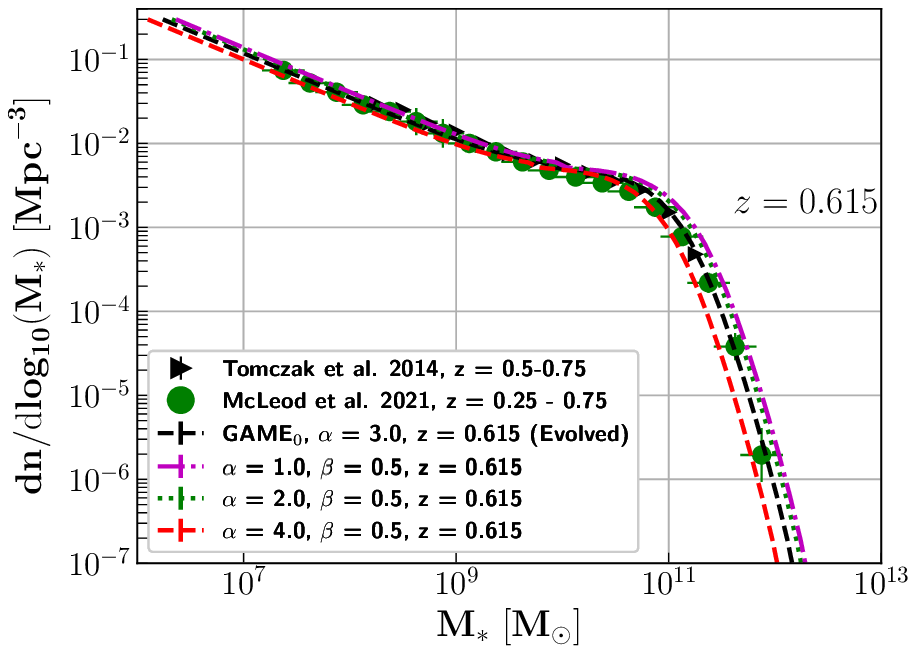}
\vspace{-1.0cm}
\includegraphics[scale=0.55]{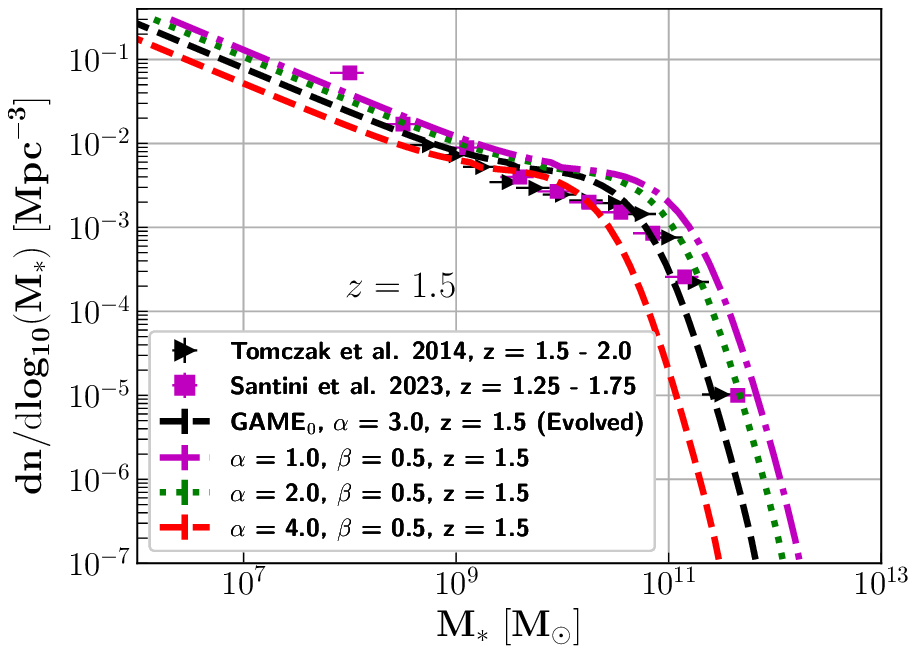}
\includegraphics[scale=0.55]{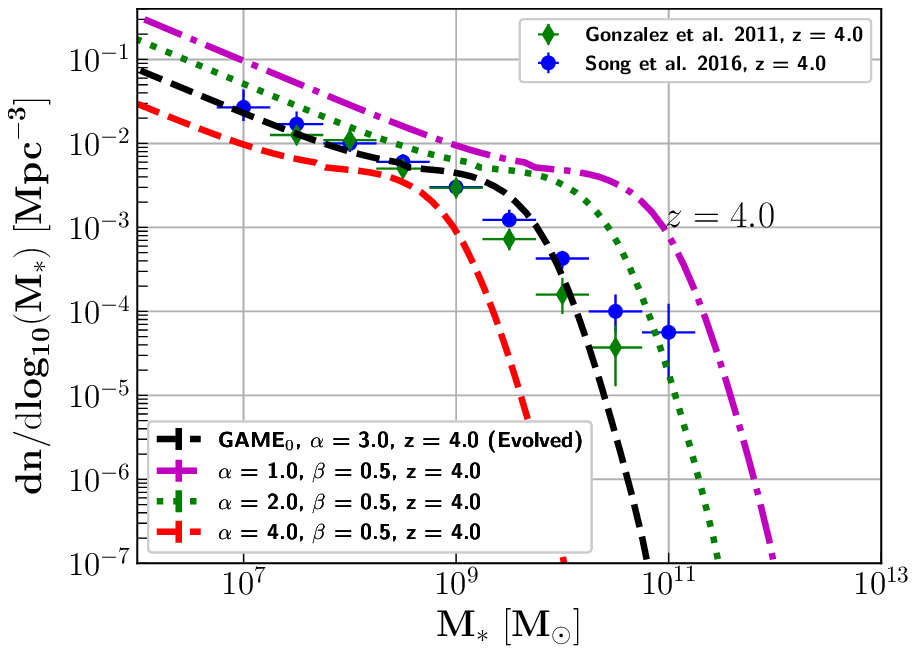}
\includegraphics[scale=0.55]{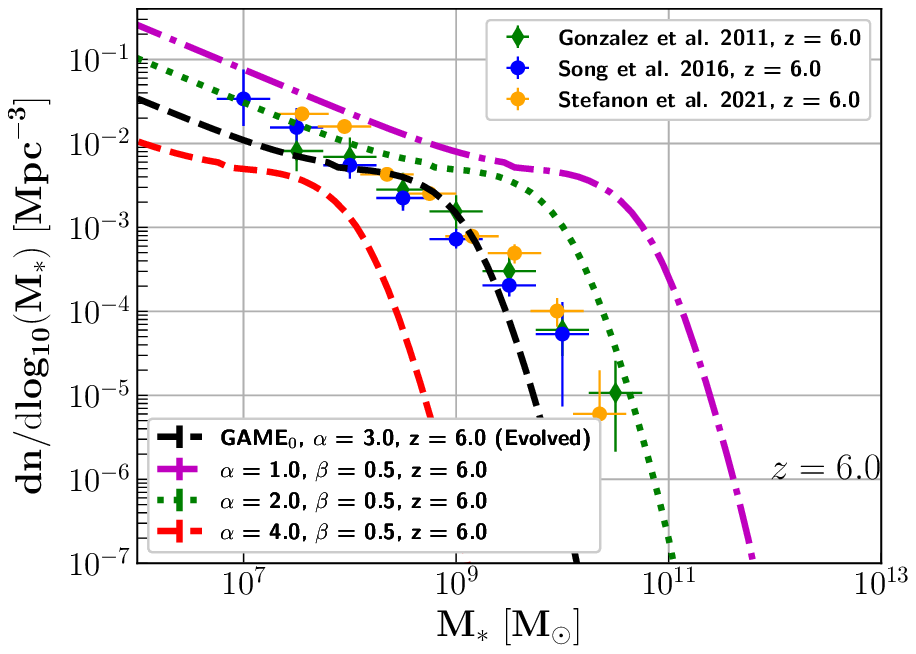}
\vspace{-1.0cm}
\caption{Comparison of the GAME$_{0}$ model with  $\alpha = 3.0$ (black dashed line) and predictions of equation \ref{eq:2} when $\alpha = 1.0$ (magenta dot-dashed line), $\alpha = 2.0$ (green dotted line) and $\alpha = 4.0$ (red dashed line).}
\label{GSMF0obs1AlphaInvestigate}
\end{figure*}

\subsection{Early growth, power law and the $\alpha$ parameter}
\label{Early}

Power law growths ($t^n$) have been used to model empirically the early growth of irregular and late type spiral galaxies. The values of the exponent n have been typically reported to be 0.2-4 \citep{McGaugh1997,Papovich2011,Behroozi2013,Behroozi2019,Haslbauer2023,Abdurro2023}. However, power-law growth motifs extend in the field of Astronomy well beyond galaxies like for example for the much smaller scales of  molecular clouds. The observational work of \citet{Caldwell2018} focused on the temporal and spatial distribution of the star formation rates of four well-studied star-forming regions (molecular clouds of Taurus, Perseus, p Ophiuchi, and Orion A). The authors demonstrated that the rate of growth over the last 10 million years has been accelerating and is roughly consistent with a $t^2$ power law. Additionally,  \citet{Clark2021} studied the growth of the mass of a star cluster's system where its stars are born in the dense gas near the bottom of the gravitational potential of the parent cloud . The authors demonstrated that the mass growth of the cluster also occurs at early times as a power law with the same exponent of 2. Slightly different values (0.5, 1 and 1.5)  were explored in \citet{Dib2013}.
 
Recently, we have reported a similar power law growth pattern\footnote{We remind that in our formalism we have $t^{a-1}$ with $\alpha = 3$ and this can be written as $t^2$. A power-law growth at high redshifts with an exponent of 1.7 was similarly  suggested by \citet{Papovich2011}. \citet{Behroozi2019} also noted that the early evolution of the CSFRD can be well reproduced by an exponent of 2 but this choice yields some inconsistencies with the evolution of the observed cosmic sSFR which would be more consistent with a power law of exponent equal to 4 (Fig., C3 of their work). Furthermore, the commonly adopted delayed SFH with a form of $t \times e^{-b \times t}$ \citep{Carnall2019,Hamed2023} implies power laws with an exponent equal to 1. } of $t^2$ for the rate of growth for the much larger scales of galaxies and the cosmic star formation rate density in \citet{Katsianis2021b}. In addition, \citet{Katsianis2023} demonstrated that the average MAH of dark matter halos and the growth of dark matter halos of $10^{11.0} \, M_{\odot}$ to $10^{13} \, M_{\odot}$, which dominate the CSFRD, also resemble broadly a power law growth pattern with an exponent of $ \sim 2$ to 3. A natural question that rises is why does a power-law growth motif appears physically / mathematically among all these different scales (from molecular clouds to the CSFRD). A second questions is why these scales have power laws with similar exponents ?

There is already some basis to understand the emergence of a power law growth pattern from ab initio theory when star formation within molecular clouds is considered  \citep{Myers2014,Murray2017,Li2018}.  We start by referring to the analytic turbulent collapse model of \citet{Murray2015} which was developed to study star formation in self-gravitating turbulent gas. The authors employ the analytical model of turbulent collapse of \citet{Lee2015} to describe the mass growth rate $\dot{M}$ of the accretor as:
\begin{eqnarray}
\label{eq:PGgastconsumptionratesimplea}
\dot{M} = f \times 4 \pi r^2  \rho(r, t) \times u,
\end{eqnarray}
where $\rho(r,t)$ is the density of the gas and $f$ is the fraction of mass that eventually accretes to the accretor. \citet{Murray2015} argue that the density around a collapsing region should approach an attractor solution of $\rho(r,t) => \rho(r) \propto r^{-3/2}$ and the velocities inside the sphere of influence (a term borrowed by galactic dynamics) should be just controlled by gravity, i.e. $u = \sqrt{G \frac{M}{r}}$. Combined, the relation of  $\rho(r)$ to r and $u$ to r result in a simple form of growth rate for equation \ref{eq:PGgastconsumptionratesimplea} of $\dot{M} = {\rm Constant} \times M^{1/2}$. Interestingly,  an integration of the latter yields a $t^2$ growth pattern for the mass growth, a value consistent with observations focusing on molecular clouds \citep{Clark2021}. 

We have to note that different $\dot{M}$-$M$ relations,  caused by different distributions of $\rho$(r), result in power law growths too but with slightly different exponents. For example, in \citet{Haemmerle2021}, where accretion of matter onto supermassive stars is studied we see a relation of $\dot{M} = {\rm Constant} \times M^{3/4} $, while in  \citet{Bonnell2001} and \citet{Bonnell2001b}  we see a relation of $\dot{M} = {\rm Constant} \times M^{2/3}$ . In contrast, Bondi–Hoyle accretion dominates when the cluster
potential is completely dominated by the accretor, and the resulting accretion rate is then given by $\dot{M} = {\rm Constant} \times M^{2}$ \citep{Ballesteros-Paredes2015}.  We note that for the case of a spherically symmetric, centrally condensed cloud the resulting accretion rate $\dot{M}$ on to the proto-system of mass M is given by  $\dot{M} \propto  M^{2/3}$ \citep{Clark2021} , and by integration we expect  a power law growth of $ M  \propto t^{3}$ (i.e. $\dot{M}  \propto t^{2}$).

A more detailed way to investigate  how a different $\rho$(r) distribution for the collapsing gas can change the exponent but still results in a power law growth and to understand better $\alpha$ is given by \citet{Quataert2012}, \citet{Dexter2013} and \citet{Wang2018} who analytically discuss the accretion into stars that are supernova progenitors. In this context, the material accreted at early times to a star originates from the slowly accreting inner shells of the nearby matter. Assuming the density profile of the matter follows a power law ($\rho(r) = \rho_0 (r/r_0)^{Y-3}$, set by Y\footnote{This is is a sensible assumption since in the presence of self-gravity-driven turbulence, a collapsing cloud naturally develops a power-law density profile.}, and  $\rho$$_0$ is the density of the shell at radius r$_0$), the accretion rate is:
\begin{eqnarray}
\label{eq:PGgastconsumptionratesimple}
\frac{dM(T)}{dT} = \frac{8 \, \pi}{3-Y}  \rho_0 \frac{r_0^{3}}{t_{0}} \frac{t}{t_0}^{\frac{3 \times (Y-1)}{3-Y}},
\end{eqnarray}
where $t_{0} = \frac{\pi \times Y}{32 \times G \times \rho_{0}}$. Equating the power law of equation \ref{eq:PGgastconsumptionratesimple}, $t^{\frac{3 \times (Y-1)}{3-Y}}$, to $t^{\alpha-1}$, we can infer the relation between the $\alpha$ parameter and Y as $\alpha = 3 \times \frac{Y-1}{3-Y}+1 = \frac{2Y}{3-Y}$. 

In conclusion, following the analysis of \citet{Murray2017} and \citet{Wang2018}, we understand that power law growth motifs can be explained by the Physics of gravitational collapse. According to the literature, the exponent is set by the relation of the accretion rate ($\dot{M}$) and the mass of the accretor ($M$), which is set by $\rho$(r)/Y. The same arguments outlined in equations \ref{eq:PGgastconsumptionratesimplea} and \ref{eq:PGgastconsumptionratesimple} can be used to describe the growths of accreting objects as long as gravitational collapse is involved. For the average SFHs of galaxies (and the CSFRD) adopting a value of $\sim 2$ \citep[describing molecular clouds,][]{Caldwell2018} or $\sim 3$ \citep[describing a spherically symmetric, centrally condensed cloud,][]{Clark2021} is sensible. Our current analysis via GAME$_{0}$ favors a mass growth rate exponent of  $\sim 2$ and an $\alpha = 3$. In Fig. \ref{GSMF0obs1AlphaInvestigate} we present how a different choice of $\alpha$ impacts the results. Important differences between the different configurations are notable at high redshifts. The power law growth is well motivated but it is evident that growth cannot persist forever because the available resources for growth at some point will start being depleted. But at what rate/timescale and following which pattern?

\begin{figure*}
\centering
\includegraphics[scale=0.55]{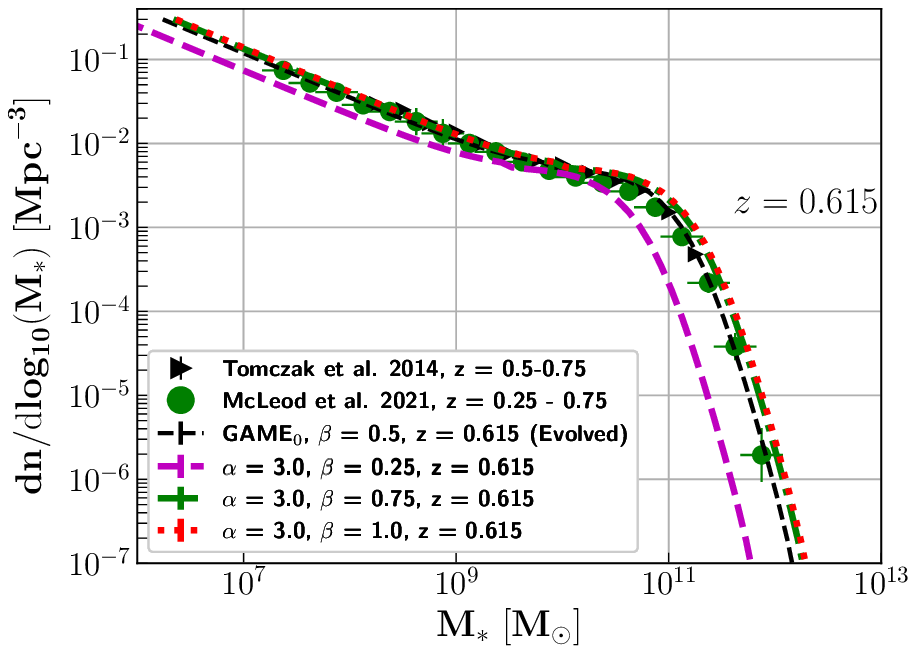}
\vspace{-1.0cm}
\includegraphics[scale=0.55]{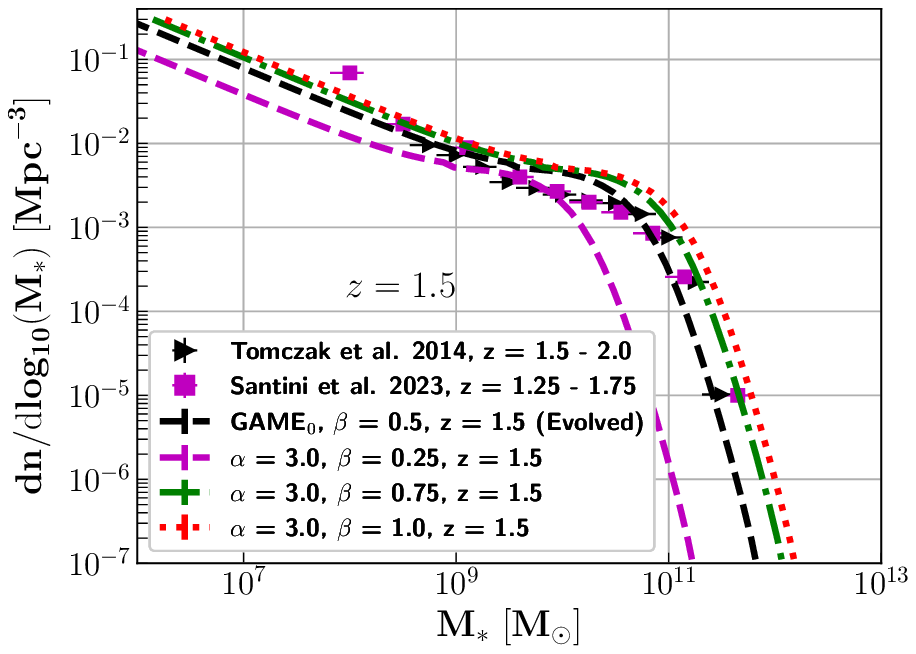}
\includegraphics[scale=0.55]{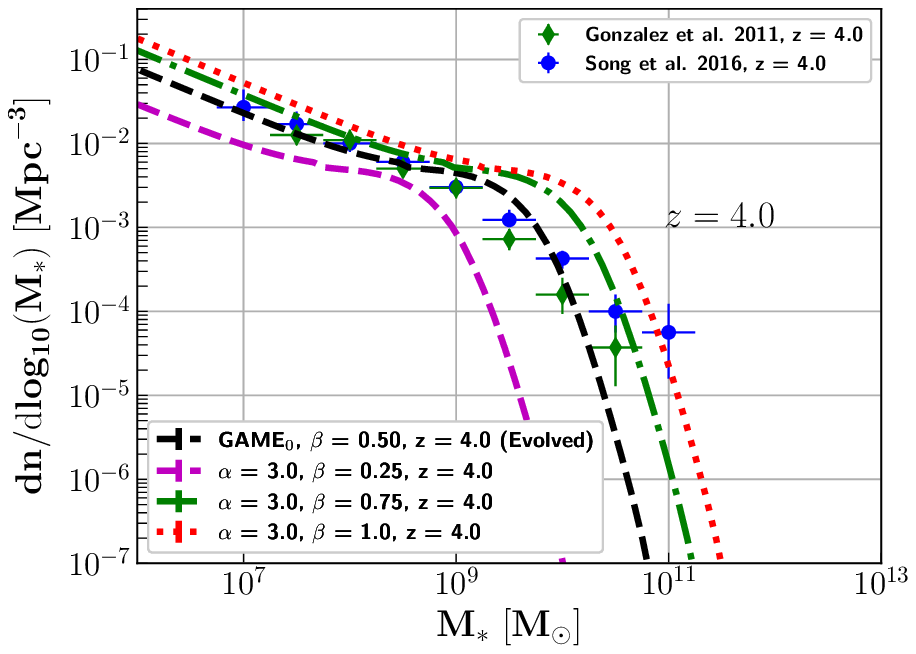}
\includegraphics[scale=0.55]{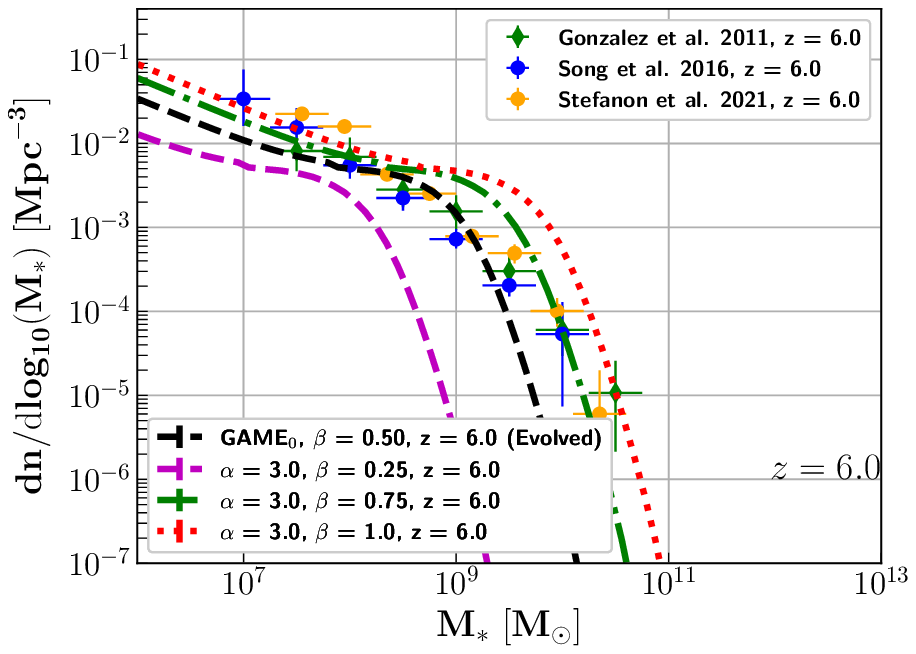}
\vspace{-1.0cm}
\caption{Comparison of the GAME$_{0}$ model with  $\beta = 0.5$ (black dashed line) and predictions of equation \ref{eq:2} when $\beta = 0.25$ (magenta dashed line), $\beta = 0.75$ (green dot-dashed line) and $\beta = 1.0$ (red dotted line).}
\label{GSMF0obs1BetaInvestigate}
\end{figure*}

\subsection{Exponential decline and the $\beta$ parameter}
\label{exponential}

An exponentially declining growth describes a growth rate that is more and more decreasing as the system grows  i.e. a growth dependent deceleration occurring  as $\dot{M} = -\beta \times M$  with an e-fold timescale of $\tau = 1/\beta$. It is a very common pattern observed when the resources for growth are finite. The more the system grows by accumulating the nearby available resources the less resources remain for the future growth (appendix in \citet{Katsianis2023} discusses the implications of this fact from colonies of bacteria growing within tubes containing finite resources to the cosmic star formation rate density). For the case of star formation, the finite resource that is being consumed is the available gas and $\tau_{\star}$ describes the timescale at which the available gas for star formation is depleted \citep{McGaugh1997,Katsianis2021}. We remind to the reader that the parameter  $\tau_{\star} = 1/\beta_{\star}$ regulates the delay of the maximum star formation rate (SFR) and the steepness of its decay \citep{Kim2010,Katsianis2014}.

 $\tau_{\star}$ is well studied in observational studies and there is a cencus in the literature for values between 0.3 to 4 Gyr. \citet{Wang2022} pointed to values of 0.3 to 0.5 Gyr. \citet{Carnall2019b} suggested that the cosmic SFH timescale is 1.4 Gyr once an exponentially declining parameterization is adopted. \citet{Kelkar2019} explored values between 0.3 to 2 Gyr, while \citet{Rowlands2018} suggested a timescale of 2.6 Gyr. \citet{Gobat2022} adopted values of 1-3 Gyr. Larger values of 4 Gyr were found in \citet{Guo2017} and \citet{Phillipps2019}. \citet{Kruijssen2019} pointed out that the timescale for the depleting gas due to star formation in galaxies is observed to be $\tau_{\star}$ of around 2 Gyr in agreement with other studies \citep{Simha2014,Ilbert2015,Katsianis2021b}. \citet{Kruijssen2019} argue that due to radiation, stellar winds and SNe explosions, the timescales governing the molecular gas consumption and galaxy star formation are spatially de-correlated on GMC scales. But then what sets this timescale $\tau_{\star}$ for galaxies? 

\citet{Katsianis2023} pointed out that the consumption of the available resources for star formation in galactic scales should be related to how gas collapses within dark matter halos, and thus the average $\langle \tau_{\star} \rangle$ and $\langle \beta \rangle$ among galaxies should be related to the dynamical timescales of halos $\tau_{dyn, z}$. We note that \citet{Torrey2018} explored the star formation timescales in the simulated IllustrisTNG galaxies and also found that these are related with the halo dynamical time. The above findings of both \citet{Torrey2018} and \citet{Katsianis2023} are not totally unexpected, as it has been a common practice to relate star formation to the dynamical timescales of the collapsing gas. For example, \citet{Henriques2015} adopts a model in which star formation is built as: ${\rm \dot{M} =  a_{sfr} \times \frac{Mgas-Mcrit}{\tau_{dyn(disk)}}}$ where ${\rm a_{sfr}}$ is a constant, ${\rm \tau_{dyn(disk)}}$ is the dynamical timescale of the disk and $M_{crit}$ is the threshold mass for SF\footnote{This is essentially the Kennicutt-Schmidt law:
$\dot{M} =  \epsilon \times \frac{M_{gas}}{\tau_{dyn(disk)}}$,
where $\epsilon$ is the star formation efficiency and we note that an integration of this form upon averaging the $\tau_{dyn(disk)}$ or adopting small time intervals at which $\tau_{dyn(disk)}$ is considered constant would result in an exponentially declining star formation history with a timescale of $\tau_{\star} = \frac{\tau_{dyn(disk)}}{\epsilon}$ \citep{Leja2018,Torrey2018,Phillipps2019}.}. 

\citet{Katsianis2023} suggested that the average star formation timescale $\langle \tau_{\star} \rangle$ of the CSFRD and its inverse $\beta$ should be directly related to the dynamical timescales of halos. We remind to the reader that the dynamical time for a matter dominated Universe  can be written as $t_{dyn}(z) = \frac{R_{halo}}{V_{circ}} = \frac{R_{halo}^{3/2}}{(G \, M_{halo})^{1/2}} = 2 \, \times (1+z)^{-3/2} \,  {\rm Gyr} \sim H(z)^{-1})$ reflecting that the density scales as $\rho^{-1/2}$ and $\rho$ scales as $(1+z)^3$ in a matter dominated universe \citep{Robertson2005}. We note that according to the above equation for z =0, $\tau_{dyn, 0} = 2 $ Gyr. \citet{Katsianis2021} and \citet{Katsianis2023} pointed to the direction that the $ \langle \tau_{dyn, z} \rangle$ for the CSFRD and average cosmic mass accretion history is actually:
\begin{eqnarray}
\label{eq:beta}
\langle 1/\beta \rangle = \langle \tau_{\star} \rangle = \tau_{dyn, z =0} = 2 \, {\rm Gyr}.
\end{eqnarray}
Thus, $\langle \tau_{\star} \rangle$ is equal to $\tau_{dyn, z = 0}$ ($\sim 2$ Gyr, the dynamical time scale of halos at z = 0) which provides the physical motivation of our adopted $\beta = \frac{1}{\tau_{dyn, z =0}} = 0.5 $ Gyr$^{-1}$ in GAME$_{0}$. \citet{Guo2017} also pointed out that the quenching timescale is comparable to the dynamical timescale of the host dark matter halo and equal to 2 Gyr. 

The question that arises is why the value  of z = 0 works so well for the case of the CSFRD \citep{Katsianis2021b}, the average MAH \citep{Katsianis2023} and the GAME$_{0}$ formalism, and there is not a strong redshift evolution for this parameter. The depletion timescales indeed are found observationally to have a very shallow/small dependence on the redshift \citep{Birkin2021}. A uniform depletion timescale across redshifts is also observed in  \citet{PadmanabhanLoeb2020} and this  aligns with other observational findings from \citet{Bigiel2011}, \citet{Genzel2015}, \citet{Tacconi2013,Tacconi2018} and \citet{Aravena2019} as well as semi-empirical models \cite{Popping2015}, which indicate a very mild evolution of  depletion timescales\footnote{We note that a stronger evolution for the depletion timescales is found by \citet{Tacconi2020} with an exponent of $-0.9$ and depletion timescales of 1-3 Gyr among different epochs but with the ``expected'' dependence of $(1+z)^{-3/2}$  still not being observed.}.

\citet{Genzel2015} and \citet{Tacconi2018} suggested that there are two main reasons for depletion timescales to remain redshift independent. These are as follow: 1) the Hubble parameter in a $\Lambda$CDM Universe instead of a matter dominated Universe changes slower  than $(1+z)^{-3/2}$ as $ (\Omega_{\Lambda}+\Omega_m \times (1+z)^{3})^{-1/2}$ . 2) the concentration parameter, which is expected to affect as well the collapse of gas within dark matter halos\footnote{For example, \citet{Genzel2015} suggested that the molecular depletion timescale can be written as $\tau_{dep} = \frac{L}{n \times C_h} \times 0.1 \times H(z)^{-1}$, L is the angular momentum, n is the efficiency, $C_h$ is the ratio of the disk’s
rotation velocity to the halo circular velocity, which depends on the halo concentration.},  is smaller at high-z. \citet{Katsianis2023} suggested that these two effects significantly decrease the effective redshift dependence of $\tau_{\star}$/$\beta$ and make it loosely time independent and equal to 2 Gyr, that is, equal to the value z = 0 at any z. 

%In conclusion, adopting that the available resources for star formation are depleted following an exponential decline, within a star formation history timescale that is equal to the dynamical timescale of halos is well supported. Due to the shallow redshift dependence our approximation of adopting $<1/\beta> = <\tau_{\star}> = \tau_{dyn, z =0} = 2 \, {\rm Gyr}$ within GAME$_{0}$ is a logical approximation, especially since we have proven that $\Gamma$ growths with $<\tau_{\star}>$ = 2 Gyr are good fits for the CSFRD/Average MAH at z = 0-9 \citep{Katsianis2021b,Katsianis2023}. Of course 
In Fig. \ref{GSMF0obs1BetaInvestigate} we present how a different choice of $\beta$ changes the results within the GAME formalism. Adopting a value for  $\beta = 0.5$ is a sensible choice, but a slightly larger value (e.g. 0.75) for the high mass end and low mass end has the potential to improve further the comparison with observations at high redshifts.  Thus,  in section \ref{GAMEba}  we explore  a  mild redshift dependence/mass tuning of $\tau_{\star}$. According to Figs. \ref{GSMF0obs1} and \ref{GSMF0obs2} this tuning is not required to be particularly strong, as GAME$_{0}$ and its simple Physics related to gravitational collapse/accretion explain already most of the GSFM evolution, excluding the high mass end at $z>1.5$.

\subsection{Final form and normalization}
\label{FinalFormNorm}

\citet{Shamshiri2015} described the CSFRD fitting an empirical functional form featuring a power-law increase in star formation at early times followed by an exponential decline of the form: 
\begin{eqnarray}
\label{eq:PGgastconsumptionratesimple222}
\frac{dM(T)}{dT} =  A \,  x^{p}  e^{-x},  x =  \frac{t-t_{Lag}}{\tau},
\end{eqnarray}
where A is a parameter, p sets the rate at which star formation builds up, and $\tau$ is the characteristic timescale over which star-formation declines. \citet{Shamshiri2015} adopted p = 1.5, $\tau$= 2.0 Gyr and $t_{Lag}$ reflects the age of the galaxy in good agreement with our adopted parameters. Furthermore, \citet{Iza2022} used 30 simulations from the Auriga Project to determine the dependency of the inflow, outflow, and net accretion rates of Milky Way-like galaxies. The authors demonstrated that Schechter ($\Gamma$) functions provide a good description of the net accretion rate while double Schechter functions describe successfully the galaxy growth. In addition, \citet{So2014} using cosmological simulations adopted that in time $\Delta$t , the amount of mass from a star particle converted into newly formed stars is given by: 
\begin{eqnarray}
\label{eq:PGgastconsumptionratesimple2}
\Delta M =  M_{\star} \, \frac{\Delta t}{\tau_{dyn, z}} \, \frac{t-t_{Lag}}{\tau_{dyn}, z} \, e^{-\frac{t-t_{Lag}}{\tau_{dyn}, z}},
\end{eqnarray}
adopting a delayed exponential SFH \citep{Sandage1986,Carnall2019}. In agreement to our work the authors relate the $\tau$ to the $\tau_{dyn}, z$ of the gas.

We enhance all the above forms following the points discussed in subsection \ref{Early} (power law - $\alpha$) and subsection \ref{exponential} (exponential decline-$\beta$) to  $\frac{dM(T)}{dT}= M_{z,0}\times\frac{\beta^{\alpha}}{\Gamma(\alpha)}(T-T_{Lag})^{\alpha-1}e^{-\, \frac{T-T_{Lag}}{\tau_{dyn, z = 0}}}$ addressing the normalization factor A and relating it to the present day stellar mass following \citet{Katsianis2021b} and \citet{Katsianis2023}. This is mathematically the result of the integration of the above equation, assuming that the available resources for star formation are consumed at z = 0 either because of mass-dependent depletion or mass-dependent feedback as outlined in \citet{Katsianis2021b}. Relating the parameter A to the present $M_{z,0}$ allows the evolution of the GSMF to unfold and the GAME$_{0}$ to be constructed. The $T_{Lag}$ marks the time of the  birth of the first stars and is set at 0.17 Gyr in accordance with the $\Lambda$CDM theory \citep{Villanueva2018,Laporte2021}.

\begin{figure*}
\centering
\includegraphics[scale=0.55]{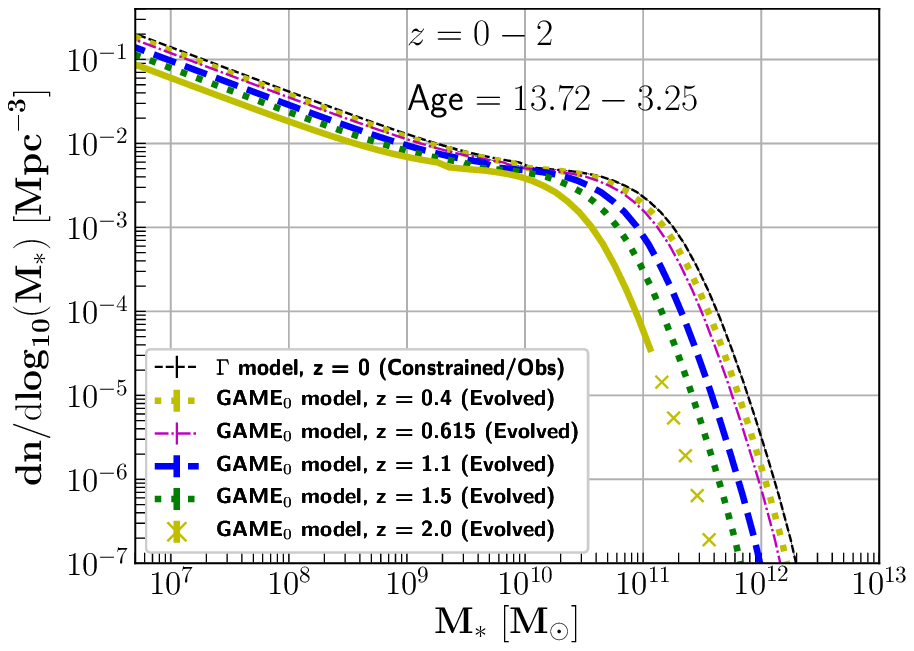}
\vspace{-1.0cm}
\includegraphics[scale=0.55]{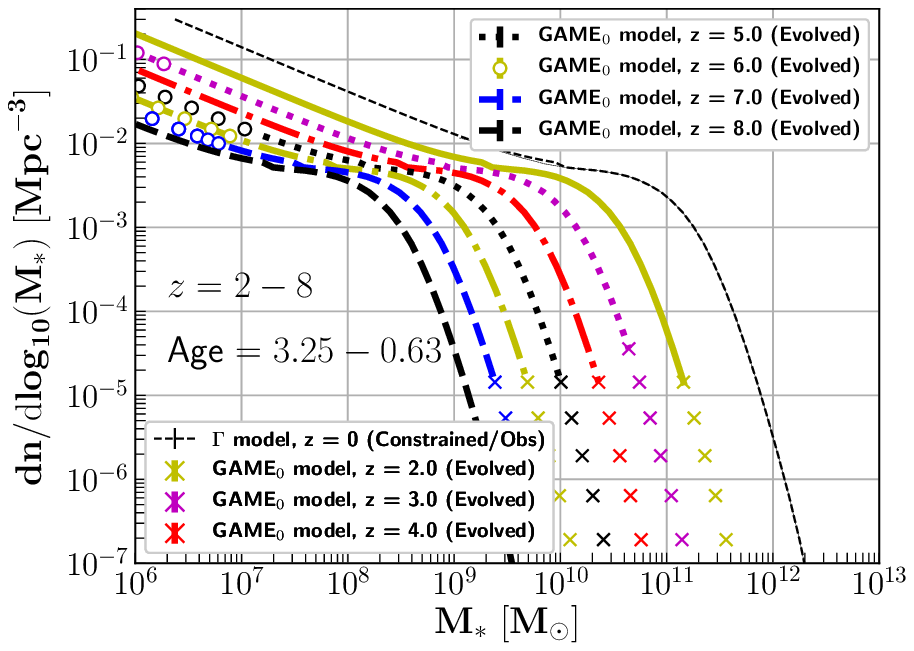}
\includegraphics[scale=0.55]{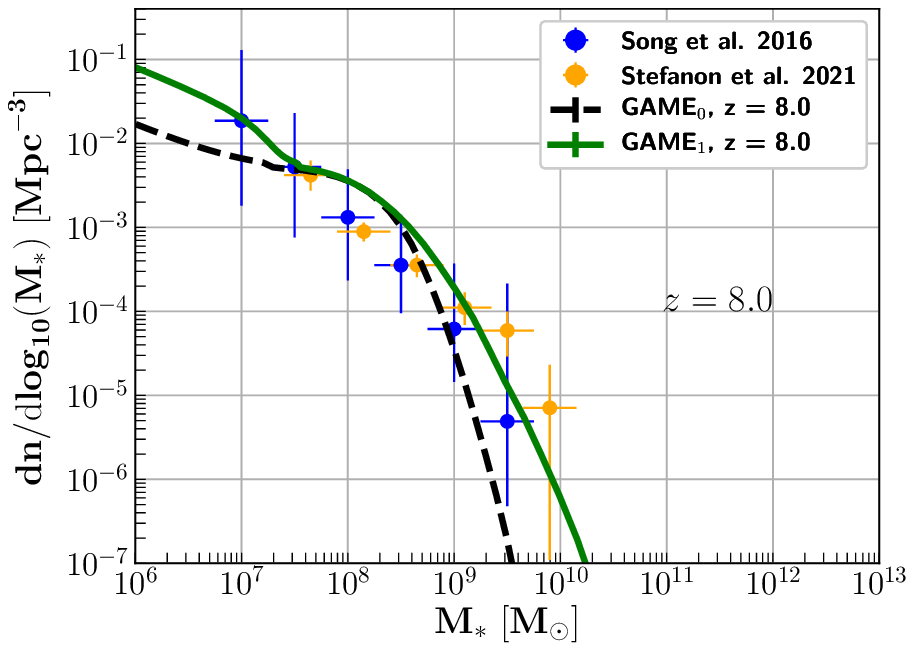}
\includegraphics[scale=0.55]{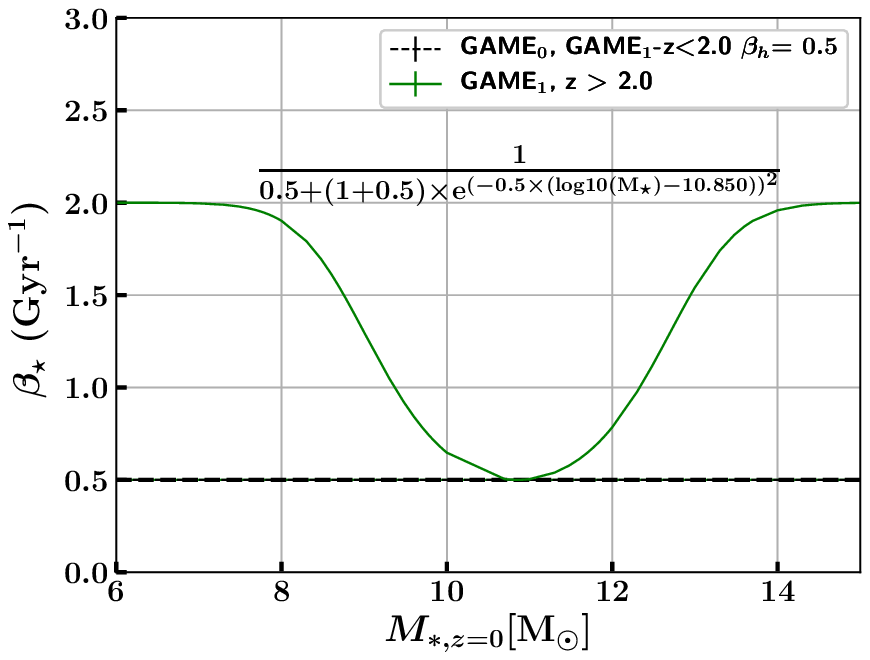}
\vspace{-1.0cm}
\caption{Top panels: The evolution of GAME$_{0}$ from redshift 0 to 8. The formalism besides its simplicity is able to reproduce 88$\%$ of the data points present in this work. The x symbols represent the mass bins that are not well captured by GAME$_{0}$ while the open circles represent the mass regime where consistency exists with respect observations but improvements can be done. Left bottom panel: Tuning the parameter $\beta$ at z $>$ 2 using the GSMF at z = 8. Right bottom panel: The $\beta$ parameter follows a log-normal shape. The timescales of consumption for the high mass end and low mass end decrease for the Physical reasons discussed in \ref{GAME12vsObservations}.}
\label{Conbeta}
\end{figure*}

\begin{figure*}
\centering
\includegraphics[scale=0.55]{SFRF8SAMstar2.ps}
\vspace{-1.0cm}
\includegraphics[scale=0.55]{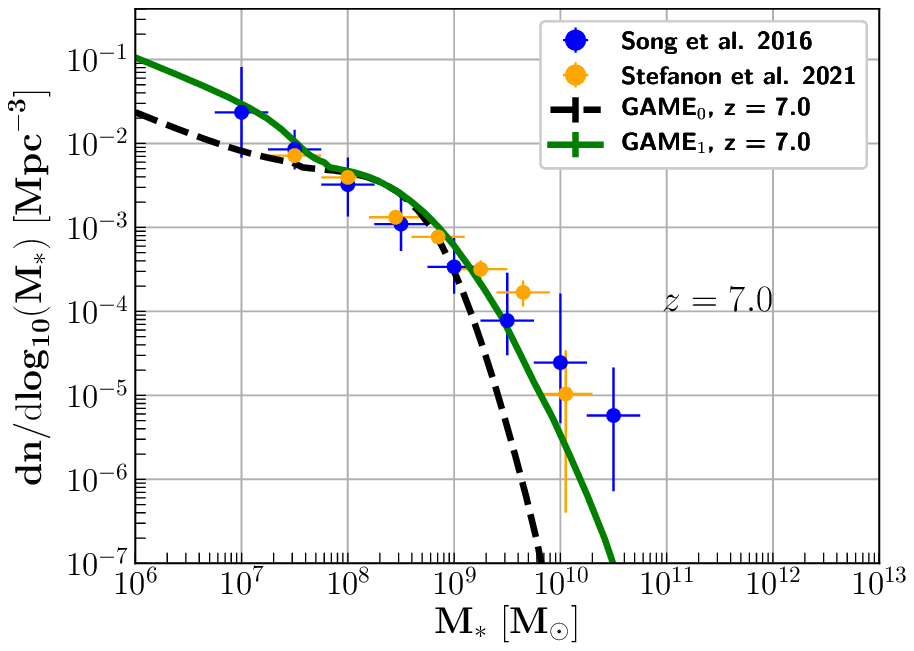}
\includegraphics[scale=0.55]{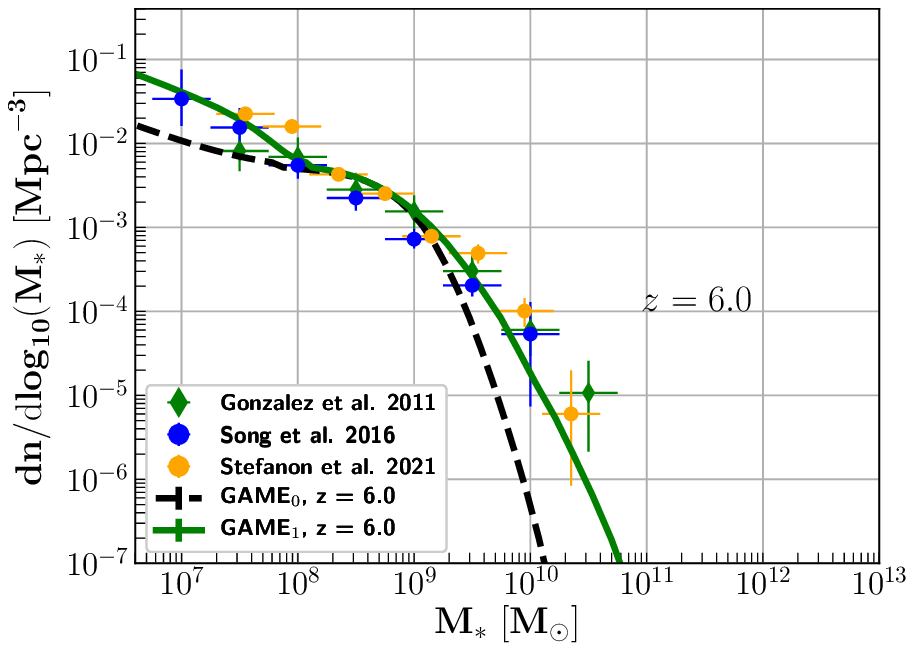}
\vspace{-1.0cm}
\includegraphics[scale=0.55]{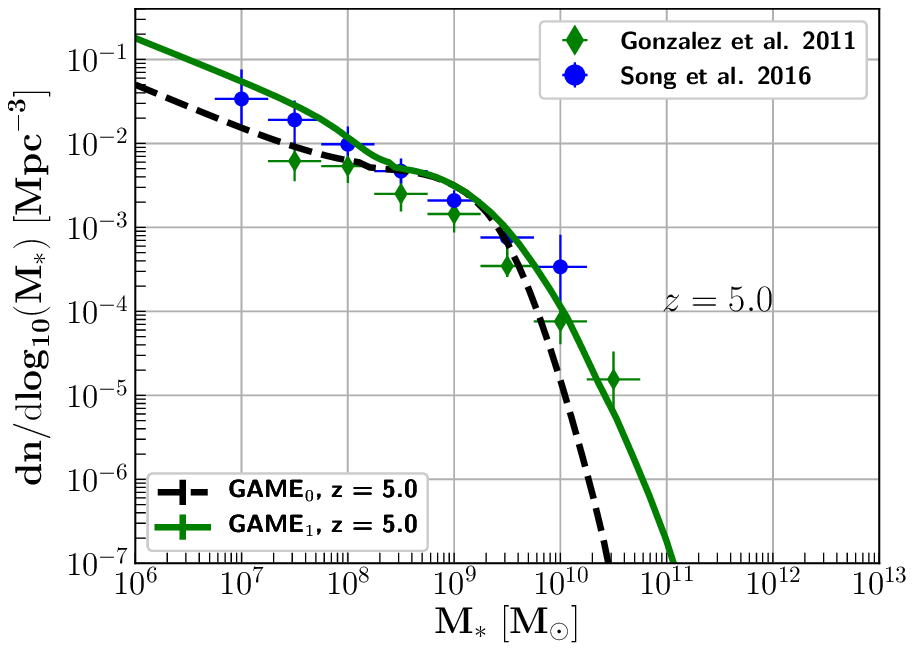}
\includegraphics[scale=0.55]{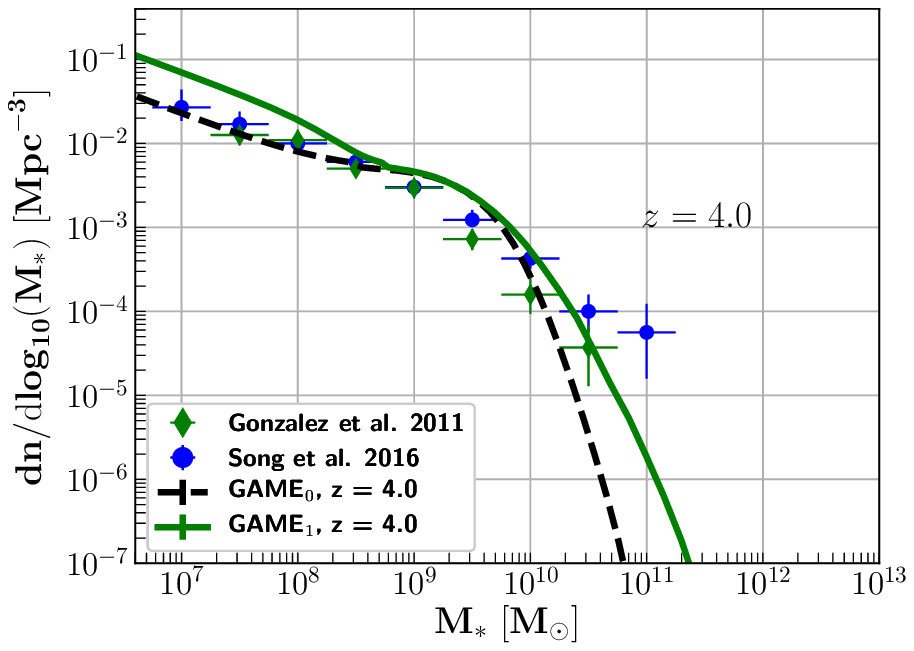}
\includegraphics[scale=0.55]{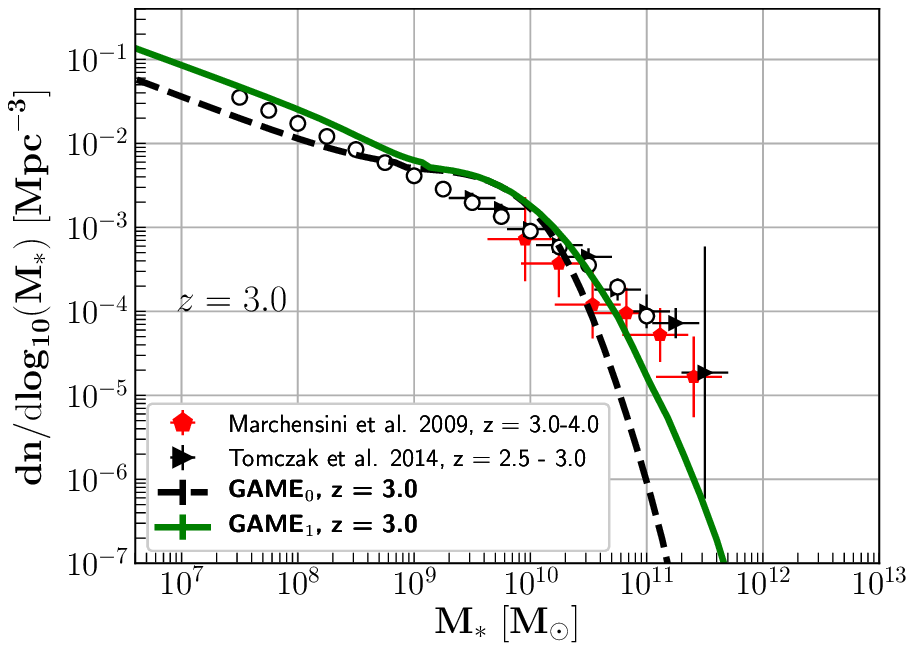}
\vspace{-1.0cm}
\caption{Comparison of GAME$_{0}$ (black dashed line) and GAME$_{1}$ (sold green line) at z $>$ 3. The small tuning of the $\beta$ parameter improves the comparison at the low and high mass ends.}
\label{Conbeta38}
\end{figure*}

\begin{figure*}
\centering
\includegraphics[scale=0.55]{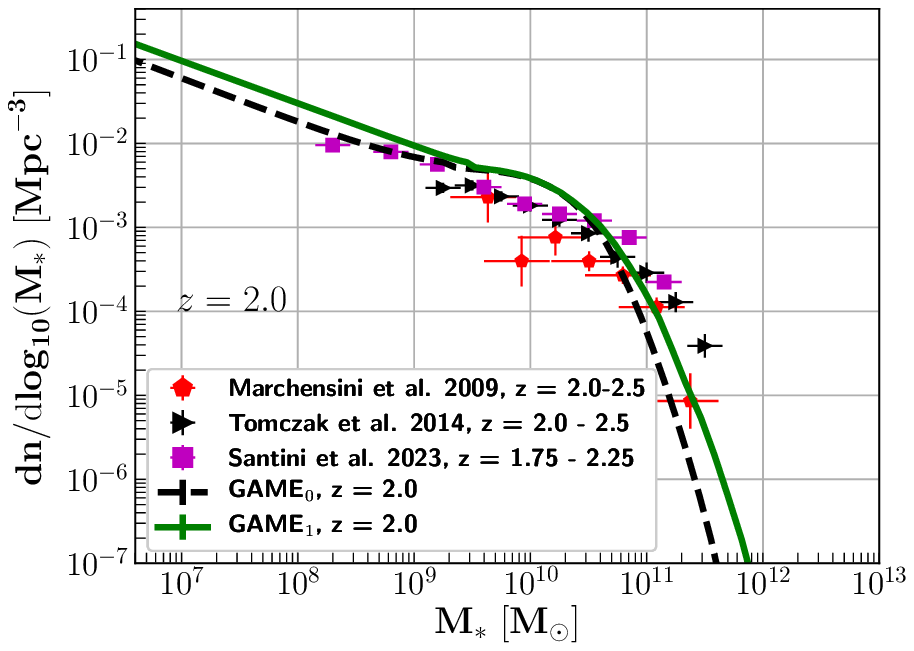}
\vspace{-1.0cm}
\includegraphics[scale=0.55]{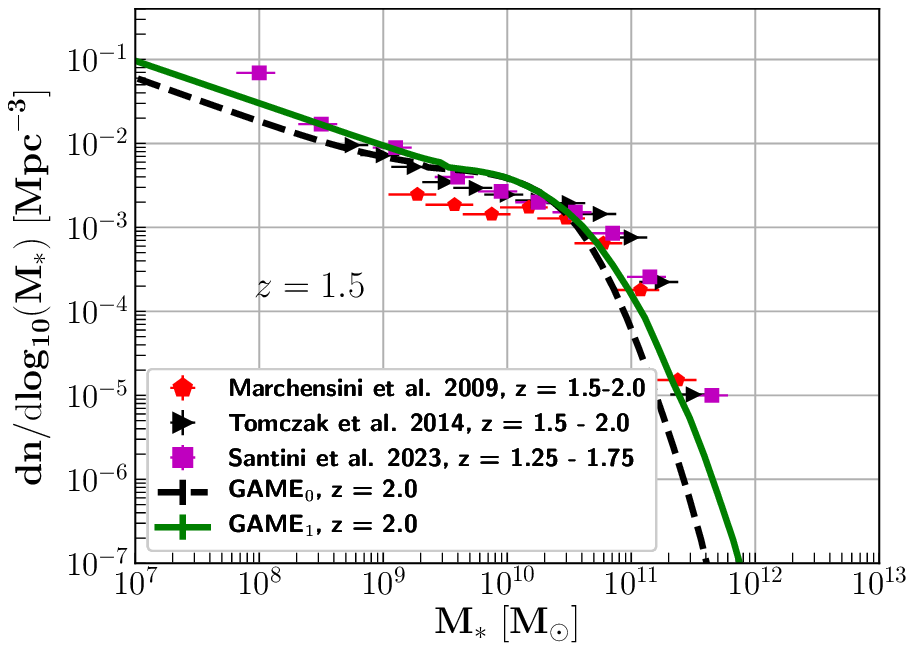}
\includegraphics[scale=0.55]{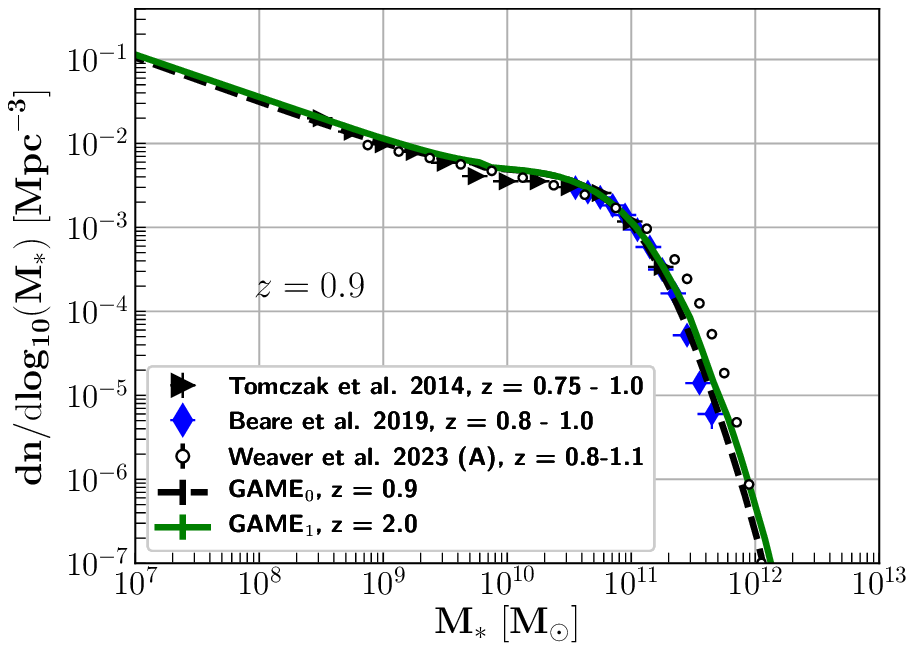}
\vspace{-1.0cm}
\includegraphics[scale=0.55]{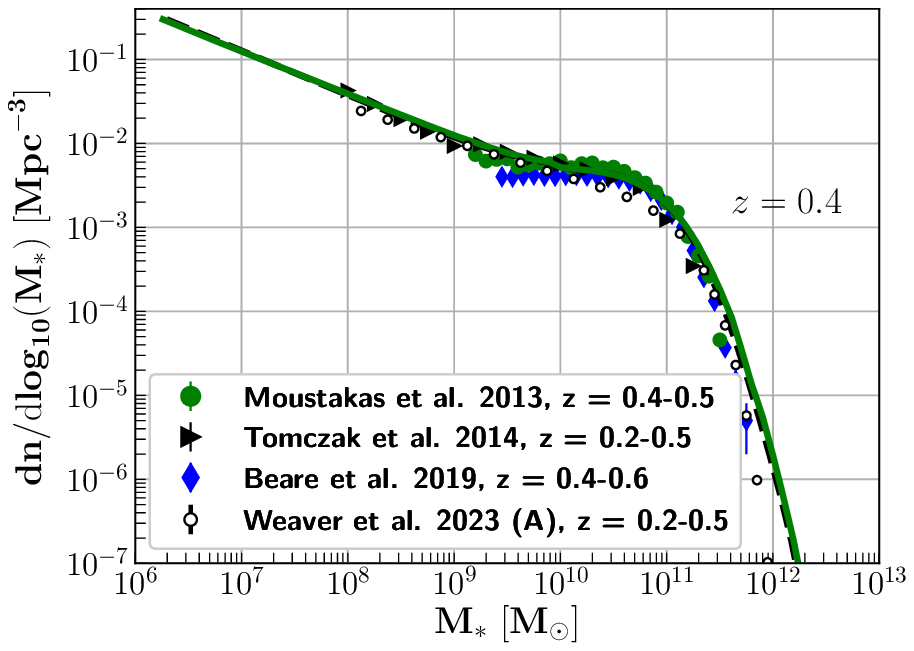}
\includegraphics[scale=0.55]{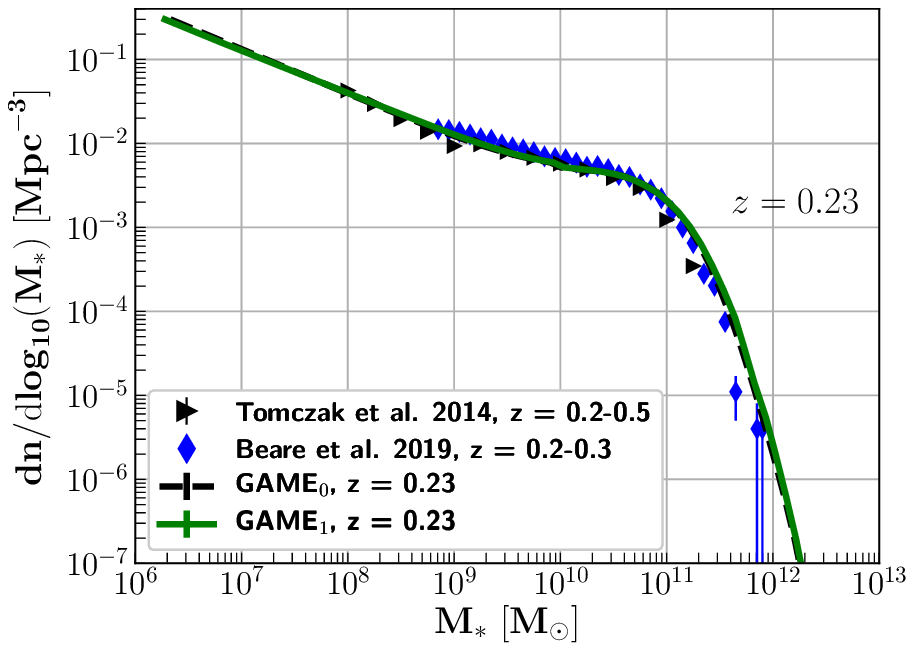}
\includegraphics[scale=0.55]{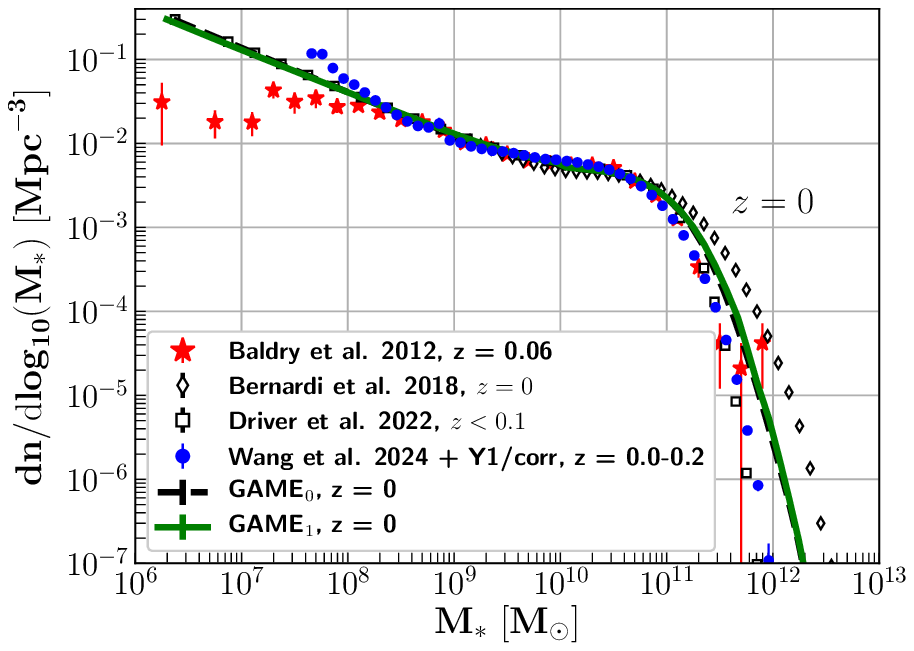}
\vspace{-1.0cm}
\caption{Comparison of GAME$_{0}$ (black dashed line) and GAME$_{1}$ (sold green line) at z $<$ 2.0. The small tuning of the $\beta$ parameter improves the comparison at the low and high mass ends while the predictions of both models are the same at z $<$ 0.9.}
\label{Conbeta00}
\end{figure*}

\begin{figure*}
\centering
\includegraphics[scale=0.55]{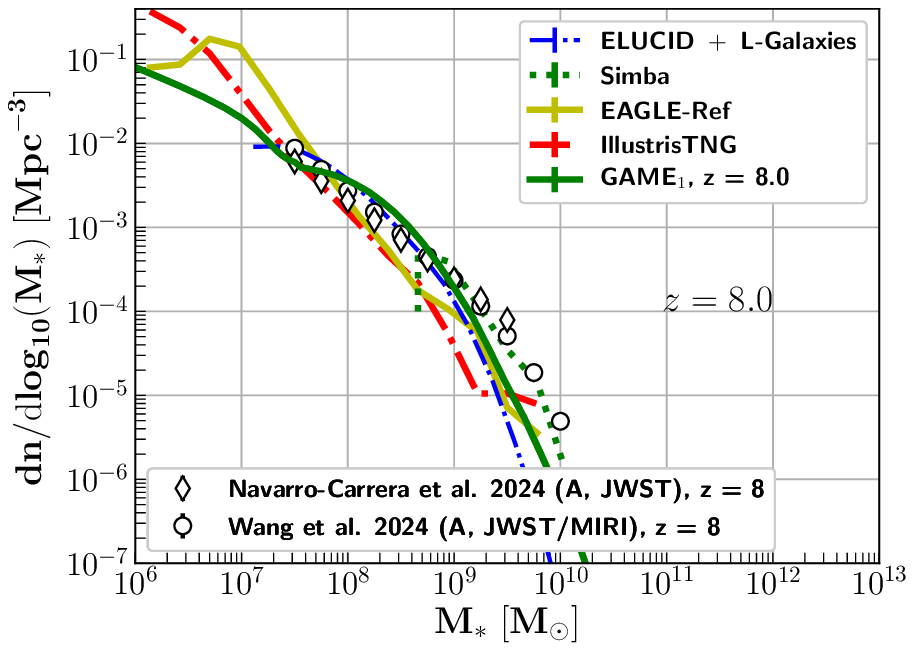}
\vspace{-1.0cm}
\includegraphics[scale=0.55]{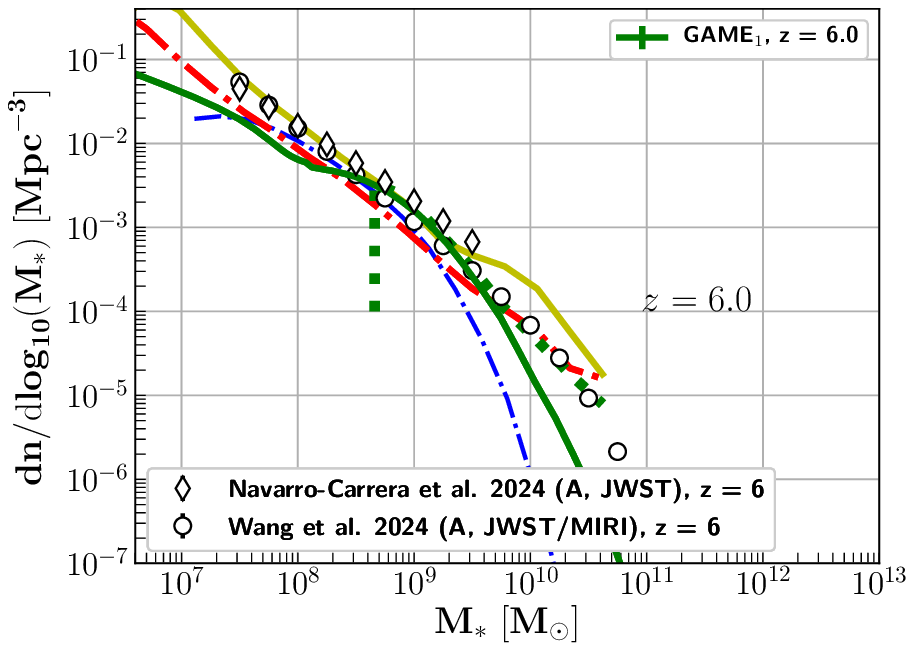}
\includegraphics[scale=0.55]{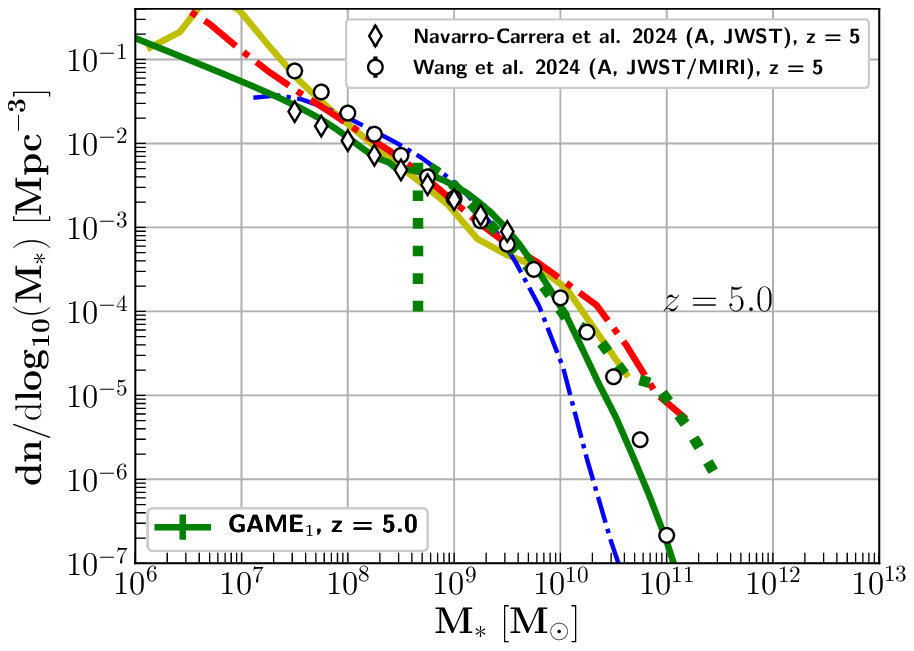}
\includegraphics[scale=0.55]{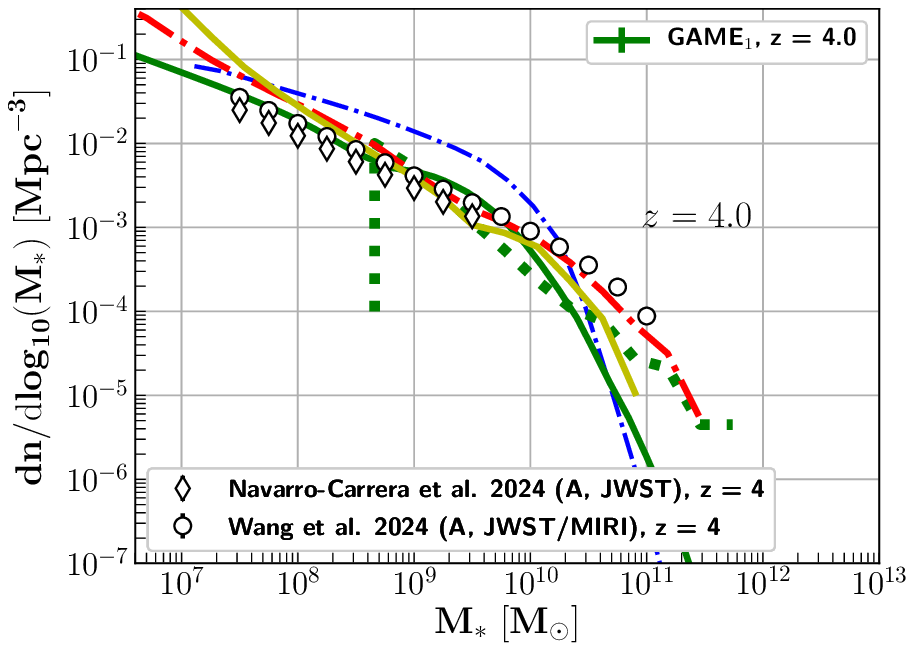}
\vspace{-1.0cm}
\caption{Comparison of GAME$_{1}$ (sold green line) with respect recent JWST observations \citet[open diamonds]{Wang2024b} and \citet[open circles]{Navarro-Carrera2024} and cosmological simulations. }
\label{Conbeta00Sims}
\end{figure*}

\begin{figure*}
\centering
\includegraphics[scale=0.55]{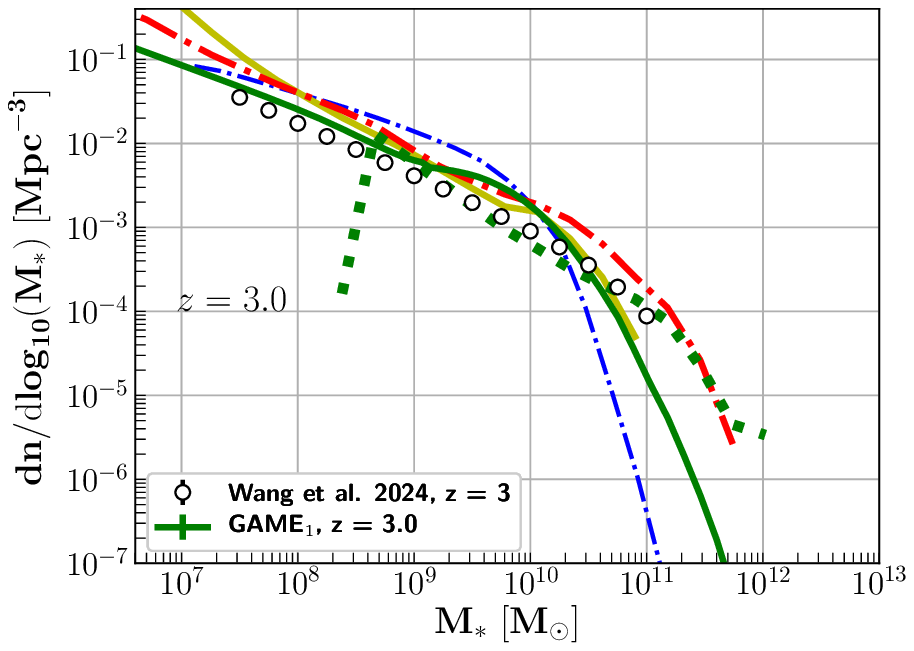}
\vspace{-1.0cm}
\includegraphics[scale=0.55]{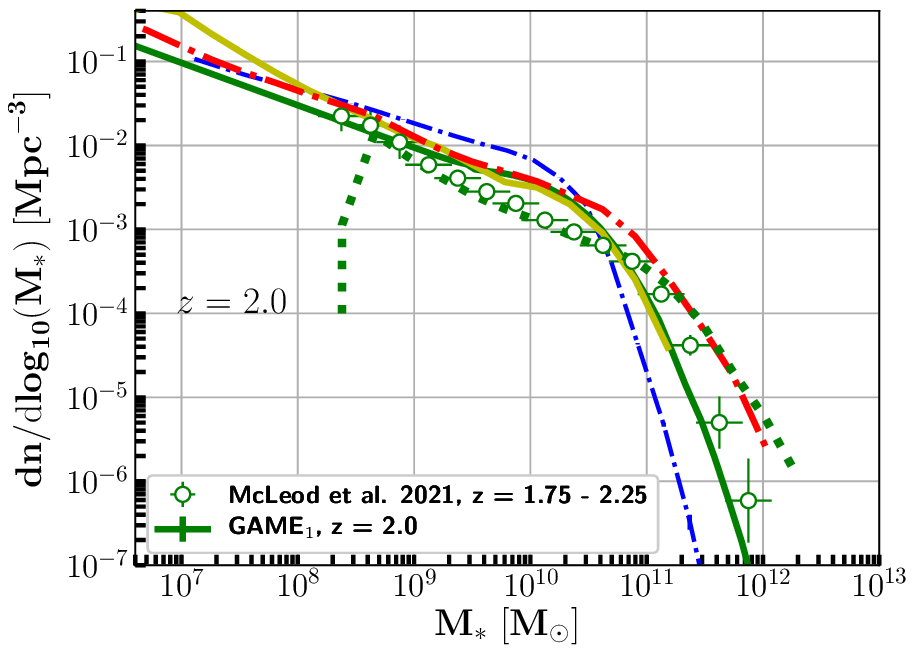}
\includegraphics[scale=0.55]{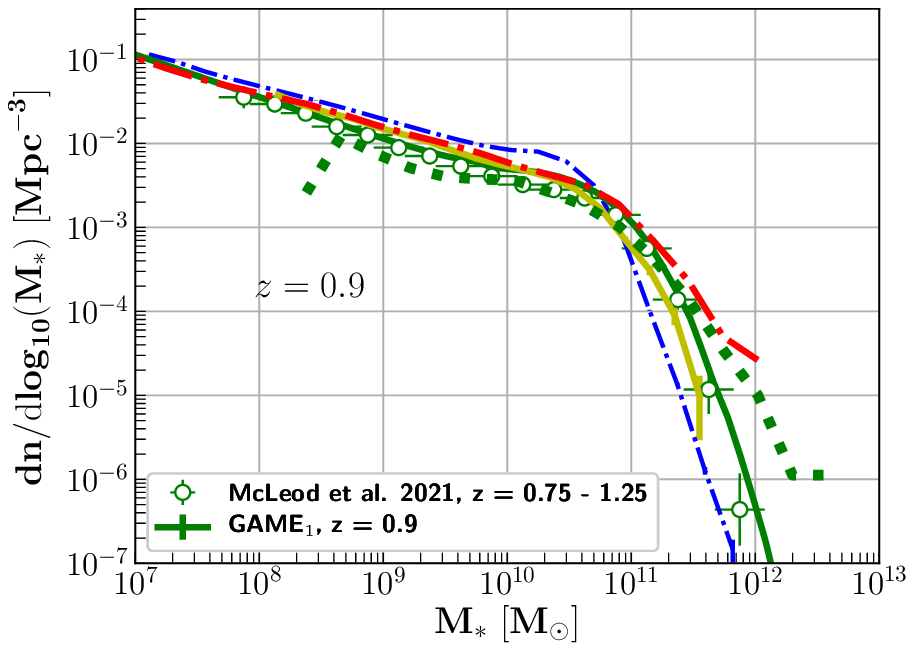}
\includegraphics[scale=0.55]{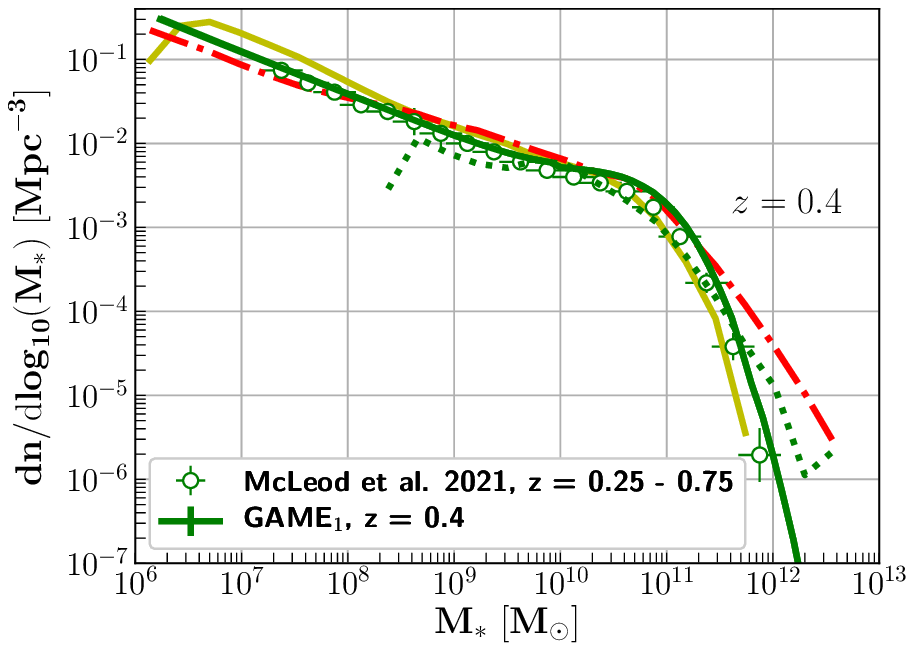}
\vspace{-1.0cm}
\caption{Comparison of GAME$_{1}$ (sold green line) with respect recent observations and IllustisTNG (red dot-dashed line), EAGLE  (orange solid line) Elucid+L-galaxies (blue dashed line) and Simba (green dotted line).}
\label{Conbeta00Simslow}
\end{figure*}

In conclusion, we just demonstrated how simple arguments related to gas accretion and gravitational collapse are enough to describe the evolution of the GSMF via the GAME$_{0}$ formalism. It seems that other mechanisms, besides of their importance, play a mild role in comparison to the physics of accretion/depletion of gas reservoirs to the GSMF evolution. Besides its success (Fig. \ref{GSMF0obs1}, \ref{GSMF0obs2} and \ref{figEpicG0}) and its strong theoretical motivation, GAME$_0$ needs a small re-tuning to precisely capture the observations of high mass galaxies at high redshifts. We remind once again that the inconsistency could always be due to the fact that ``observations'' of the GSMF are flawed. In the last decade, it is confirmed that the assumed SFH, adopted IMF, and employed SED fitting code can affect the results reported by observational studies especially at high redshifts \citep{Katsianis2019,Pacifici2023,Woo2024,Harvey2025} while the Eddington bias and photometric redshift errors that are not usually considered can also impact the high mass end. In addition,  we have to note that besides there is a census in the literature about an increasing CSFRD at $z > 2$ and that parameterizations which combine power laws and exponential decline,  reasonably capture the Cosmic SFH, some studies may suggest different forms like exponentially declining, constant, double power law or log-normal,  \citep{Simha2014,Carnall2019b,Arteaga2024}.

\section{The GAME$_1$ model, varying $\beta$ with $M_{\star,0}$}
\label{GAMEba}

\citet{Tacconi2018} observed that the timescale for gas consumption in galaxies at z = 0 averages around 2 Gyr, in accordance with GAME$_{0}$. However, the authors suggested that this value decreases by a factor of 2 at z=2.5 for their set of galaxies, indicating faster gas consumption at higher redshifts \citep{Saintonge2011,Tacconi2013,Bicalho2019}. In addition, \citet{Inayoshi2022} using JWST data proposed that higher star formation efficiencies should be attained at high redshifts for high mass galaxies. The above point out that galaxies at the high mass end possibly consume their gas faster to form stars at a higher pace at the early Universe.

The success of GAME$_{0}$ suggests that on average, galaxies maintain a nearly constant $\tau{\star} = 2$ Gyr for most mass bins close to the knee of the GSMF. However, the ``x'' symbols in the top panels of Fig. \ref{Conbeta} indicate where GAME$_{0}$ could be improved. The extremely high mass and extremely low mass bins deviate from  the simple $\Gamma$ growth pattern with $\alpha = 3.0$ and $\beta = 0.5$ gyr$^{-1}$  and thus it seems some improvements within our formalism should be applied preferably at extremely low masses/high masses and high redshifts.  According to Fig. \ref{GSMF0obs1BetaInvestigate} moderately larger/smaller values for $\beta$/star formation timescales are the key for this improvement and the $\beta_{\star}$ parameter  should slowly increase at high redshifts the more we approach the low and high mass ends. What kind of galaxies dominate these two regimes of the z = 0 GSMF ?

It is evident that elliptical galaxies are abundant both at the high and low mass ends (dwarf ellipticals) of the galaxy stellar mass function at z = 0  \citep{Aguerri2009,Scott2020,Pinter2023}. These objects have star formation histories that are expected to peak at higher redshifts with respect to galaxies that are located at the knee of the GSMF \citep{DeLucia2006}. The time of the peak of the SFH of a galaxy is set by the SFH timescale and thus the $\beta$ parameter in our formalism. What is the Physical motivation for the shorter SFH timescales (larger $\beta$) at high redshifts for these  galaxies ?

Major mergers and tidal interactions (which occur commonly at high redshifts) are known to increase the amount of dense gas, consequently accelerating star formation \citep{Paswan2018,Li2022b,Otero2022,Andersson2024} i.e., leading to shorter depletion times. Cosmological simulations and observational studies indeed indicate that major mergers affect the SFH of galaxies at high redshifts, leading to higher star formation rates at z $>$ 2 \citep{Fakhouri2008,Stewart2008,Asada2023}. Thus, mergers are expected to contribute approximately 20-30$\%$ of the baryonic mass assembly history of the high mass galaxies \citep{Dekel2009,Perret2014,Pearson2019}.  Additionally, present day elliptical galaxies are expected to have low angular momentum at higher redshifts. In this case gas is expected to collapse fast / being depleted at higher rates at high redshifts (i.e. smaller star formation history timescales are expected) \footnote{We note that besides mergers other factors may contribute to higher star formation efficiencies for high redshift galaxies. For example, the smooth cold gas accretion along cosmic filaments at z $>$ 2 \citep{Perret2014,Zhu2022}.}. \citet{Hensler2004}, \citet{Gutcke2022} and \citet{Seo2023} also noted that present day low mass elliptical galaxies possibly experienced a rapid collapse of gas which produced a high star formation activity at early times (bursts), followed by an extended and declining SFH.

GAME$_{0}$ considers growth through accretion and the accumulation of the available closeby resources. The scheme considers the same SFH (consistent with the CSFRD)  regardless of mass and it does not involve the contribution of mergers, tidal interactions or enhanced efficiencies (bursts) at high redshifts, which all combined could change the $\beta$ parameter at the low and high mass end.  However, the nature of the correction is quite interesting and seems to be increasing symmetrically on logarithmic mass scales (Fig. \ref{Conbeta}).

If we know that a variable is positive and that its logarithm has a certain mean (e.g. the $M_{\star}$ $=$ $\sim$ $10.85$ $M_{\odot}$ has a value of $\tau_{\star} = 2$ Gyr (or $\beta = 0.5$ Gyr$^{-1}$)) and a symmetrical standard deviation $\sigma$ in log scale, the least biased assumption we can make about the distribution of that variable is that it follows a log-normal distribution. \citet{Mutch2013} suggested that the efficiency in dark matter halos follows a log-normal distribution centered around the halo virial mass ${\rm M_{peak}}$, and with a standard deviation $\sigma_{M_{vir}}$ as follows: 
\begin{eqnarray}
\label{eq:PGgastconsumptionratesimpleLognorm2}
F_{Phys}(M_{vir}) = \epsilon_{M_{vir}} \times exp(-\frac{\Delta M_{vir}}{\sigma_{M_{vir}}})
\end{eqnarray}
where $\Delta M_{vir} = \log_{10}(M_{vir})-\log_{10}(M_{peak})$ and the $\epsilon_{M_{vir}}$ parameter represents the maximum possible efficiency. In addition, \citet{Sargent2012} and \citet{Bethermin2013} suggested a double log-normal distribution for the SFR of galaxies. Furthermore, \citet{Hazlett2024}  show that the distribution of the star formation efficiencies is log-normal and centered around a mean efficiency. Last, \citet{Haslbauer2024} gas depletion timescales are also fitted with a log-normal function. It seems it is quite common practice to adopt similar solutions in analytical frameworks to address star formation efficiencies. We note that besides a log-normal form, other reasonable approaches could be applied like a power law mass dependence \citep{Dave2017,Chen2020} or a  double power law dependence with respect a characteristic mass \citep{Shankar2006,Boettner2025}. 

In our formalism, we wish to keep the number of parameters low and find that choosing a double log normal \citep{Sargent2012,Bethermin2013} $\tau_{\star}$ for z $>$ 2 is indeed a valid starting point. Besides for the $z = 0$ GSMF, this time, for the case of GAME$_{1}$ we pick as our constrain the stellar mass function at $z = 8.0$ to constrain the high redshift correction (bottom left panel of Fig. \ref{Conbeta}). Taking into account all the above points, we explore the following change for $\tau_{\star}$ at $z > 2$ with respect GAME$_{0}$.
\begin{eqnarray}
\label{eq:PGgastconsumptionratesimpleLognorm}
\begin{split}
\tau_{\star} = 0.5 + \\ 
0.5 \times e^{(-0.5 \times (log10(M_{\star, 0})-10.850))^2} \, + \\ 
e^{(-0.5 \times (log10(M_{\star,0})-10.850)^2)},
\end{split}
\end{eqnarray}
which can be written as $\tau_{\star} = 0.5 + 1.5 \times e^{(-0.5 \times (log10(M_{\star,0})-10.850)^2)}$ while $\beta$ can be written as:
\begin{eqnarray}
\label{eq:PGgastconsumptionratesimpleLognorm24}
\beta_{\star} = \frac{1}{0.5 + 1.5 \times e^{(-0.5 \times (log10(M_{\star,0})-10.850)^2)} }.
\end{eqnarray}
Equation \ref{eq:PGgastconsumptionratesimpleLognorm} ensures that $\tau_{\star} = 2$ Gyr for the Knee of the GSMF, building upon the success of GAME$_{0}$ while the only additional parameter is the mass of $10^{10.850}$ $M_{\odot}$. Interestingly, we note that instead of adding a prescription (like AGN or SNe feedback) to decrease the SF efficiency at low redshifts in GAME, we actually required a prescription to increase the SF efficiency at high redshifts, in accordance with both theoretical and observational arguments \citep{Saintonge2011,Tacconi2013,Bicalho2019}.

\subsection{GAME$_1$ and Observations}
\label{GAME12vsObservations}

In Fig. \ref{Conbeta38} we present the evolution of the GSMF predicted from the GAME$_{1}$ model (represented by the black dashed line) from redshift z = 3.0  to z = 7.0. In addition, we present the observations of \citet{Song2016} and \citet{Stefanon2021}. At z = 7, z = 6, z  = 5 we are able to reproduce the observed results.  We note that  the overall shape of the GSMF produced by the GAME$_{1}$ model has a mild distinctive bump accompanied by a notable upturn for the low mass end. These characteristics for the GSMF are sometimes encountered in the literature  \citep{McLeod2021,Kelvin2014,2024arXiv241204553P,Boettner2025}, with some studies using a double Schechter function to describe their results.  At z = 4, GAME$_{1}$ performs better at the high mass end with respect GAME$_0$ but predicts a larger number of low mass objects. This maybe reflects that the $\beta$ parameter has to be slightly reconsidered or that the observations at z = 4 are incomplete at the low mass end. At redshift 3 we see that the observations of \citet{Wang2024b} are between GAME$_{0}$ and GAME$_{1}$. We remind the reader that observational studies, especially at high redshifts, are uncertain and that these small deviations could always be attributed to limitations in observations.
In Fig. \ref{Conbeta00} we present the evolution of the GSMF predicted from the GAME$_{1}$ model (represented by the black dashed line) from redshift z=0.0  to z = 2.0. The results as expected are very similar to the GAME$_{0}$ model. In the appendix \ref{RMSDPrecision} we demonstrate that GAME$_{1}$ has an impressive RMSD of 0.27 at z = 0-8 with respect observations at the low and high mass end, performing slightly better than EAGLE (0.4 dex) and IllustrisTNG (0.34 dex). In addition, excellent AIC quality checks are demonstrated in appendix \ref{AICquality}. We plan to consider further modifications of GAME$_{0}$, beyond the log-normal approach used in GAME$_{1}$,  in future work when additional observations beyond the GSMF will be considered. Treating the low mass end and high mass ends separately, while adopting more parameters, has the potential to establish a better comparison with  observations but this is beyond the scope of our current work which focus on reproducing the GSMF in a straightforward way, employing few parameters.

\subsection{GAME$_1$, Simulations and JWST observations}
\label{GAME1vsSimulations}

A recent challenge in the study of galaxy formation is the early assembly of massive galaxies \citep{Weibel2024}. Observational data suggest that the number density and maximum stellar mass of these galaxies in the early universe exceed the predictions of existing models, potentially posing a significant challenge to $\Lambda$ CDM cosmology. However, \citet{Wang2024b} pointed out that a common limitation in many previous studies is the substantial uncertainty in estimating stellar masses, due to the absence of constraints in the near-infrared part of the rest frame of the spectral energy distribution, which is vital for accurately determining the properties of M$_{\star}$ and dust \citep{Katsianis2020,Kouroumpatzakis2021,Xie2023,Qin2024}. Thus, \citet{Wang2024b} used data from an extensive JWST/MIRI survey in the PRIMER program to perform a systematic analysis of massive galaxies at $z \sim 3 - 8$. The results of the authors (using analytical Schecter functions) are represented by filled circles in Fig. \ref{Conbeta00Sims}. At a similar redshift range \citet{Navarro-Carrera2024} analyzed a sample of 3300 galaxies between redshifts $z \sim 3.5$ and $z \sim 8.5$ selected from JWST images in the Hubble Ultra Deep Field (HUDF) and UKIDSS Ultra Deep Survey field, including galaxies with stellar masses as low as $\sim$ 10$^{8}$ $M_{\odot}$. The combination of depth and wavelength coverage of the data enabled construction of the low-mass end and the knee of the GSMF. The analytical results of \citet{Navarro-Carrera2024} are represented by the open diamonds of Fig. \ref{Conbeta00Sims}.

In this section, we compare GAME$_1$ with observations and other models. At this point, it is important to recall some important aspects of how star formation is addressed in state-of-the-art simulations at high redshifts. \\

{\bf EAGLE:} Among the different cosmological simulations considered in this work, EAGLE is the simplest. The simulation reproduces realstic SFRs and stellar masses while it employs a very straightforward thermal feedback scheme for both the SNe and AGN feedback  \citep{Crain2015,Katsianis2017,Trayford2019,Manuwal2024}. The parameters within the simulations were tuned to reproduce the GSMF, the BH-stellar mass relation, and the stellar mass-stellar size relation at z = 0. The simulation performs very well with respect to the observations at high redshifts, besides that it was not tuned to do so. However, we stress that the simulation does not pass the strong convergence test \citep{Schaye2015} \footnote{We note that this test is not passed by any of the cosmological simulations considered in this work.} and does not produce sufficient star-burst galaxies, thus is in disagreement with observations \citep{Castillo2024}. Our GAME$_{1}$ model performs similarly to EAGLE, while GAME$_1$ usually demonstrates better RMSD values with respect observations (tables \ref{B3} and \ref{B4}). Our scheme performs slightly better at the low mass end at z = 8.0 to z = 4.0. The knee of the GSMF is well reproduced by GAME at the same red-shift interval. Going to redshifts z = 3 to 0.4 we see once again a good consistency of GAME$_{1}$ and observations. However, both GAME$_{1}$ and EAGLE predict a slightly higher value for the GSMF at the knee of the distribution. Both models have very similar values for the high mass end. In general, GAME$_{1}$ and EAGLE produce similar results for all the mass bins of the GSMF but this was done by using a {\it completely different approach and motivated by different physics/mechanisms}. This highlights how strong the degeneracies between models are when solely GSMFs are investigated. \\

{\bf Simba}: \citet{dave2019} has shown that Simba is capable of successfully reproducing the observations of \citet{Tomczak2013} and \citet{Song2016}. However, the authors pointed out that a reduction for the mass loading factor ($\eta$) was required at high redshifts to allow for early galaxy growth, which would not be possible because of the poor resolution of Simba. Thus, the authors tuned the results on the basis of the resolution. {\it \citet{dave2019} explicitly stated that due to the ad hoc nature of this correction, the results for the galaxy stellar growth at high redshifts from Simba should be considered as tuned rather than predictive.} In this work, we find that Simba performs excellently with respect to the observations of \citet{McLeod2021} and \citet{Wang2024b} at z = 2.0 and 3.0. However, going to higher redshifts, we see that the knee of the GSMF is better reproduced by GAME$_{1}$. We note that Simba reproduces very well the high mass end of the GSMF of \citet{Wang2024b}. The impressive average RMSD of Simba at z = 0-8 of 0.35 is a bit worse than the performance of GAME$_{1}$ (0.276 dex).\\

{\bf IllustrisTNG}: The parameter selection ($33$ for SF, AGN and galactic winds) is discussed in \citet{Weinberger2017} and \citet{Pillepich2018}. We must note that the IllustrisTNG model has been calibrated/tuned to reproduce the following observations \citep{Nelson2019}:
\begin{itemize}
\item the galaxy stellar mass function at z = 0
\item the stellar-to-halo mass relation at z = 0
\item the total gas mass content within the virial radius of massive groups at z = 0
\item the stellar mass—stellar size at z = 0
\item the BH–galaxy mass relation at z = 0
\item {\it the cosmic star formation rate density (CSFRD) at z = 0-10}
\end{itemize}
This is a very large number of constrains and IllustrisTNG results (like in the case of Simba) should be considered at an extent tuned rather than predictive for high redshift SFRs, considering especially the CSFRD constrain. However, we note that IllustrisTNG performs very well at the high-mass and low-mass end of the GSMF. The Knee of the distribution is slightly over-predicted, a characteristic that is found both in GAME and EAGLE. Interestingly, we find a good consistency of IllustrisTNG with respect to the observations of \citet{Navarro-Carrera2024}, even though the authors pointed out very large inconsistencies between IllustrisTNG and their observations. The authors reported a good agreement for z = 4 and z = 5. However,  the normalization of the simulated GSMF was significantly lower with respect to their observations. Thus, IllustrisTNG was reported to have a lower density of galaxies at all stellar mass bins. However, we believe that actually IllustrisTNG has the potential to reproduce \citet{Navarro-Carrera2024} observations (see Fig. \ref{Conbeta00Sims}) and that the inconsistency reported in \citet{Navarro-Carrera2024} could be related to how limits are imposed to calculate the total stellar mass within a galaxy. We find that IllustrisTNG has an impressive average RMSD from z = 0 to 8 of 0.34 dex, performing a bit worse than GAME$_{1}$. However, the simulation was tunned extensively with respect observations and this inevitably invloved a very high number of parameters, dedicated to reproduce the CSFRD and GSMF at z=0. This results in a much lower AIC quality with respect GAME$_{1}$ {\it and} GAME$_{0}$ (tables \ref{B5}, \ref{B6}, \ref{B7} and \ref{B8}).

In conclusion, GAME$_{1}$ has a different philosophy with respect to the cosmological simulations described above, similar to other analytical frameworks \citep{Sargent2012,Bethermin2013}. The backbone of the model (GAME$_{0}$) required few parameters, which are derived from the first principles of accretion/gravitational collapse (Section \ref{Why?}) within a multi-scale context. No modification, tuning or additional complicated prescriptions were required to have a good grasp of the evolution of the GSMF, and GAME$_{0}$ was actually quite successful considering its simplicity, especially at $z<1.5$. An overall excellent performance for low and intermediate mass galaxies was evident for GAME$_{0}$ which had an average RMSD of 0.17 dex at z =0-8. At high redshifts, we explored a rather modest tuning for the parameter $\beta$ via the GAME$_{1}$ model. Besides that cosmological simulations, like Simba and IllustrisTNG, have to as well tune their parameters to reproduce the high redshift star formation the main advantages of tuning $\beta$/$\tau_{\star}$ in GAME$_{1}$ are the following:
\begin{itemize}
\item Far fewer parameters are needed to be added/tuned.
\item We adjusted the results against very few observables (GSMF at z = 8).
\item The motivation for tuning high redshift $\tau_{\star} $ is physically motivated (mergers and low angular momentum at the low- and high-mass ends that decrease $\tau_{\star}$ as expected from the theory, see section \ref{GAMEba}) rather than resolution motivated (as in the case of Simba).
\end{itemize}
However, unlike GAME$_{0}$, similar to state-of-the-art models  GAME$_{1}$ results for {\it the high mass end} should be considered also as tuned rather than predictive at z = 4-8.

%In conclusion, GAME$_{1}$ as an improvement of GAME$_{0}$ is expected to be successful at reproducing the observed GSMF. Surprisingly, we report a very good consistency of state-of-the-art simulations like IllustrisTNG with respect new JWST data (which tend to reproduce results very similar to past observational studies). 
We also note that besides that we demonstrated in our work that cosmological simulations have a somewhat consistent GSMF with new JWST data this does not guarantee that other observables are reproduced. Actually, JWST observations demonstrate much more quenched galaxies at high redshifts \citep{Weller2024}, different quenched fractions \citep{Kartaltepe2023} and galaxies residing already in galaxy clusters \citep{Sun2024} indicating that cosmological simulations have to reconsider their star formation and feedback prescriptions or even consider alternatives to $\Lambda$CDM \citep{Liu2024}. Regardless, we are looking forward to exploring and improving the performance of our formalism and comparing state-of-the-art simulations with respect to additional properties of galaxies and even higher redshifts in future work.

\section{Conclusions}
\label{sec_concl}

Our work presents a simple yet physically motivated formalism to describe the evolution of the galaxy stellar mass function (GSMF) across cosmic time. The main conclusions are as follows.

\begin{itemize}
\item  While valuable for exploring different subgrid models , current state-of-the-art cosmological simulations (IllustrisTNG, EAGLE, Simba)  suffer from severe limitations, including: lack of consistency in subgrid physics schemes across groups, reliance on empirical recipes lacking strong multi-scale physical justification, continuous re-calibration of numerous  free parameters needed to describe star formation and feedback after new observations emerge, severe resolution effects, and limitations to probe the high redshift Universe (Sections \ref{sec_intro} and \ref{GAME1vsSimulations}). 
\item  The $\Gamma$ growth pattern encompasses important elements of growth when 1) small scale interactions that are responsible for accretion/accumulation of nearby resources are involved, 2) finite resources are available  and 3) many random factors (usually described by exponential or $\Gamma$ distributions) aggregate \citep{Vazquez2021,Lippiello2022}. The GAME$_{0}$ growth motif which employs a minimum number of parameters ($\alpha$ =3.0, $\beta$ =0.5 Gyr$^{-1}$) and a $\Gamma$ form, successfully reproduces the observed GSMF from $z = 0$ to $z = 8$  using a growth pattern arising from the fundamental physics of gravity and accretion (Section \ref{GAME0Msparam}), {\it without any tuning or adopting additional constrains}. Our model, which adopts an early power-law increase followed by an exponential decline, arises naturally from the growth-dependent consumption of available resources and the accumulation of nearby material, a pattern seen across many scales (from molecular clouds to dark matter halos) and disciplines in nature (e.g. epidemiology). The GAME models perform slightly better than state-of-the-art simulations like EAGLE and IllustrisTNG in reproducing the GSMF evolution up to M*, despite employing a completely different approach with far fewer free parameters (Section \ref{GAMEba}). In addition, our formalism reproduce the total distribution of the GSMF for a wide redshift range involving the last 9.5 billion years. The parameters of GAME$_{0}$ are directly connected to ab initio theory with $\alpha$ describing the distribution of the accreting matter and $\beta$ being the dynamical timescales of halos (Section \ref{Why?}).
\item Despite these successes deviations occur for the most massive/rare galaxies at high redshifts (z $>$ 1.5) for the GAME$_{0}$ model, these can be resolved by a modest adjustment of the $\beta$ parameter resulting in the GAME$_{_1}$ model. The above tuning is physically motivated by expected changes in star formation timescales due to early mergers, tidal interactions, and low angular momentum in elliptical galaxies that dominate the low and the high mass ends. However, the results of GAME$_{1}$ at high redshifts for {\it the high mass end}  should still be considered as "tuned" rather than purely predictive, a limitation shared with cosmological simulations (Section \ref{GAMEba}). We note that our findings suggest an increase in SF efficiency at high and low mass ends and has a different approach to the idea of using feedback mechanisms at lower redshifts to decrease the star formation rate. This result is in agreement with recent studies that suggest high redshift galaxies have high star formation efficiencies.
\item Interestingly, our GAME models and state-of-the-art cosmological simulations like IllustrisTNG demonstrate good consistency in reproducing the evolution of the galaxy stellar mass function at high redshifts with respect to recent observations from the James Webb Space Telescope, which tend to align closely with past studies. This suggests that both analytical frameworks like GAME and numerical simulations are capable of capturing the broad growth of galaxy growth even at high redshifts, despite their differing approaches and levels of complexity (Section \ref{GAMEba}).
\end{itemize}
In summary, the GAME models highlight an elegant simplicity underlying the process of the growth of galaxy stellar mass. We are looking forward to further developing our formalism and exploring its ability to replicate other key observables relevant to galaxy formation and evolution (SFRs, gas contents and quenched fractions) in future work and disentangle the important factors that govern galaxy growth. 

\section*{Acknowledgments}
We thank the anonymous reviewer for carefully reading our manuscript and providing us with useful suggestions and comments that improved significantly the quality of our work.
A.K has been supported by the 100 talent program of the Sun Yat-sen University,  the Guangdong Basic and Applied Basic Research Foundation with No 2025A1515012670 and the Yang-Yang faculty award from the Shanghai Jiao Tong University. X.Y. is supported by the national science foundation of China (grant Nos. 11833005,11890692,11621303) and Shanghai Natural Science Foundation,grant No.15ZR1446700 and 111 project No. B20019. We acknowledge the science research grants from the China Manned Space Project with NO.  CMS-CSST-2021-A02.  A.K. would like to express his  gratitude to Claudia Lagos and Houjun Mo for  valuable discussions. WC is supported by Atracci\'{o}n de Talento fellowship no. 2020-T1/TIC19882 granted by the Comunidad de Madrid and by the Consolidación Investigadora award no. CNS2024-154838 granted by the Agencia Estatal de Investigación (AEI)  in Spain. He also thanks the Ministerio de Ciencia e Innovaci\'{o}n (Spain) for financial support under Project grant PID2021-122603NB-C21, ERC: HORIZON-TMA-MSCA-SE for supporting the LACEGAL-III Latin American Chinese European Galaxy Formation Network) project with grant number 101086388, and  the science research grants from the China Manned Space Project

\section*{Data Availability Statement}

No new data were generated or analyzed in support of this research.

\bibliographystyle{mn2e}	
\bibliography{KatsianisMNRAS10.bib}
%\bibliography{Katsianis_MNRAS9}

\appendix

\begin{figure*}
\centering
\includegraphics[scale=0.55]{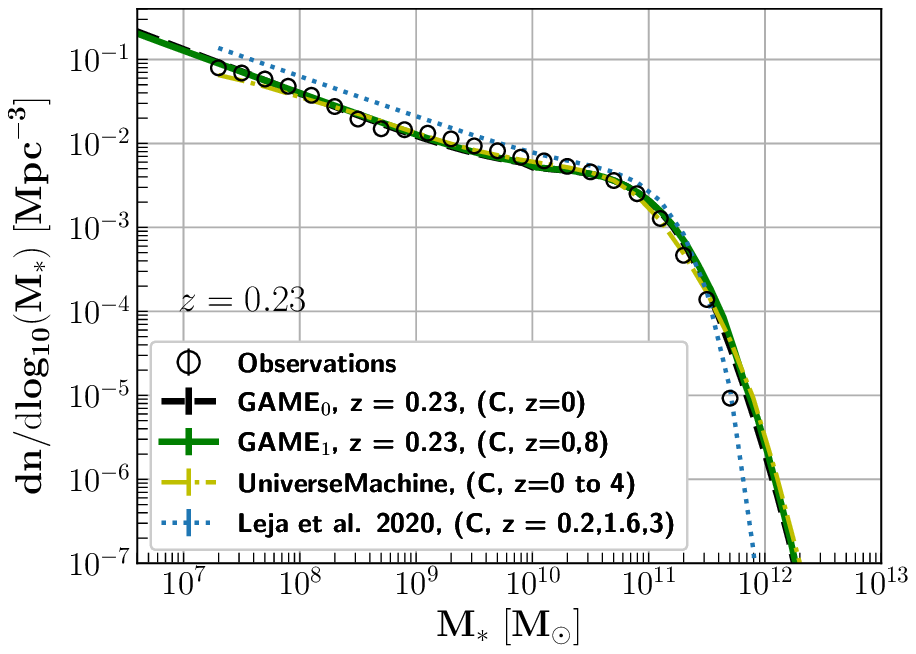}
\vspace{-1.0cm}
\includegraphics[scale=0.55]{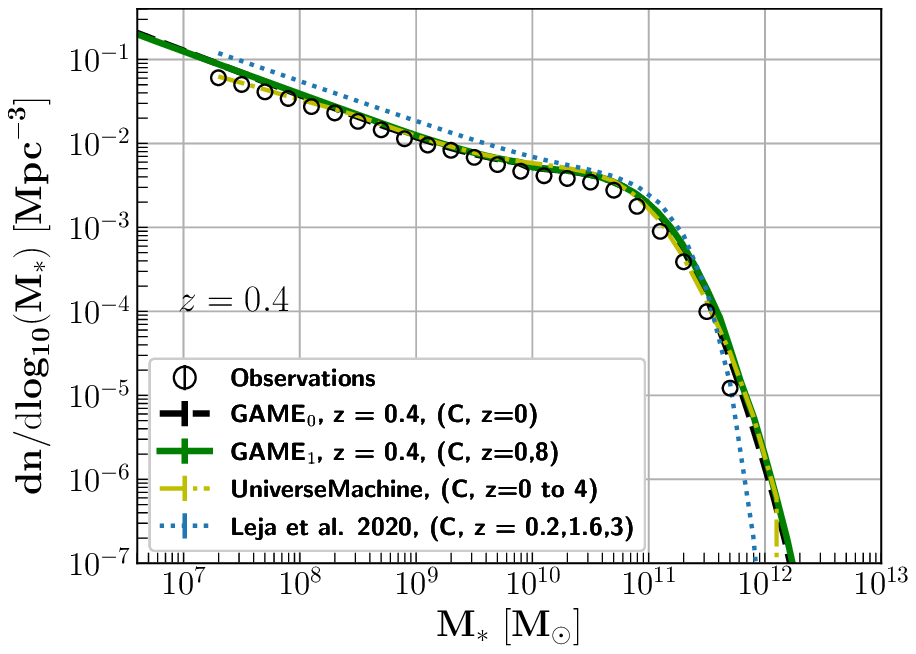}
\includegraphics[scale=0.55]{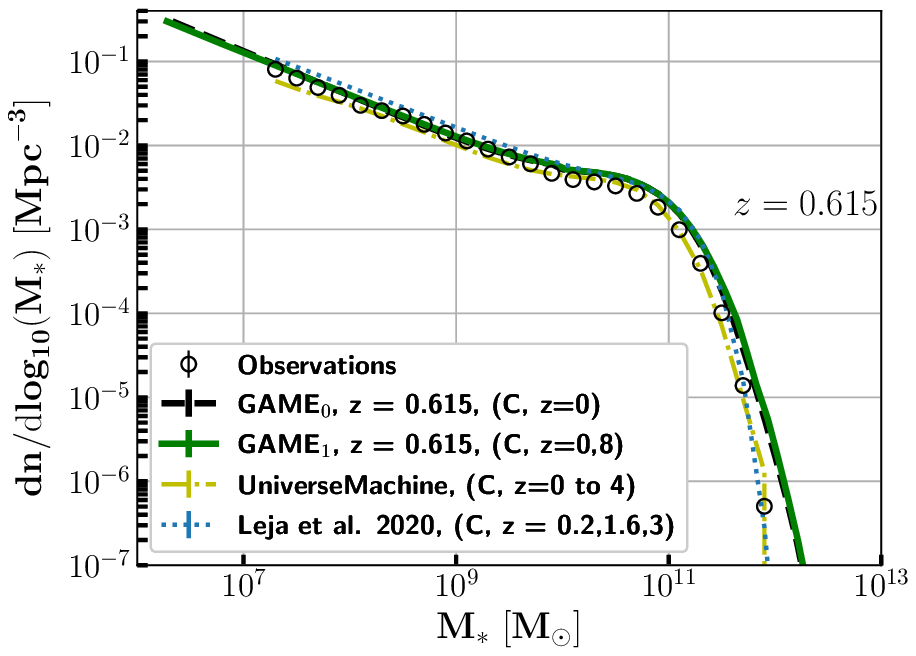}
\vspace{-1.0cm}
\includegraphics[scale=0.55]{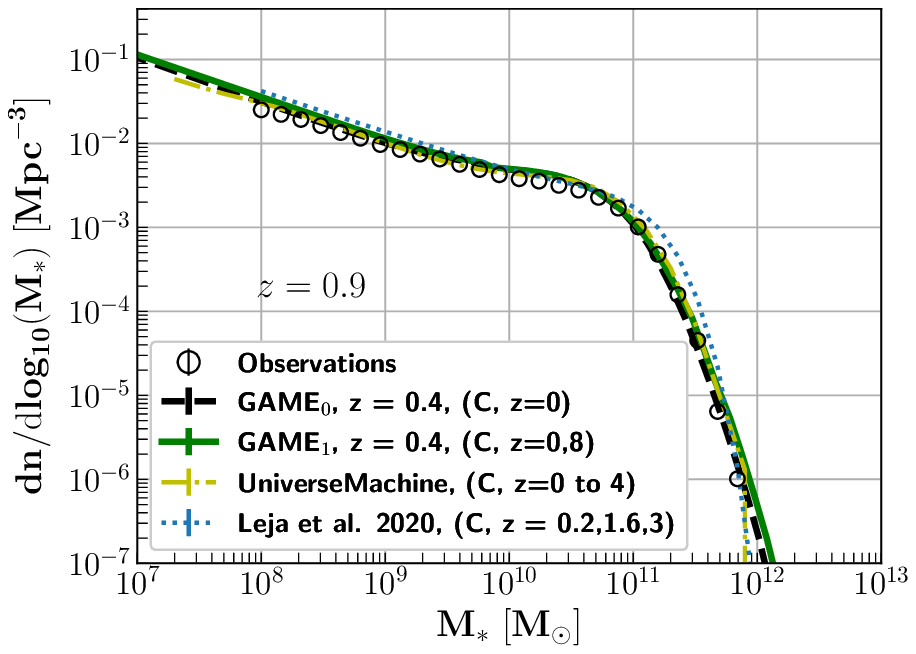}
\includegraphics[scale=0.55]{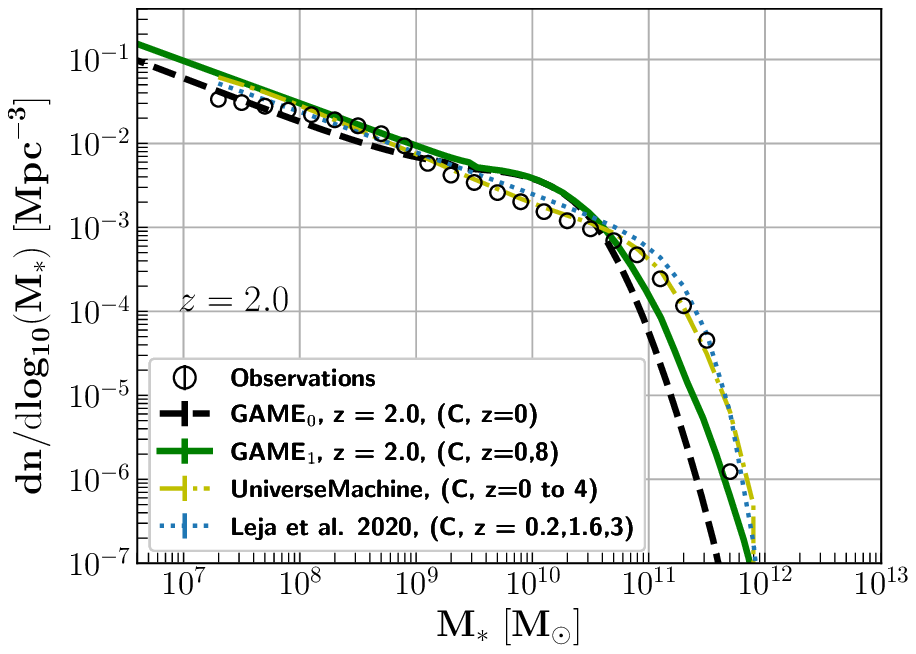}
\includegraphics[scale=0.55]{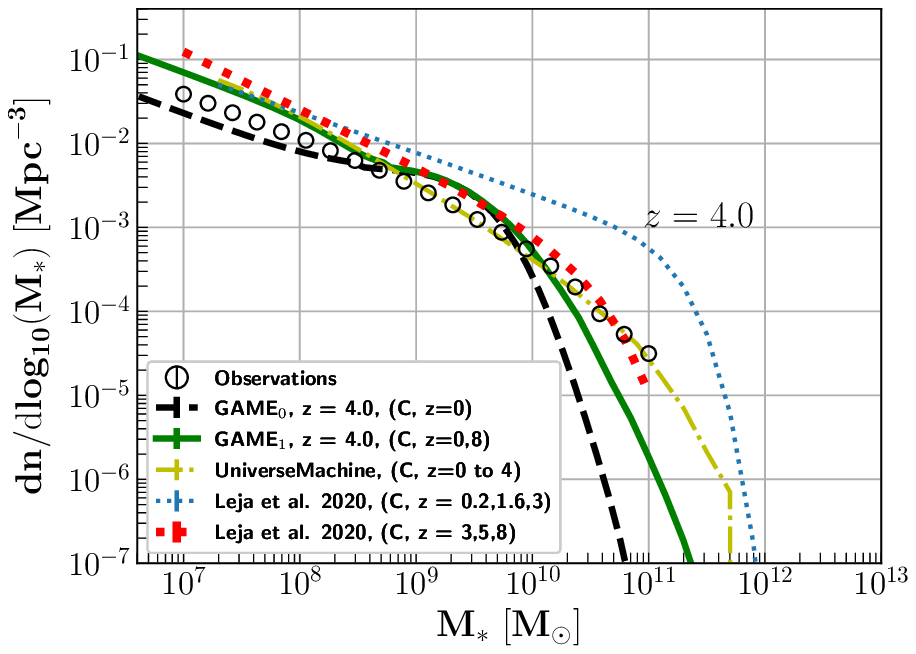}
\vspace{-1.0cm}
\caption{Comparison of the GSMFs produced by the models GAME$_{0}$ (black dashed line), GAME$_{1}$ (green solid line), \citet{Leja2020} (blue dotted line) and UniverseMachine (yellow dot-dashed line). We note that UniverseMachine was constrained to reproduce the GSMF at z = 0-4 while the empirical model of \citet{Leja2020} was constrained at z = 0.2, 1.6 and 3.0. }
\label{UniLeja}
\end{figure*}

\begin{figure*}
\centering
\includegraphics[scale=0.55]{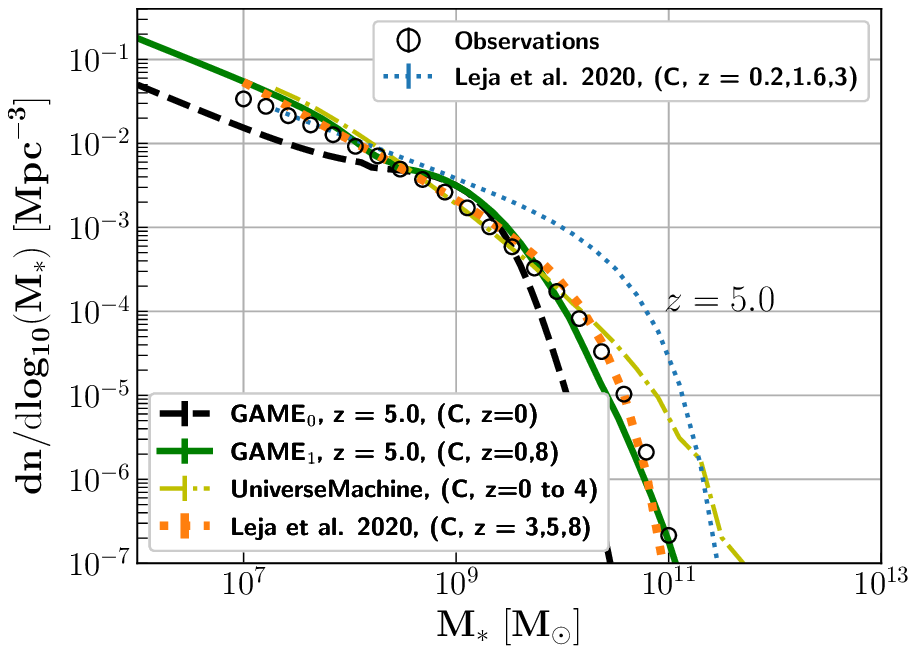}
\vspace{-1.0cm}
\includegraphics[scale=0.55]{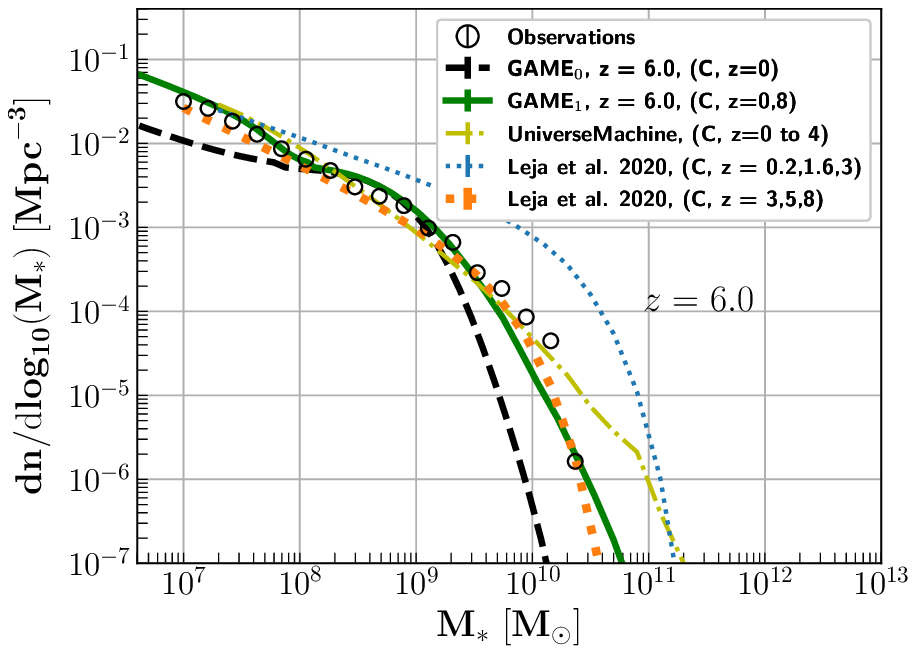}
\includegraphics[scale=0.55]{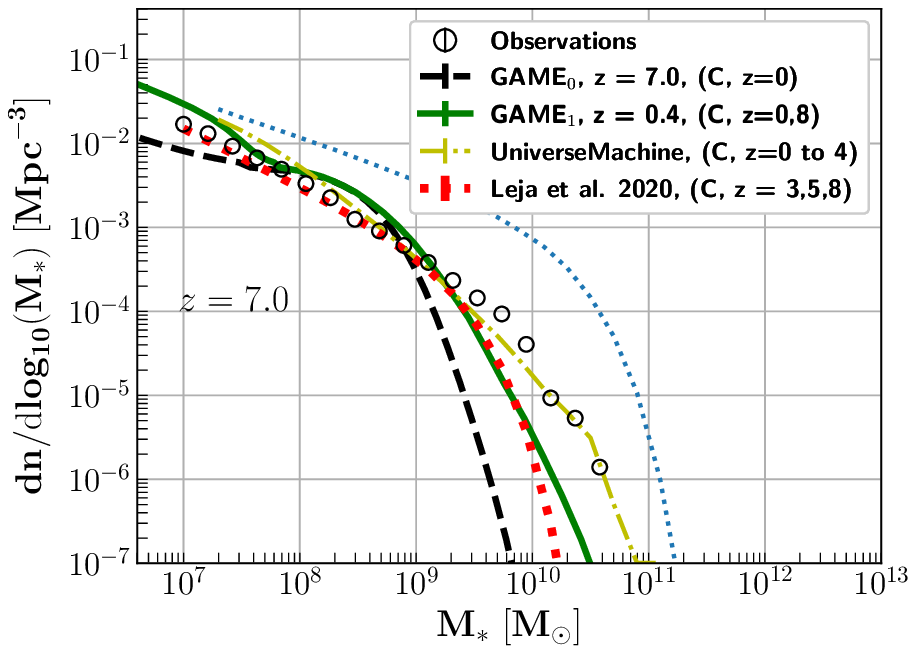}
\includegraphics[scale=0.55]{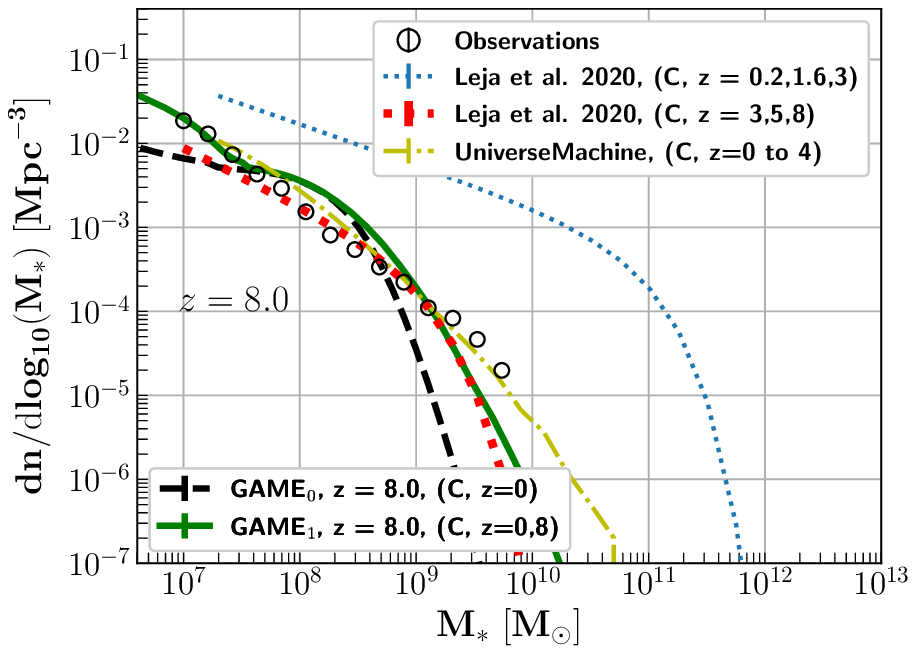}
\vspace{-1.0cm}
\caption{Comparison of the GSMFs produced by the models GAME$_{0}$ (black dashed line), GAME$_{1}$ (green solid line), \citet{Leja2020} (blue dotted line) and UniverseMachine (yellow dot-dashed line). We note that UniverseMachine was constrained to reproduce the GSMF at z = 0-4 while the empirical model of \citet{Leja2020} was constrained at z = 0.2, 1.6 and 3.0. The model of \citet{Leja2020} can reproduce the z = 4-8 GSMF as long as it is further constrained at z =3, 5 and 8 (represented by the orange dotted line).}
\label{UniLeja2}
\end{figure*}

\section{A comparison with empirical models.}
\label{comparisonEmpirical}

In  our work we compare GAME$_0$ and GAME$_1$ with state-of-the-art models like EAGLE, IllustrisTNG, Simba and L-Galaxies (sections \ref{GAME0} and \ref{GAMEba}) . We demonstrated that each model has different strengths and shortcomings but overall most models considered produced a sensible evolution for the GSMF. Besides cosmological hydrodynamic simulations, the evolution of the GSMF has also been followed by empirical models.  For example, two baseline models are described in \citet{Behroozi2019} and \citet{Leja2020}.

UniverseMachine is a sophisticated computational framework designed to follow {\it precisely} the SFH and the GSMF of galaxies  within the $\Lambda$CDM paradigm. To do so it uses a comprehensive list of empirical relations that parametrize the correlations between galaxy evolution and halo assembly. UniverseMachine is constrained by various observational data, including:
\begin{itemize}
\item {\it the GSMF at z = 0-4}, spanning 90$\% $ of the history of the Universe,
\item the cosmic star formation rate density at z =0-10,
\item the specific star formation rates at z = 0-8,
\item the UV luminosity function at z = 4-10,
\item quenched fractions at z = 0-4,
\item Autocorrelation functions for quenched/SF/ galaxies from SDSS (z = 0)
\item cross-correlation functions from SDSS at z = 0,
\item autocorrelation functions for quenched/SF galaxies from PRIMUS at z = 0.5,
\item quenched fraction of primary galaxies as a function of neighbour density at z =0,
\item median UV–stellar mass relations at z =4-7,
\item and the IRX–UV relation at z =4-7
\end{itemize}

In Fig. \ref{UniLeja} and \ref{UniLeja2} we present the GSMFs generated by UniverseMachine, represented by the yellow dot-dashed line, at z = 0.23-4 and 5-8, respectively. The open circles represent the median of all the observed GSMFs considered in this work.  As expected UniverseMachine can reproduce precisely (the RMSD performance is presented in appendix \ref{Quantify}) the observed GSMF at z = 0-4. This is due to the fact that the observed GSMF at this redshift interval has been used as a key constraint for the development of the model and the tuning of the {\it 44 parameters} involved. The high number of complexity and the additional observational constraints considered (e.g. the CSFRD at z = 0-10 and the UV luminosity functions at z= 4-10) ensured that the simulated SFHs resemble closely the observed SFHs even at higher redshifts (z = 5-8).  The main goal of the framework to {\it precisely} capture the observed evolution of galaxies and study the relation between galaxies and dark matter halos at the expense of adding many parameters is well motivated, as accuracy is important for the generation of a robust mock catalogue.  However, we note that GAME has different objectives. In our work we try to decrease the necessary parameters to describe the evolution of the GSMF, relying on the factorial growth pattern (i.e. 2 parameters) described in \citet{Katsianis2021b,Katsianis2023}.  In addition, our aim is to explore the underlying Physics of each parameter (section \ref{Why?}) going beyond a pure empirical approach . Last, similarly to other efforts (like EAGLE) we only tune our results for the GSMF at z = 0 in order to evaluate if our model is actually predictive at higher redshifts. We then explore, what further elements we need to add (GAME$_{1}$ employs 1 more parameter) to improve further the comparison with observations.
 
\citet{Leja2020} presented a comprehensive analysis of the GSMF over the redshift range z = 0.2 to 3.0 using the Prospector SED-fitting code. The study was able to resolve some important discrepancies between observed star formation rate densities and stellar mass buildup by incorporating more flexible star formation histories. Some key findings include:
\begin{itemize}
\item The inferred stellar masses are typically higher than those from traditional methods, primarily due to the older stellar ages inferred by Prospector. This results in a higher number density of galaxies at a fixed stellar mass than previous measurements, leading to a higher total stellar mass density (~50$\%$ higher than prior studies),
\item The low-mass end slope remains approximately constant over time, while the massive end exhibits minimal evolution below z$<1$,
\item A continuity model is introduced to describe the evolution of the GSMF.
\end{itemize}
The continuity model of \citet{Leja2020} is provided in the appendix B of their work. The python rootine  is designed to compute the evolution of the GSMF at z = 0-3 by providing the parameters of the double Schechter functions of the GSMF at z = 0.2, z= 1.6 and z = 3.0. Thus, the number of necessary parameters needed (excluding errors) is 11, since a constant low-mass end slope is adopted.   We present the predicted GSMFs  at z = 0.23-4 and 4-8, respectively. The empirical model, represented by the blue dotted line in Fig. \ref{UniLeja} is able to reproduce successfully the GSMF at z = 0-3. This can be attributed to both to the fact that the scheme was tuned at z =0.2, 1.6 and 3.0 and that it properly captures the evolution of different mass bins. The small offsets at z =0.23 and z = 0.4 can be attributed to the fact that the Prospector SED-fitting code combined with more flexible SFHs reproduced slightly different results than other methods at the low mass-end.  However, in order to reproduce the observations at z = 4-8 (Fig. \ref{UniLeja2}) a further input of parameters is required\footnote{If we do not consider additional parameters and tuning with respect the high redshift GSMF the continuity model would produce inconsistent results with respect observations (blue dotted line of Fig. \ref{UniLeja2})}. We performed the necessary tuning by fitting the results at z = 5 and 8 (represented by the orange dotted line). First, the low mass end slope has to be reconsidered and change from -1.48  to -1.68 . In addition, the parameters of the double Schechter functions for two additional redshifts (5 and 8) have to be added (+6 parameters) . Thus, {\it a total of 18 parameters}  and a constraining at {\it z = 0.2, 1.6, 3, 5 and 8} would be required to capture the evolution of the GSMF at z=0-8 via the continuity model of  \citet{Leja2020} at z =0-8.

\section{Quantifying success.}
\label{Quantify}
 
Quantifying the success of a model is a useful step in understanding its precision and shortcomings. Several statistical metrics can be used to evaluate how well a model describes the observed data. Among these, the Root Mean Squared Distance (RMSD) and the Akaike Information Criterion (AIC) stand out as two essential tools. 

\subsection{Using RMSD to evaluate the precision of a model.}
\label{RMSDPrecision}

The RMSD is a widely used metric that quantifies clearly the difference between observed and predicted values \citep{REVA1998141,Carugo2001,Fukutani2020}. In our analysis, we compute the RMSD between the observations and model values using logarithmic differences. Specifically, the RMSD is calculated as the square root of the average squared differences between the logarithms of the model and observation values as follows:
\begin{eqnarray}
\label{RMSD}
\text{RMSD} = \sqrt{\frac{1}{n} \sum_{i=1}^{n} \left(\log_{10}(\text{model}_i) - \log_{10}(\text{observation}_i)\right)^2},
\end{eqnarray}
where \(n\) is the total number of data points. Thus, the units of the RMSD in our work are expressed in dex. 

RMSD is a straightforward way to evaluate how {\it precisely} a model reproduces the observations. and a low value of RMSD indicates a good performance.  The RMSD is an absolute measure, meaning its value alone conveys meaningful information about how well the model reproduces the data. For example, a RMSD of 0.15 dex reflects that the typical deviation between the model and observations is 0.15 dex (i.e. a factor of 1.4 difference). We remind that differences of 0.3 dex between observations and models can be attributed to the uncertainty of the observed stellar masses.

\begin{figure*}
\centering
\includegraphics[scale=0.55]{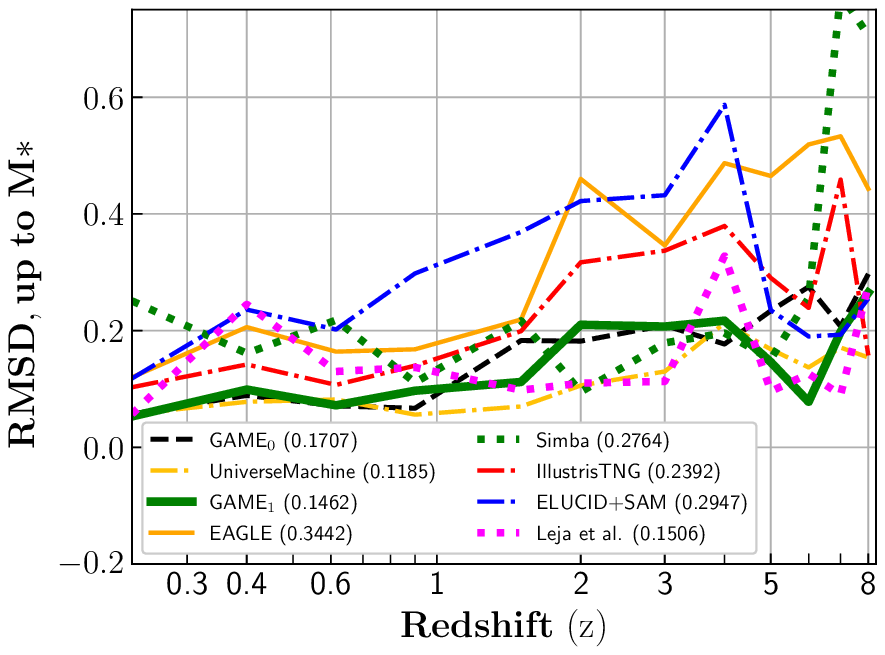}
\includegraphics[scale=0.55]{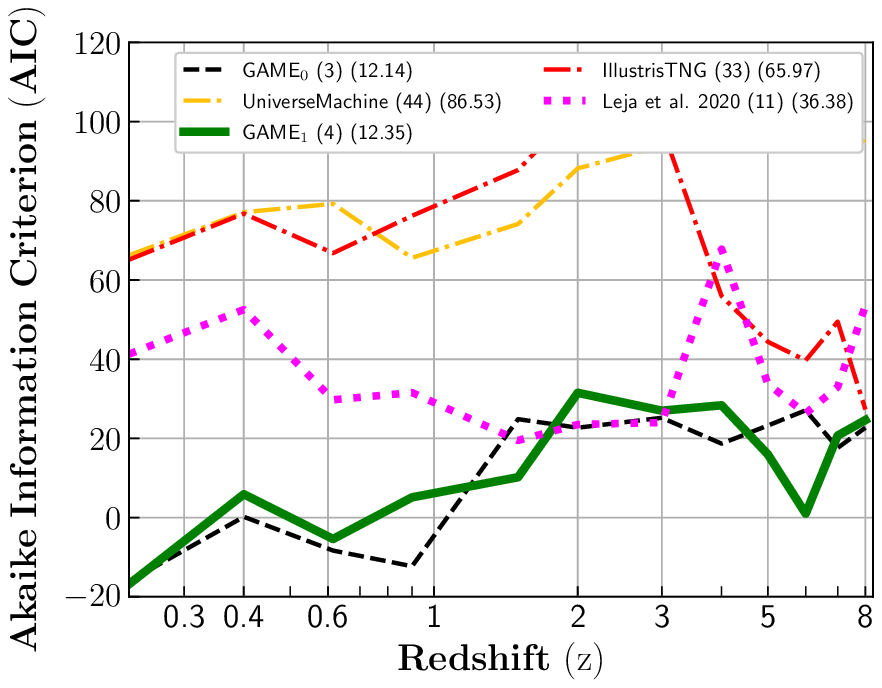}
\vspace{-1.0cm}
\caption{Left: Evolution of the RMSD up to M* of each model from z= 0.23 to 8.0. The average RMSD of all redshifts  is also noted. GAME$_0$ (with average RMSD of 0.171 dex) and GAME$_1$ (with average RMSD of 0.146 dex) reproduce the observed GSMF with {\it high precision} on par with models which use a significantly higher number of parameters/complexity. Right. Akaike Information Criterion (AIC) used to evaluate the quality of the different models.  A low AIC value for a model means that it is able to achieve an optimal balance between model complexity and goodness-of-fit. Both GAME$_0$ and GAME$_1$ have the highest AIC quality among the models considered.}
\label{GSMF0obsRMSDuptoMstar}
\end{figure*}

In tables \ref{B1} and \ref{B2} we summarize the RMSD of various models compared to our compilation of observations across the redshift range $z=0.23-8.0$. At first we limit the comparison at the low mass end and intermediate mass galaxies (i.e. we exclude the high mass end). At lower redshifts ($z\leq3$), the GAME$_0$ and GAME$_1$ models consistently rank among the most precise models, with RMSD values below 0.1 dex in several cases. Notably, UniverseMachine also performs exceptionally well in this regime, though it is constrained to reproduce the GSMF at several redshifts (indicated by the letter 'C'). Reproducing the observations at these redshifts should not be considered as a prediction, but rather successful tuning of the parameters. Other models, such as EAGLE, Simba, and IllustrisTNG, generally show higher RMSD values, suggesting larger deviations from the observed GSMF. In the left panel of Fig. \ref{GSMF0obsRMSDuptoMstar} we present the evolution of the RSMD with redshift, dedicated for the low mass and intermediate mass galaxies.

At higher redshifts ($z>3$), the relative performance of the models varies significantly. GAME$_0$, GAME$_1$, and UniverseMachine continue to provide some of the most precise predictions, with UniverseMachine achieving particularly low RMSD values at $z=7$ and $z=8$. The continuity model from \citet{Leja2020} also performs well, especially at selected redshifts where it was constrained. Interestingly, the Simba simulation, which exhibits relatively high RMSD values at lower redshifts, shows competitive performance at $z=4$ and $z=5$. On the other hand, models like EAGLE, IllustrisTNG and ELUCID+SAM tend to exhibit the highest RMSD values across all redshifts, indicating some significant discrepancies with the observed GSMF of almost 0.5 dex. While certain models provide precise predictions at specific redshifts, the GAME models appear to offer consistently strong performance across a wide redshift range without being extensively constrained to fit the GSMF at particular epochs. In the left panel of Fig.  \ref{GSMF0obsRMSDuptoMstar} we present the average RMSD across all redshifts. 

GAME$_{0}$ and GAME$_1$ exhibit very impressive comparisons with the observations. The average RSMD among all redshifts show that the models are deviating only by $\sim 0.17$ $\sim$ 0.15 dex, respectively. This underscores the ability of GAME to involve the key elements governing the growth of low mass and intermediate mass galaxies at z = 0 to 8.

When including the high-mass end in the RMSD calculations (tables \ref{B5} and \ref{B6}), the performance of GAME$_0$ noticeably declines, particularly at higher redshifts. While GAME$_0$ remains among the most precise models at lower redshifts ($z<1.5$), its RMSD values increase significantly at $z>1.5$, exceeding 1 dex at $z=3.0$. This suggests that GAME$_0$ is more optimized for low-mass/intermediate-mass galaxies and struggles to accurately capture the high-mass end evolution in the early Universe. We note that reproducing the total GSMF {\it for 9.5 billion years (i.e., at $z = 0-1.5$)} should not be overlooked. As we discuss in section \ref{Comparison0}, this is a valuable realization and can have severe implications for how feedback in high-mass galaxies is treated. However, the strong increase in RMSD for GAME$_0$ at high redshifts suggests that a pure $\Gamma$ growth pattern cannot fully capture the processes governing massive galaxy growth, highlighting the need for improved modeling of high-mass systems in the early Universe. In contrast, GAME$_1$ demonstrates much better consistency across redshifts (including $z > 1.5$). Simba and IllustrisTNG perform well at certain redshifts, while UniverseMachine and \citet{Leja2020} remain among the most precise models. Considering $z = 0-8$, the average performance of GAME$_1$ for the total GSMF has an average RMSD of 0.27 dex. This is not only within the uncertainties of observations (0.3 dex) but also makes GAME$_1$ one of the best-performing models considered in this work, second only to UniverseMachine. This success was achieved using a particualrely low number of parameters and limited tuning with respect observations.

\begin{table*}
    \centering
    \caption{Root Mean Square Distance (RMSD) of different models with respect our compilation of observations at different redshifts (z=0.23-3.0) up to the characteristic mass $M^*$ (i.e. low mass end+intermediate galaxies included). The models with the lowest RMSD are the most {\it precise} in reproducing the observations. The two models providing the most precise predictions at each redshift (i.e. results that are not constrained) with respect observations are in bold. We remind that UniverseMachine was tuned to reproduce the GSMF at z = 0-4 while the continuity model of  \citet{Leja2020} at z =0.2, 1.6, 3.0, 5 and 8.  If a model is constrained to reproduce the the GSMF at a specific redshift ($\pm 0.1$) this is noted by the letter C.  We remind that uncertainties of 0.3 dex can be attributed to the uncertainty of observations.}
    \begin{tabular}{lccccccc}
        \hline
        Model & $z=0.23$ & $z=0.4$ & $z=0.615$ & $z=0.9$ & $z=1.5$ & $z=2.0$ & $z=3.0$  \\
        \hline
        \textbf{GAME$_0$} & \textbf{0.057} & \textbf{0.089} & \textbf{0.071} & \textbf{0.067} & \textbf{0.183} & 0.182 & 0.209 \\
        UniverseMachine & 0.057 (C) & 0.078 (C) & 0.082 (C) & 0.056 (C) & 0.070 (C) & 0.106 (C) & 0.130 (C)  \\
        \textbf{GAME$_1$} & \textbf{0.053} & \textbf{0.099} & \textbf{0.072}  & \textbf{0.097} & \textbf{0.112} & 0.210 & \textbf{0.207} \\
        EAGLE & 0.120 & 0.206 & 0.164 & 0.168 & 0.219 & 0.460 & 0.346 \\
        Simba & 0.251 & 0.161 & 0.218  & 0.111 & 0.217 & \textbf{0.095} & \textbf{0.179} \\
        IllustrisTNG & 0.103 & 0.142 & 0.107  & 0.140 & 0.199 & 0.317 & 0.337 \\
        ELUCID+SAM & 0.118 & 0.236 &  0.202 & 0.298 & 0.369 & 0.422 & 0.432 \\
        Leja et al. 2020 & 0.057 (C) & 0.245 & 0.130 & 0.137 & 0.098 (C) & \textbf{0.110}  & 0.113 (C)  \\
        \hline
    \end{tabular}
\label{B1}
\end{table*}

\begin{table*}
    \centering
    \caption{Continuation of table \ref{B1} at z=4.0-8.0.}
    \begin{tabular}{lcccccccc}
        \hline
        Model & $z=4.0$ & $z=5.0$ & $z=6.0$ & $z=7.0$ & $z=8.0$ \\
        \hline
        GAME$_0$ & \textbf{0.177} & 0.234 & 0.275 & 0.2075 & 0.297 \\
        \textbf{UniverseMachine} & 0.213 (C) & 0.168 & 0.137 & \textbf{0.171} & \textbf{0.154} \\
        \textbf{GAME$_1$} & 0.217 & \textbf{0.146} & \textbf{0.078} & 0.198 & 0.265 (C) \\
        EAGLE & 0.487 & 0.465 & 0.519 &  0.533 &  0.443 \\
        Simba & \textbf{0.196} & \textbf{0.151} & 0.256 & 0.776 &  0.706 \\
        IllustrisTNG & 0.3793 & 0.2904 & 0.2389 & 0.4588 & \textbf{0.1577} \\
        ELUCID+SAM & 0.587 & 0.235 & 0.190 & 0.193 & 0.255 \\
        Leja et al. 2020 & 0.328 & 0.096 (C) & \textbf{0.126} & \textbf{0.090} & 0.277 (C) \\
        \hline
    \end{tabular}
\label{B2}
\end{table*}

\begin{table*}
    \centering
    \caption{Root Mean Square Distance (RMSD) at different redshifts  (z=0.23-3.0), including the high mass end. The two models providing the most precise predictions (i.e. results that are not constrained) with respect to observations are in bold. If a model is constrained to reproduce the GSMF at a specific redshift  ($\pm 0.1$) this is noted by the letter C. We remind that uncertainties of 0.3 dex can be attributed to the uncertainty of observations.}
    \begin{tabular}{lccccccc}
        \hline
        Model & $z=0.23$ & $z=0.4$ & $z=0.615$ & $z=0.9$ & $z=1.5$ & $z=2.0$ & $z =3.0 $\\
        \hline
        GAME$_0$ & \textbf{0.147} & \textbf{0.108} & \textbf{0.145} & \textbf{0.067} & 0.600 & 0.608 & 1.201\\
        UniverseMachine & 0.168 (C) & 0.083 (C) & 0.088 (C) & 0.082 (C) & 0.079 (C) & 0.105 (C) &  0.247 (C) \\
        \textbf{GAME$_1$} & 0.185 & \textbf{0.125} & 0.176 & \textbf{0.091} & 0.347 & 0.344 & 0.488 \\
        EAGLE & 0.177 & 0.198 & 0.187 & 0.195 &  0.270 & 0.437 &  \textbf{0.332} \\
        \textbf{Simba} & 0.329 & 0.156 & 0.236 & 0.146 & \textbf{0.228} & \textbf{0.103} &  \textbf{0.331} \\
        IllustrisTNG & 0.333 & 0.178 & 0.236 & 0.188 & \textbf{0.184} & 0.301 &  0.424 \\
        ELUCID+SAM & \textbf{0.152} & 0.250 & 0.238 & 0.421 & 0.602 & 0.822 & 1.314 \\
        Leja et al. 2020 & 0.341 (C) & 0.254 & \textbf{0.152} & 0.202  & 0.134 (C) & \textbf{0.133} &  0.186 (C) \\
        \hline
    \end{tabular}
\label{B3}
\end{table*}

\begin{table*}
    \centering
    \caption{Continuation of table \ref{B5} at z=4.0-8.0}
    \begin{tabular}{lcccccccc}
        \hline
        Model & $z=4.0$ & $z=5.0$ & $z=6.0$ & $z=7.0$ & $z=8.0$ \\
        \hline
        GAME$_0$ & 0.794 & 1.26 & 0.956 & 1.744 & 1.294 \\
        \textbf{UniverseMachine} & 0.185 (C) & 0.423 & \textbf{0.145} & \textbf{0.176} & \textbf{0.140} \\
        \textbf{GAME$_1$} & \textbf{0.290} & \textbf{0.192} & 0.244 & \textbf{0.492} & 0.340 (C) \\
        EAGLE & 0.421 &  0.726 & 0.489 & 0.829 & 0.472 \\
        Simba & \textbf{0.194} & 0.444 & 0.226 & 1.209 & 0.578 \\
        IllustrisTNG & 0.355 &  0.527 & 0.219 & 0.706 & \textbf{0.389} \\
        ELUCID+SAM & 0.579 & 0.889 & 0.696 & 1.347 & 0.493 \\
        Leja et al. 2020 & 0.298 & 0.2010 (C) & \textbf{0.186} & 0.878 & 0.565 \\
        \hline
    \end{tabular}
\label{B4}
\end{table*}

\begin{figure*}
\centering
\includegraphics[scale=0.55]{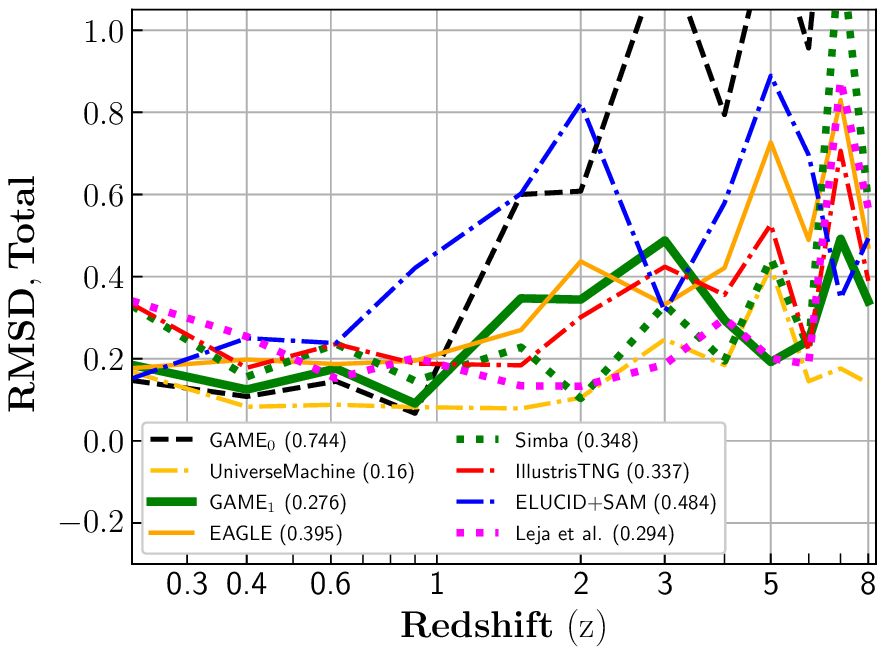}
\includegraphics[scale=0.55]{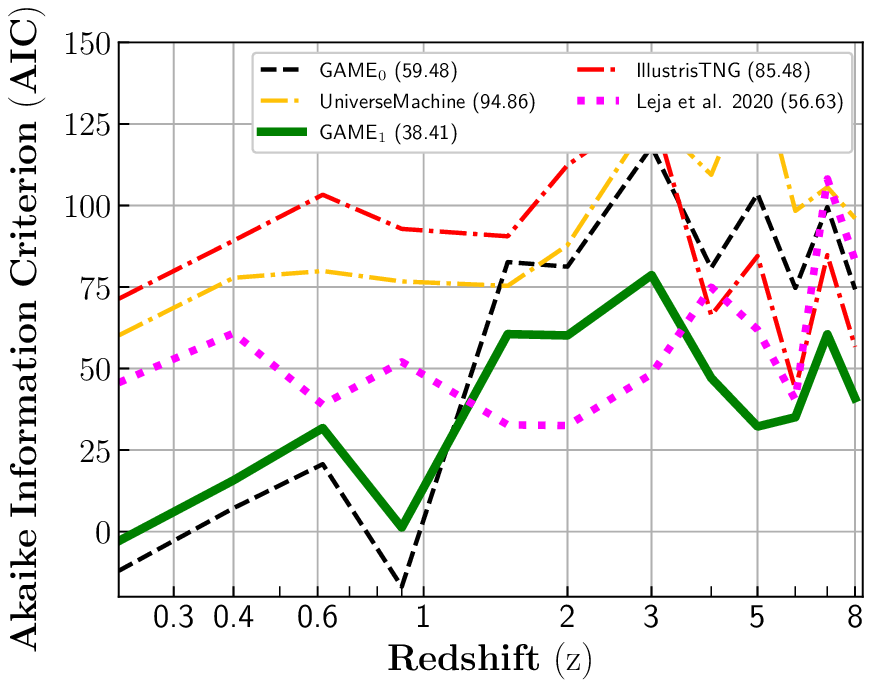}
\vspace{-1.0cm}
\caption{Left: Evolution of the RMSD of all galaxies considered (including the high mass end) of each model from z= 0.23 to 8.0. The average RMSD of all redshifts  is also noted. GAME$_0$ is not successful at reproducing the high mass end of the GSMF and a high RMSD is evident. GAME$_1$ (with average RMSD of 0.276 dex) reproduces the total observed GSMF (including the high mass end) with {\it high precision} on par with models which use a significantly higher number of parameters/complexity. Right. Akaike Information Criterion (AIC) used to evaluate the quality of the different models.  A low AIC value for a model means that it is able to achieve an optimal balance between model complexity and goodness-of-fit. Both GAME$_0$ and GAME$_1$ have the highest AIC quality among the models considered.}
\label{GSMF0obsRMSDALL}
\end{figure*}

\begin{table*}
    \centering
    \caption{Akaike Information Criterion (AIC) for  the total GSMF (low mass end+intermediate galaxies) at z=0.23-3.0 for models demonstrating good RMSD.  The free parameters used for each model are included in the parenthesis. The two models with the highest AIC quality (lowest AIC values) are in bold. }
    \begin{tabular}{lccccccc}
        \hline
        Model & $z=0.23$ & $z=0.4$ & $z=0.615$ & $z=0.9$ & $z=1.5$ & $z=2.0$ & $z =3.0 $ \\
        \hline
        \textbf{GAME$_0$} (3) & \textbf{-16.065} & \textbf{0.206} & \textbf{-8.324} & \textbf{-12.280} & 24.859 & \textbf{22.704} & \textbf{25.207} \\
        UniverseMachine (44) & 66.239 & 77.124 & 79.236 & 65.596 & 74.153 & 88.188 & 93.928 \\
        \textbf{GAME$_1$} (4) & \textbf{-16.817} & \textbf{5.902} & \textbf{-5.396} & \textbf{5.119} & \textbf{10.201} & 31.520 & 27.000 \\
        IllustrisTNG (33) & 65.132 & 76.773 & 66.793 & 76.225 & 87.766 & 103.546 & 98.609 \\
        Leja et al. 2020 (11) & 41.283 & 52.533 & 29.747 & 31.573 & \textbf{19.492} & \textbf{23.521} & \textbf{24.008} \\
        \hline
    \end{tabular}
\label{B5}
\end{table*}

\begin{table*}
    \centering
    \caption{Continuation of table \ref{B3} at $z=4.0-8.0$.}
    \begin{tabular}{lcccccc}
        \hline
        Model & $z=4.0$ & $z=5.0$ & $z=6.0$ & $z=7.0$ & $z=8.0$ \\
        \hline
        \textbf{GAME$_0$} (3) & \textbf{18.683} & \textbf{23.251} & \textbf{27.094} & \textbf{17.605} & \textbf{22.706} \\
        UniverseMachine (44)  & 107.729 & 99.189 & 94.393 & 97.739 & 94.846 \\
        \textbf{GAME$_1$} (4)  & \textbf{28.363} & \textbf{15.962} & \textbf{0.994} & \textbf{20.717} & \textbf{24.630} \\
        IllustrisTNG (33)  & 55.939 & 44.398 & 39.717 & 49.478 & 27.309 \\
        Leja et al. 2020 (18)  & 67.831 & 33.734 & 26.321 & 32.993 & 53.499 \\
        \hline
    \end{tabular}
\label{B6}
\end{table*}

\begin{table*}
    \centering
    \caption{Akaike Information Criterion (AIC) for  the total GSMF (including the high mass end) at z=0.23-3.0 for models demonstrating good RMSD.  The free parameters used for each model are included in the parenthesis. The two models with the highest AIC quality are in bold. }
    \begin{tabular}{lccccccc}
        \hline
        Model & $z=0.23$ & $z=0.4$ & $z=0.615$ & $z=0.9$ & $z=1.5$ & $z=2.0$ & $z =3.0 $ \\
        \hline
        \textbf{GAME$_0$} (3) & \textbf{-12.034} & \textbf{7.189} & \textbf{20.677} & \textbf{-16.895} & 82.650 & 81.227 & 118.043 \\
        UniverseMachine (44) & 60.165 & 77.814 & 79.889 & 76.701 & 75.347 & 87.810 & 127.386 \\
        \textbf{GAME$_1$} (4) & \textbf{-2.901} & \textbf{15.725} & \textbf{31.743} & \textbf{1.243} & \textbf{60.555} & \textbf{60.147} & \textbf{78.657} \\
        IllustrisTNG (33) & 71.381 & 89.194 & 103.294 & 92.818 & 90.549 & 112.360 & 130.262 \\
          Leja et al. 2020 (11) & 45.662 & 60.837 & 38.905 & 52.056 & \textbf{32.704} & \textbf{32.496} & \textbf{48.316} \\
        \hline
    \end{tabular}
\label{B7}
\end{table*}

\begin{table*}
    \centering
    \caption{Continuation of table \ref{B7} at $z=4.0-8.0$.}
    \begin{tabular}{lcccccc}
        \hline
        Model & $z=4.0$ & $z=5.0$ & $z=6.0$ & $z=7.0$ & $z=8.0$ \\
        \hline
        GAME$_0$ (3) & 80.873 & 103.557 & 74.679 & 99.517 & 74.311 \\
        UniverseMachine (44) & 109.418 & 143.781 & 98.347 & 105.594 & 96.074 \\
        \textbf{GAME$_1$} (4) & \textbf{47.093} & \textbf{32.208} & \textbf{35.064} & \textbf{60.510} & \textbf{40.915} \\
         \textbf{IllustrisTNG} (33) & 66.315 & 84.508 & 43.578 & \textbf{84.782} & \textbf{56.705} \\
        Leja et al. 2020 (18) & \textbf{74.942} & \textbf{61.934} & \textbf{40.358} & 108.172 & 83.127 \\
        \hline
    \end{tabular}
\label{B8}
\end{table*}

\subsection{Using the AIC to evaluate the quality of a model.}
\label{AICquality}

When assessing a model's performance, the number of parameters/constraints used for its development must be taken into account. While the RMSD provides a measurement of precision, it alone is not sufficient to determine a model's success, as extensive parameter tuning to match observations can be used and this can obscure the key factors/patterns involved. The AIC serves as a valuable tool in this context, offering a widely used metric \citep{Bozdogan1987,Cavanaugh2019,Portet2020} that quantifies the balance between precision and model complexity (parameters employed, necessary for tuning). The AIC is computed using the formula:

\begin{eqnarray}
\label{AIC}
\text{AIC} = -2 \ln(L) + 2k,
\end{eqnarray}
where \(L\) is the likelihood of the model, and \(k\) is the number of free parameters employed by the model. A lower AIC indicates a better description of the data. The term of $2k$  penalizes a model for having too many parameters, thereby preventing over-fitting/tuning with respect observations. In our analysis, we use the following approach to compute the AIC. The log-likelihood is given by:
\begin{eqnarray}
\label{AIC2}
\ln(L) = -\frac{n}{2} \left( \ln(2 \pi s^2) + 1 \right)
\end{eqnarray}
where \(n\) is the number of data points and $(s^2)$ is calculated from the residuals as $s^2 = mean(ln(model)- ln(observed))^2)$.

It is important to note for the AIC that the models are compared relative to a specific dataset (in our case the GSMF at a specific redshift) but unlike RMSD values, which have direct interpretability, the absolute AIC values are not meaningful on their own. Instead, we are interested about the comparative differences between the different models considered.  The question that the AIC is answering is: From the models of interest which one contains the necessary elements to explain the data, using a sensible number of parameters ?  Models with the lowest AIC values demonstrate a balance between precision and complexity. 

According to our analysis GAME$_0$ (tables \ref{B3} and \ref{B4}) demonstrates strong AIC values, for the low mass-end and intermediate mass galaxies from z = 0 to 8.  Similarly, GAME$_1$ maintains also strong AIC performance across the whole redshift range and shows particularly competitive AIC suggesting that it offers a robust description for the data, with only a marginal increase in complexity due to one additional parameter.  The two GAME models have the lowest AIC values (i.e. highest AIC quality) and this is also reflected at the right panel of Fig. \ref{GSMF0obsRMSDuptoMstar}. The model by \citet{Leja2020} also stands out at higher redshifts, with some of the lowest AIC values observed, particularly at z = 1.5 - 3.0. The model was tuned to reproduce the observations at this redshifts but this was well-suited to capture the evolution of the galaxy stellar mass function with a relatively sensible number of parameters. 

In contrast, UniverseMachine generally presents significantly  worse AIC values across all redshifts, which could be indicative of over-fitting, given the high number of free parameters (44). For example, at z = 4.0-8.0, the AIC values for UniverseMachine show a significant decline, suggesting that the model is able to capture precisely the observed evolution of the GSMF (demonstrated by the RMSD in subsection \ref{RMSDPrecision}) but at an expense of adding a large number of parameters. Surprisingly, GAME$_{0}$ actually demonstrates a more reasonable combination of goodness of fit and complexity at z = 0-8.  We note that UniverseMachine is able to reproduce many other observations beyond the GSFM, including quenched fractions and the relationship between dark matter halos and SFH. Our analysis in this appendix just demonstrates that the GSMF can be captured reasonably with far less parameters and that is why GAME$_0$ and GAME$_1$ have better AIC values. IllustrisTNG demonstrates bad AIC values (similar to UniverseMachine) at low redshifts but its performance increases at $z>3$.  Here we need to state that IllustrisTNG is a state-of-the art simulation that has been used to study many properties of galaxies, including the SMBH evolution and gas contents of galaxies.  However, the 33 parameters employed were mostly focusing on reproducing the GSMF at z = 0 {\it and} the CSFRD at z = 0-10.  Our work just suggests that the evolution of the GSMF can be reproduced using far less complexity.

According to our analysis present in  Fig. \ref{GSMF0obsRMSDALL}, table \ref{B7} and table  \ref{B8}, GAME$_{0}$ provides still a competitive AIC , even when high mass galaxies are included. However, GAME$_{1}$ is a significant improvement and the additional parameter considered  is well justified.  We note that the model of \citet{Leja2020} has a good AIC value, similar to that of GAME$_{0}$. In summary, our statistical analysis reveals that GAME$_0$ and GAME$_1$ offer an excellent description of the evolution of the GSMF at z = 0-8  balancing well between model complexity and precision. GAME$_0$ performs particularly well at $z <1.5$ and low mass/intermediate mass galaxies. GAME$_1$ provides a more consistent performance across the entire range of masses and redshifts. Further analysis and modifications of GAME will be considered in future work to reproduce more properties and statistics for galaxies.

\label{lastpage}
\end{document}